\documentclass[twocolumn, showpacs,prb,amsmath,amssymb,amsthm,superscriptaddress]{revtex4}

\pdfoutput=1

\usepackage{graphicx}% Include figure files
\usepackage{bm}% bold math
\usepackage{bbm}
\usepackage{color}

\newtheorem{definition}{Definition}
\newtheorem{theorem}{Theorem}
\newtheorem{lemma}{Lemma}

\begin{document}

\title{Quantum Decoherence of the Central Spin in a Sparse System of Dipolar Coupled Spins}
\author{Wayne M. \surname{Witzel} \email{wwitzel@sandia.gov}}
\affiliation{Sandia National Laboratories, New Mexico 87185 USA}
\author{Malcolm S. \surname{Carroll}}
\affiliation{Sandia National Laboratories, New Mexico 87185 USA}
\author{{\L}ukasz \surname{Cywi{\'n}ski}}
%\affiliation{Condensed Matter Theory Center, Department of Physics, University of Maryland, College Park, Maryland 20742-4111, USA}
\affiliation{Institute of Physics, Polish Academy of Sciences,
 Al.~Lotnik{\'o}w 32/46, PL 02-668 Warszawa, Poland}
\author{S. \surname{Das Sarma}}
\affiliation{Condensed Matter Theory Center, Department of Physics,
University of Maryland, College Park, Maryland 20742-4111, USA}

\begin{abstract}
The central spin decoherence problem has been researched for over 50 years in the context of both nuclear magnetic resonance and electron spin resonance. Until recently, theoretical models have employed phenomenological stochastic descriptions of the bath-induced noise. During the last few years, cluster expansion methods have provided a microscopic, quantum theory to study the spectral diffusion of a central spin. These methods have proven to be very accurate and efficient for problems of nuclear-induced electron spin decoherence in which hyperfine interactions with the central electron spin are much stronger than dipolar interactions among the nuclei.  We provide an in-depth study of central spin decoherence for a canonical scale-invariant all-dipolar spin system.  We show how cluster methods may be adapted to treat this problem in which central and bath spin interactions are of comparable strength. 
Our extensive numerical work shows that a properly modified cluster theory is convergent for this problem even as simple perturbative arguments begin to break down.
By treating clusters in the presence of energy detunings due to the long-range (diagonal) dipolar interactions of the surrounding environment and carefully averaging the effects over different spin states,
we find that the nontrivial flip-flop dynamics among the spins becomes
effectively localized by disorder in the energy splittings of the
spins. This localization effect allows for a robust calculation of the
spin echo signal in a dipolarly coupled bath of spins of the same kind, while considering clusters of no more than six spins. 
We connect these microscopic calculation results to the existing stochastic models.  We, furthermore, present calculations for a series of related problems of interest for candidate solid state quantum bits including donors and quantum dots in silicon as well as nitrogen-vacancy centers in diamond.
\end{abstract}
\pacs{
03.65.Yz; % decoherence
76.30.-v; % Electron paramagnetic resonance and relaxation
76.60.Lz; % spin echoes
03.67.Lx % quantum computation architectures and implementations 
}
\maketitle

\section{Introduction}
% Some history of spectral diffusion
The problem of decoherence of a spin interacting with a bath of spins (the ``central spin problem'') has its roots in classic works on electron and nuclear magnetic Resonance (see Ref.~\onlinecite{deSousa_PRB03} and references cited therein). In these early works, the dynamics of an ensemble of spins being resonant with external control field (spin species A), and interacting with a larger ensemble of off-resonant spins (species B), was considered. The fluctuations of the B spins (due to their mutual spin-spin interactions and due to spin-lattice relaxation) leads to precession frequency fluctuations of the A spins (the \emph{spectral diffusion}), which were then modeled as a classical stochastic process. Spin echo (SE) signals of A spins were calculated using different assumptions about the statistical properties of this process.\cite{Klauder_PR62,Chiba_JPSJ72,Zhidomirov_JETP69}  

% Intro to current motivation for working on this problem
The central spin decoherence problem has received renewed attention
due to emergence of ideas for using localized spins in solid state
systems  as qubits in a quantum computer. The currently studied
systems include gate-defined quantum dots,\cite{Hanson_RMP07}
self-assembled quantum dots,\cite{Liu_AP10,Morton_ARCMP11} phosphorous
donors in silicon,\cite{Morton_ARCMP11} and nitrogen-vacancy (NV)
centers in diamond.\cite{Wrachtrup_JPC06} In all of these systems, the
coupling of the central (qubit) spin to a bath of other spins is the
dominant process of the loss of coherence in a  superposition of spin
up and down states (i.e., dephasing). 

% Hyperfine-induced decoherence in Si:P and dots
A lot of attention has been recently devoted to the problem in which
the electron spin is coupled by a contact hyperfine (hf) interaction
to a bath of nuclear spins. For large magnetic fields only the
longitudinal  part of this interaction should be relevant (due to a
large Zeeman splitting mismatch between the electron and nuclear spins
suppressing their mutual flip-flops), and the decoherence of the qubit should occur due to intrinsic fluctuations of the nuclear spins caused by their mutual dipolar coupling.
Quantitative comparison between
theory\cite{deSousa_PRB03,Witzel_PRB05,Witzel_PRB06,Saikin_PRB07} and
experiments for spin echo in Si:P (see
Refs.~\onlinecite{Tyryshkin_PRB03,Abe_PRB04,Tyryshkin_JPC06,Abe_PRB10})
and Si:Bi (see Ref.~\onlinecite{George_PRL10}) systems has shown that
for natural concentration of spinful isotope of $^{29}$Si this is
indeed the case. The same origin of spin echo decay was predicted for
electron spins in III-V compound based quantum dots in the regime of
large magnetic fields.\cite{Witzel_PRB06,Yao_PRB06,Witzel_PRB08}
Recent experiments\cite{Bluhm_NP11} in GaAs singlet-triplet qubit
agree with these calculations for magnetic fields higher than $\sim \!
0.5$ T. At lower fields, the electron-nuclear spin flip-flops cannot be completely ignored, and  the SE decay is dominated by the contact hf interaction,\cite{Cywinski_PRL09,Cywinski_PRB09,Neder_PRB11} with dipolar dynamics being only a correction.

% Transition to all-dipolar systems: experimental status and a remark on relevance for qubit architectures
The case of a purely dipolarly coupled system, more closely analogous
to the original spectral diffusion problem, was also recently brought
back into focus by developments in spin qubit physics. One motivation
is the fact that silicon can be isotopically enriched to reduce the
concentration of spinful $^{29}$Si. Below a certain concentration
threshold, one can expect that the dipolar interactions between the
electron spins themselves will limit the coherence time. In fact it
was pointed out years ago that at large P concentrations the decay of
the observed SE signal of the donor-bound electrons might decay due to
dipolar interactions between these electron spins.\cite{Chiba_JPSJ72}
A theory addressing the range of currently studied small
concentrations of both P donors and the $^{29}$Si nuclei was proposed
recently.\cite{Witzel_PRL10} In that work, it was shown that (1) the SE
decay time in Si:P is bounded by a few seconds due to long-range
dipolar interactions between electron spins for realistically small
donor concentrations (about $10^{13}$ cm$^{-3}$) and (2) the presence of some $^{29}$Si can actually increase the $T_{2}$ time considerably by suppressing donor-induced decoherence. The latter effect is due to the nuclei providing quasi-static Overhauser shifts of electron spin splittings, which increase the detunings between the electron spins, and suppress the dipolar flip-flop dynamics in the bath.\cite{deSousa_considerations_PRB03} The predictions of Ref.~\onlinecite{Witzel_PRL10} for SE decay times have been recently confirmed experimentally.\cite{Tyryshkin_2011}

% A separate paragraph for NV center
Another system for which both the qubit-bath and the intrabath
  couplings are of the dipolar origin is the nitrogen-vacancy (NV)
  center in
  diamond.\cite{Wrachtrup_JPC06,Childress_Science06,Hanson_Science08,Balasubramanian_NM09,deLange_Science10,fuchs_quantum_2011,huang_observation_2011}
  In this case, decoherence of the qubit (which is made out of two
  levels of an electronic spin triplet) is dominated either by
  interaction with a bath of electron spins of nitrogen atoms
  (so-called P1 centers), as in
  Refs.~\onlinecite{Hanson_Science08} and \onlinecite{deLange_Science10}, or, in the
  case of purer samples, by interaction with a bath of nuclear spins
  of $^{13}$C atoms, as in
  Refs.~\onlinecite{Childress_Science06,Balasubramanian_NM09}, and \onlinecite{huang_observation_2011}.
% electron bath: T2^*~ 300-500 ns,  T_SE~3 microseconds (de Lange)
% 13C bath (high-purity diamond), T2*~2 microseconds, T_SE ~ 13 microseconds (Childress 06)
% Huang: 13C bath - "type IIa sample"
% Balasubramanian, G. et al. Ultralong spin coherence time i  isotopically engineered diamond. Nature Mater. 8, 383-387 (2009).  -  T_SE = 1.8 ms 

% More on modern theories of spectral diffusion
In this paper we present a detailed description of a cluster-based
theory applicable to a sparse dipolarly coupled
system,\cite{Witzel_PRL10} and we give multiple examples of
applications of the theory. In order to put this work into context, let us
briefly review the modern \emph{microscopic}
approaches to spectral diffusion
(for an attempt at pedagogical introduction to these theories see Ref.~\onlinecite{Cywinski_APPA11}). 
A method using a \emph{cluster expansion} of bath dynamics was developed in
Refs.~\onlinecite{Witzel_PRB05} and \onlinecite{Witzel_PRB06} and applied
in the context of spin qubit decoherence in semiconductors.  This
theory produced
results in remarkable agreement with experimental spin echo decay
measurements using only well-known microscopic (no fitting)
parameters.~\cite{Witzel_AHF_PRB07}
Various theories of this type have been applied to
problems in which an electron spin decoheres due to contact hf
interactions with a dynamical nuclear spin bath.\cite{Witzel_PRB05, 
Witzel_PRB06,Yao_PRB06, Yao_PRL07,Liu_NJP07,Witzel_AHF_PRB07,Maze_PRB08}
In all of these works based upon cluster expansions of some form, 
the contributions of the bath dynamics to central spin
decoherence were grouped according to the number of bath spins
participating in a nontrivial way (e.g., undergoing flip-flop
processes). 
In all of these nuclear-induced spectral diffusion problems, the coupling
of the central electron spin to nuclear spins is typically much larger
than dipolar interactions that couple the nuclear spins to each other.
The cluster expansions are essentially perturbative expansions in the
intra-bath coupling, (related to a diagrammatic linked cluster
expansion\cite{Saikin_PRB07} but less cumbersome to compute
numerically) and are therefore well suited to problems in which these
interactions are relatively weak. 
Problems in which the
interactions among bath spins are comparable to their interactions
with the central spin (e.g., sparse, dipolar coupled electron spins)
present a challenge for these cluster methods.  In this article, we
show that we can adapt the cluster correlation expansion (CCE) of 
Refs.~\onlinecite{Yang_CCE_PRB08} and \onlinecite{Yang_CCE_PRB09} to treat these
problems successfully.  
The CCE of
Refs.~\onlinecite{Yang_CCE_PRB08} and \onlinecite{Yang_CCE_PRB09}
is essentially equivalent to the original cluster
expansion~\cite{Witzel_PRB05, Witzel_PRB06} but greatly simplified and
more convenient for considering large cluster corrections.
This method was recently applied to the NV center coupled to the nuclear spin
bath where it was used to
predict interesting effects related to the qubit back-action on the
bath dynamics.\cite{Zhao_PRL11,Zhao_PRB12,huang_observation_2011}
Without relying upon the large-bath approximations or
cumbersome corrections of the earlier cluster expansion
theory,\cite{Witzel_PRB06} or relying upon the clustered
grouping approximation of the disjoint cluster approach,\cite{Maze_PRB08} the CCE is
 well suited for including larger spin clusters.
These larger clusters need to be calculated when considering dynamical
decoupling of the central spin\cite{Witzel_PRL07,Lee_PRL08}, or when the bath is
sparse and multi-spin correlations build within it on the timescale
of the central spin decoherence. The latter case applies to the problem
that is the focus of this article.

% Importance of establishing a theory for a sparse bath
It may appear at first that the cluster expansion, which depends
critically on the higher order clusters making systematically weaker
contributions to decoherence in a parametrically well-behaved manner,
would be completely impractical for problems involving a sparse bath
of environmental spins and/or a bath environment containing similar
spins to the central spin. One may wonder that in either case
(i.e., sparse bath or qubit-bath interaction being the same as the
intra-bath interaction), it may simply be impossible to define
``clusters'' in any meaningful manner for a reasonable cluster
expansion technique to work.  In fact, this has inhibited the
application of the cluster expansion technique, in spite of its great
success in the standard spectral diffusion problem of spin decoherence
in Si and GaAs, to a number of important problems of increasing
experimental importance.  In the current work, we establish the
applicability of a cluster expansion technique for the central spin
decoherence problem for a sparse bath with strong intra-bath couplings
(comparable to bath-qubit couplings).  
Once such a theoretical technique is established, we can
then solve a number of central spin quantum decoherence problems of
current experimental relevance using it, and we apply the technique to
solve several problems of interest in Si and diamond quantum computing
architectures.

% First mention of the localization effect:
The key insight, which follows from a careful implementation of the
CCE to the sparse dipolar bath and from extensive numerical
calculations involving increasing cluster sizes, is our finding of an
effect of localization of flip-flop dynamics of bath spins that is not
obvious \emph{a priori}.
For any given group (cluster) of spins, their mutual energy detunings
are affected by the state of all the other spins, outside of the
cluster.  This is due to the diagonal ($S^{z}_{i} S^{z}_{j}$) part of
the dipolar interaction.  Our calculations show that in a sparse
dipolarly coupled bath these interactions are introducing strong
disorder in the energy splittings of bath spins.  This disorder 
suppresses the contribution from larger cluster sizes in an appropriately 
modified CCE.  We find that the CCE can be adapted to converge well for a sparse dipolar bath by defining cluster contributions to include these externally-induced energy splittings and to be effectively, but efficiently, averaged over internal and external spin states.  (We note that Ref.~\onlinecite{Yao_PRB06} had previously included effects of externally-induced energy splittings in its pair approximation).
The results and the convergence can be strongly dependent upon the arrangement of bath spins, but the convergence of results that are averaged over different spatial realizations of the bath is
 well controlled. 
In our modified formulation of the CCE, we can obtain convergent
results for SE decay up to times at which the coherence had decayed by
an order of magnitude while calculating clusters of at most four spins
(with six spins clusters being shown to contribute a negligible correction at this timescale).

% Relation to Dobrovitski et al: classical noise model and the single qubit / ensemble of qubits distinction
Our main focus is thus the case of a central spin coupled to the bath
spins \emph{of the same kind} by dipolar interaction, e.g.~an electron
spin coupled to other electron spins. Such a situation has been
extensively studied theoretically by Dobrovitski {\it et al.} in
Refs.~\onlinecite{Dobrovitski_PRB08,Dobrovitski_PRL09,Hanson_Science08}
and \onlinecite{deLange_Science10}.
The most important conclusion of these papers, based on extensive
comparison between exact numerical calculations, stochastic model, and
diverse experiments for NV center coupled to electron spins, is that
the decoherence of a single NV center (coupled to an electron spin
bath) can be modeled very well by replacing the bath by a source of
classical Ornstein-Uhlenbeck
noise.\cite{Hanson_Science08,Dobrovitski_PRL09,deLange_Science10} For
spin echo this means decay of the $\exp[-(t/T_{SE})^3]$ form crossing
over at long times to $\exp[-t/T_{\text{long}}]$. An important
distinction\cite{Dobrovitski_PRB08} was also made between the results
of experiments on an ensemble of qubits, and results obtained by
repeated measurement of the same qubit (as it is done in experiments
on a single NV center). This distinction is very important for our
work here: we consider both cases (the ensemble of qubits, and a
single qubit), since the first of them is important for current
experiments on Si:P, while the second is relevant for current NV
center experiments as well as considerations for addressable quantum
bits.  It is important to note that these studies using exact numerics
are limited to very small bath sizes (tens of spins with current
computing technology) and that cluster expansions do not have that limitation.

The paper is organized in the following way. In
Sec.~\ref{Sec:Canonical}, we describe the Hamiltonian of the system of
interest, a central spin dipolarly coupled to an ensemble of randomly
positioned spins, all of them also coupled by dipolar interactions,
and define the coherence measurement procedure (the spin echo) which
our theory addresses. In Sec.~\ref{Sec:Methods}, we provide a detailed
description of a variant of cluster expansion theory applicable to
such a problem. There we give all the details of the theory used in
Ref.~\onlinecite{Witzel_PRL10} to predict the $^{29}$Si and P
concentration dependence of the electron spin coherence time is Si:P
system. In Sec.~\ref{Sec:Variants}, we describe many interesting and
experimentally relevant variations of the ``canonical'' problem
defined in Sec.~\ref{Sec:Canonical}. There we discuss the role of the
intra-bath coupling strength relative to their interactions with the
central spin, the possible geometrical variants of the problem (i.e., the
case of the bath spins being localized in a plane some distance from
the central spin), and generalizations of spin echo experiment to
sequences of multiple pulses (dynamical decoupling). Finally, in
Sec.~\ref{Sec:Applications}, we present theoretical results for many
example systems: donor-bound electrons in bulk silicon or near an interface, Si-based quantum dots, and NV centers in diamond.

\section{Canonical Problem: Decoherence in a Sparse All-Dipolar Spin System}
\label{Sec:Canonical}

In this section, we describe our canonical problem of interest, the spin echo 
decoherence of a central spin in a sparse all-dipolar spin-1/2 system in a strong, homogeneous magnetic field environment.  In this canonical problem, we assume that the central spin is shifted off of resonance from the spins of the bath, but relax this assumption for one of the variants in Sec.~\ref{Sec:Variants}.  Using spin-1/2 particles is relevant for applications to electron spin systems; however, our methods equally capable of treating spins of larger magnitude.

\subsection{System of spins}

For our canonical problem, we consider a sparse system of electron
spins uniformly distributed at random in a three-dimensional continuum
(or on a lattice in which the diluteness of the spins makes the
lattice structure irrelevant).  The electrons are localized (bound to
donors, for example) and dilute to the extent that they may be treated
as point dipoles.  In fact, it is not essential that they be
electrons, but we employ conventions of notation (such as $g$-factor) that are consistent with electron particles.  We also assume the existence of a uniform magnetic field that we choose, for convention, to lie along the ``z'' direction.  Our Hamiltonian, written in atomic units ($\hbar = 1$ and $1 / 4 \pi \epsilon_0 = 1$), is thus
\begin{equation}
\label{Eq:Ham}
\hat{\cal H} = \sum_i \mu_B g_i B_i \hat{S}^{z}_i + \mu_B^2 \sum_{j>i} g_i g_j
\hat{\bf S}_i \cdot {\bf D}({\bf R}_i -  {\bf R}_j) \cdot \hat{\bf S}_j,
\end{equation}
where $\hat{\bf S}_i$ are spin operators for the spin-1/2 particles,
$\mu_B$ is the Bohr magneton, $g_i$ is the g-factor of the
$i$th electron (typically, $g_i=2$), $B_i$ is the externally applied magnetic field at each electron site, and ${\bf D}({\bf r})$ is a tensor to characterize dipolar
interactions and is defined by
\begin{equation}
\label{Eq:Dip}
D_{\alpha, \beta}({\bf r}) =
\left[ \frac{\delta_{\alpha \beta} - 3 r_{\alpha} r_{\beta} / {\bf r}^2}{{\bf r}^{3}}\right],
\end{equation}
with $\alpha, \beta = x, y, z$.
$\delta_{\alpha \beta}$ is the Kronecker delta and $r_{\alpha}$ is the $\alpha$
vector component of ${\bf r}$.  Our convention is to index the central spin as $i=0$.

For our canonical problem, we set $g_i = 2$ and take the limit of a large applied
magnetic field that is equal among all bath spins but different for the central spin.  That is, all except the central spin are on resonance with each other (neglecting at this point the energy offsets due to dipolar interactions).
If our central spin represents a quantum bit, it makes sense that we would be able to address it individually and shift its Zeeman energy to be off resonant from the bath spins.
In taking this limit of a large applied magnetic field, we should disregard interactions which do not preserve the net Zeeman energy (the secular approximation).  Thus, our effective Hamiltonian for our canonical problem becomes
\begin{equation}
\label{Eq:HeffOffRes}
\hat{\cal H}_{\mbox{\scriptsize eff}} = \sum_{i, j > 0} b_{i, j} \hat{S}_i^+ \hat{S}_j^-
- 2 \sum_{i, j} b_{i, j} \hat{S}_i^z \hat{S}_j^z,
\end{equation}
with
\begin{equation}
\label{Eqn:bnm}
b_{i, j} = -\frac{1}{4} (g \mu_B)^2 \hbar \frac{1 - 3 \cos^2
  \theta_{ij}}{R^3_{ij}},
\end{equation}
where $\theta_{ij}$ is the angle that the vector from spin $i$ to spin $j$ makes with the ``z'' unit vector (the direction of applied magnetic field) and $R_{ij}$ is the length of this vector.
By forcing the magnetic field of the central spin to be different from
the rest and taking the large field limit, we suppress any flip-flopping between the central spin (with index zero by our convention) and bath spins.  This is, then, a standard spectral diffusion problem in which the polarization of the central spin is preserved but the qubit will dephase due to bath-induced variations of its precessional frequency (i.e., its spectral line ``diffuses'').  We later treat the case in which the central spin is resonant with the bath spins in Sec.~\ref{Sec:VariedCouplingAndTemp}.

Note that the effective Hamiltonian of Eq.~(\ref{Eq:HeffOffRes}) has a
1/$R^3$ dependence entirely.  The entire Hamiltonian, therefore,
scales with the concentration of the spins ($C_E$).  The dynamics is
therefore scale invariant, with time that scales inversely with the
$C_E$.  Likewise, time scales inversely with the square of the
g-factors.  Our results, unless otherwise specified, apply to a bath
of $C_E = 10^{13}/\mbox{cm}^3$ and $g=2$.  However, adjusting these parameters only serves to rescale the time axis.

\begin{figure}
\includegraphics[width=3in]{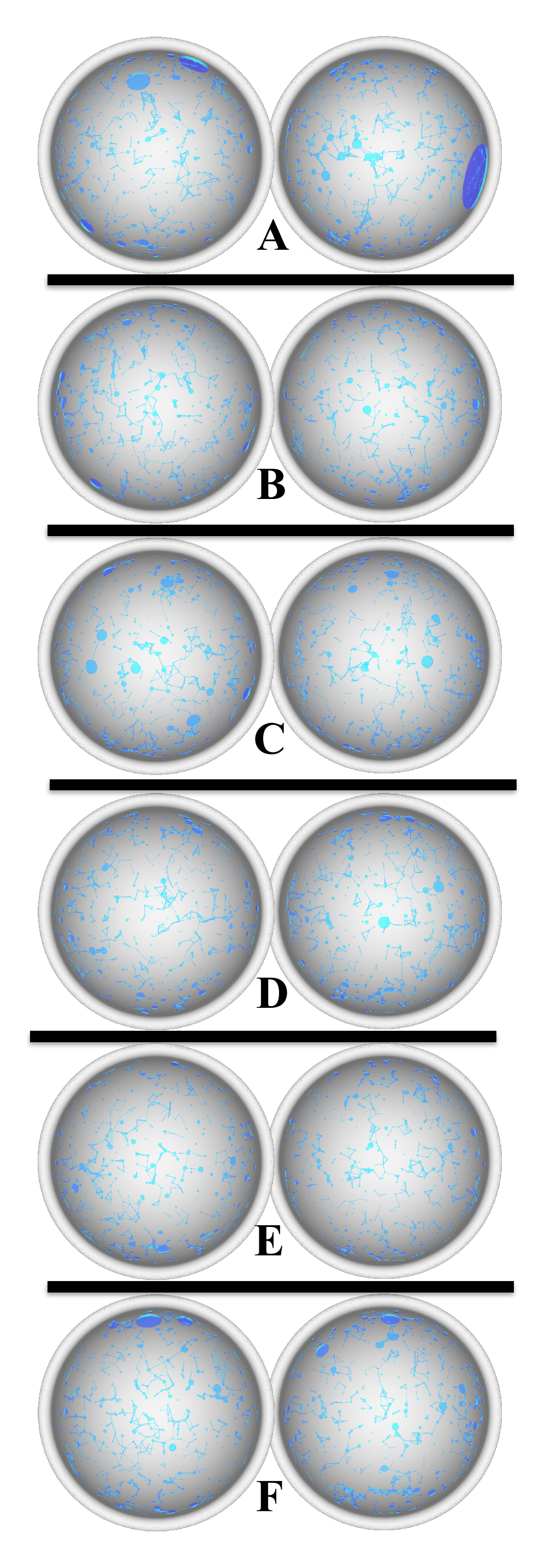}
\caption{
\label{Fig:SpatialConfigs}
``Celestial map'' representations of six randomly chosen spatial configuration instances of our canonical problem.  Each show left and right hemispheres with the applied magnetic field in the up (north) direction.
}
\end{figure}

Since we are specifying that the spins are at random positions in space, there are many spatial realizations of this problem.  As a way to visualize a particular instance of a spatial realization of the problem, we use ``celestial map'' diagrams as shown in Fig.~\ref{Fig:SpatialConfigs}.  These diagrams represent each particular ``universe'' from the perspective of a central spin.  Positions of the bath spins are projected onto a sphere (centered at the central spin) as cylinders whose size is proportional to the strength of the interaction with the central spin.  The left and right hemispheres are split apart so we can look out in any direction from the central spin.  We connect the representatives of the bath spins with rods in proportion to the strength of their mutual interaction as well as their interactions to the central spin (where these interactions are beyond some criteria in strength).  In these way, we get effective ``constellations'' of bath spins.  We chose six random instances labeled A through F depicted in Fig.~\ref{Fig:SpatialConfigs}, and will refer to these by letter throughout the text.

We consider the limit of an infinite bath temperature in which the initial bath state is random without bias.  We generate random initial bath states as product states of each bath spin being up or down with equal probability.  We consider the finite-temperature variant in Sec.~\ref{Sec:VariedCouplingAndTemp}.

\subsection{Spin echo}
\label{Sec:SpinEcho}

If one considers an ensemble of the spin systems described in the previous section, 
each with a different spatial configurations of spins and initial spin states, 
one would observe a relatively rapid dephasing occur simply due to the ensemble averaging.  
This inhomogeneous dephasing time is known as $T_2^*$.  The central spin of each
system would experience a different shift of precessional frequency as a result of
the magnetic field generated from its environment.  The standard approach to remove
this trivial effect is to apply refocusing pulses to the central spin.  The simplest 
of these is the Hahn spin echo in which one rotates the central spin by an angle of $\pi$ about an axis perpendicular to the applied magnetic field midway through the 
evolution.  A $\tau \rightarrow \pi \rightarrow \tau$ sequence, for example, will give a refocused signal at time $t = 2 \tau$.  We report, in the study of our canonical problem, 
the normalized spin echo as a function of the total time, $t$, of the sequence.  At $t=0$, no signal is lost and we report a spin echo value of one.  The general trend will be a decay of the spin echo as the pulse sequence time is increased.  We identify the dephasing time $T_{2}$ with the time at which  the signal reaches a value of $\exp{(-1)}$.

In our study, we assume that the central spin may be addressed individually, 
and that our refocusing pulses are instantaneous and ideal.  The methods we describe 
should be applicable to problems that relax these assumptions, and also consider other types of pulse sequences, but we choose to keep 
the problem simple in the scope of this work.  For ESR measurements in
which all of the spins are essentially resonant with each other, it is
possible to extrapolate the central spin echo decay by adjusting the
angle of the refocusing pulse.\cite{Tyryshkin_PRB03, Tyryshkin_2011}  Otherwise, the decay can be dominated by the inhomogeneous decay from the environment of like spins that are all flipped together (known as instantaneous diffusion).  By using a smaller angle in the refocusing pulse, the signal of the echo 
is reduced and harder to measure, but the effects of instantaneous diffusion are 
reduced.  Extrapolating to a refocusing pulse angle of zero yields the
proper spin echo decay.\cite{Tyryshkin_PRB03, Tyryshkin_2011}
% cite also Abe ?

\section{Methods}
\label{Sec:Methods}

In this section, we describe methods we use for solving the central spin decoherence problem with a particular focus on our canonical problem.
We start, in Sec.~\ref{SecCCE}, with a general description of the CCE method.  We show, through examples of instances of our canonical problem, a need to modify the original expansion, and demonstrate effective techniques that overcome the arising difficulties.

\subsection{Cluster correlation expansion}
\label{SecCCE}

Explicitly evolving bath states becomes infeasible even for baths of
moderate size,\cite{Zhang_JPC07} with number of bath spins $> \! 20$.
Recently developed cluster
techniques,\cite{Witzel_PRB05,Witzel_PRB06,Yang_CCE_PRB08, Yang_CCE_PRB09} however, can make such evaluations
possible by breaking up the problem into smaller pieces.  Our approach
will use the CCE method.
This was developed to resolve deficiencies of previously developed
spin decoherence cluster expansion
techniques\cite{Witzel_PRB05,Witzel_PRB06} in small
bath scenerios.  In hindsight, the formulation of
Refs.~\onlinecite{Witzel_PRB05} and \onlinecite{Witzel_PRB06} should be viewed as a large bath
approximation to the CCE that may be convenient where applicable (in principle, the
approximation may be systematically corrected, but if corrections are
necessary than one is better off using the CCE formalism).
% cite also Yao_PRB06 at least once (for PCA ~ two-spin clusters_

The CCE has a simple and easily generalized formulation.  In principle,
it is always exact in the large cluster limit (apart from division by
zero situations that may arise).  In practice, the expansion converges
best for sufficiently short simulation times but becomes numerically
instable for long simulation times.
Let $L$ denote the bath averaged quantity of interest.
For the spin echo dephasing problem, we choose $L = \rho_q^{+-}(t) / \rho_q^{+-}(0)$, where $\rho_q^{+-}(t)$ is the off-diagonal component of the reduced density matrix for the central spin after evolving a $t = 2 \tau$ spin echo sequence.  Since
it is not feasible to solve this directly for a moderately-sized bath,
let us define $L_{\cal S}$, where ${\cal S}$ is any subset of bath
spins, as the result of $\rho_q^{+-}(t) / \rho_q^{+-}(0)$ as computed when we
only involve spins outside of set ${\cal S}$
in a trivial manner (to be explained below).  Here we are choosing to be more general than the
original derivation of the CCE and only require that $L_{\cal S}
= L$ when ${\cal S}$ includes the full set of bath spins and leave
some flexibility in the way that we define $L_{\cal S}$ for smaller
sets.  We refer to a given
set of bath spins, ${\cal S}$, as a cluster although there is no
requirement that the constituent spins of the set necessarily be
closely spaced (or clustered).

At this point, the CCE formulation will approximate $L$ in terms of the $L_{\cal S}$
for various ${\cal S}$ up to some maximum ``cluster'' size and will be
exact when the maximum size limit reaches the size of the bath.  To do
this, we will implicitly define $\tilde{L}_{\cal S}$ such that
\begin{equation}
\label{CCEeqns}
L = \prod_{\cal S} \tilde{L}_{\cal S},~~ L_{\cal S} = \prod_{{\cal C} \subseteq {\cal S}} \tilde{L}_{\cal C} = \tilde{L}_{\cal S} \prod_{{\cal C} \subset {\cal S}} \tilde{L}_{\cal C},
\end{equation}
with products over {\it all} subsets meeting the specified criteria.
There will be, therefore, overlapping clusters.  These are not
disjoint sets as they are in the approach of Ref.~\onlinecite{Maze_PRB08}.
Explicitly, $\tilde{L}_{\cal S}$ is then
\begin{equation}
\label{DefTildeL}
\tilde{L}_{\cal S} = L_{\cal S} / \prod_{{\cal C} \subset {\cal S}} \tilde{L}_{\cal C}
\end{equation}
as a recursive definition for any $\tilde{L}_{\cal S}$.
This is simply a tautology that serves as the definition of 
$\tilde{L}_{\cal S}$.  It becomes useful when we can disregard the vast majority of the $\tilde{L}_{\cal S}$ factors for $L$ such as limiting the cluster size.  
We define the $k$th order of the CCE as the approximation of $L$ with a maximum cluster size $k$:
\begin{equation}
\label{CCEtruncated}
L_{\mbox{\scriptsize{CCE}}}^{(k)} = \prod_{\|{\cal S}\| \leq k} \tilde{L}_{\cal S}.
\end{equation}
Even though we are including overlapping sets of clusters, overcounting
of contributions are systematically corrected as increasingly large clusters are
included in the approximation.  We know this simply by the fact that
the CCE is exact when all clusters are included.

To get an intuition for why the CCE takes the form of 
a product of contributing factors, consider the idealized
scenario with our effective canonical Hamiltonian
[see Eq.~(\ref{Eq:HeffOffRes})] 
in which there are two sets of non-central spins, ${\cal A}$ and ${\cal B}$, with no cross
interactions (e.g., two clustered groupings that are far apart with 
negligible interactions to each other):
\begin{eqnarray}
\nonumber
\hat{\cal H}_{\mbox{\scriptsize eff}} &=& \sum_{i, j \in {\cal A}} b_{i,
  j} \hat{S}_i^+ \hat{S}_j^- - 2 b_{i, j} \hat{S}_i^z \hat{S}_j^z \\
&& {} + \sum_{i, j \in {\cal B}} b_{i, j} \hat{S}_i^+ \hat{S}_j^- - 2 b_{i, j} \hat{S}_i^z \hat{S}_j^z,
\end{eqnarray}
In this scenario,
with $L$ defined as $\rho_q^{+-}(t) / \rho_q^{+-}(0)$, we
may factorize $L$ as $L = L_{\cal A} \times L_{\cal B}$.
Beyond the ideal scenario, this factorization is only approximate but is
related to the perturbation theory discussed in 
Sec.~\ref{CCEconvergence}.

We report various spin echo results calculated using the CCE up to
various maximum cluster sizes: $L_{\mbox{\scriptsize{CCE}}}^{(k)}$.
In practice, we do not generally include all clusters of a given size in the calculations.  We use heuristics with cut-off parameters to select the clusters of the most potential importance.  We adjust the cut-off parameters until we are quite confident in our results.
Our cluster sampling heuristics are described in Appendix~\ref{clusterSampling}.

\subsection{Perturbation theory and the cluster correlation expansion}
\label{CCEconvergence}

The fact that $L_{\mbox{\scriptsize{CCE}}}^{(k)}$ equals $L$ in the
large $k$ limit is apparent from our derivation above.  But how well
does this expansion converge?  The key to understanding the
convergence properties of CCE is to
understand the properties of $\tilde{L}_{\cal C}$ with respect to a
perturbation in the interaction that couples the bath spins.  
Using perturbation theory, one may express $L_{\cal S}$ as an infinite power
series with respect to the coupling constants between bath spins,
denoted with $b_{i, j} = b_{j, i}$ (in reference to the
Hamiltonian of Sec.~\ref{Eqn:bnm}, but is general for any pairwise
interaction Hamiltonian and may be generalized for $n$-way
interactions).  
It is also possible to expand $\tilde{L}_{\cal C}$ into such a power
series by recursively expanding the denominators of Eq.~(\ref{DefTildeL}).  This expansion is possible
(with only positive powers of the coupling constants)
as long as $L_{\cal S}$, for any ${\cal S}$, is non-zero when all coupling constants are
taken to be zero; in the case of $L = \rho_q^{+-}(t)/\rho_q^{+-}(0)$
and using our effective Hamiltonian in Eq.~(\ref{Eq:HeffOffRes}), $L_{\cal S}=1 +
O(b_{i,j} t)$ and, by the recursive definition of Eq.~(\ref{DefTildeL}),
$\tilde{L}_{\cal C} = 1 + O(b_{i,j} t)$.

As defined and noted in Ref.~\onlinecite{Yang_CCE_PRB08}, $\tilde{L}_{\cal C} = 1
+ O(b_{i,j}^k t^k)$ with $k = \|{\cal C}\|$.  In our generalization, such
a result is conditional.  The following definition, lemma, and theorem
specify the conditional perturbative properties of $\tilde{L}_{\cal C}$.

\begin{definition}
\item We say that $L_{\cal S}$ is 
{\it factorable under disconnected interactions} if, for any ${\cal
  S}$, given ${\cal X}
\in {\cal S}$, ${\cal Y }\in {\cal S}$ such that ${\cal X} \cup {\cal
  Y} = {\cal S}$, ${\cal X} \cap {\cal
  Y} = \emptyset$ (disjoint), and that all $b_{i \in {\cal X}, j \in
  {\cal Y}} = 0$, then $L_{\cal S} = L_{\cal X} L_{\cal Y}$.
\end{definition}

\begin{figure}
\includegraphics[width=2.5in]{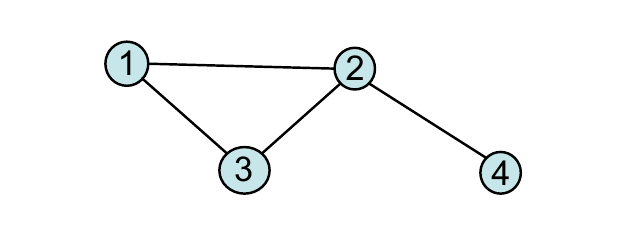}
\caption{
\label{connectedCluster}
Example of a connected cluster of Lemma~1 where edges represent the 
existence of $b_{i, j}$ factors of a given term of $\tilde{L}_{\cal C}$.
}
\end{figure}

\begin{lemma} If $L_{\cal S}$ is factorable under disconnected
interactions, all non-constant ($b_{i, j}$ dependent) terms
of $\tilde{L}_{\cal C}$ in a power expansion with respect to 
$b_{i, j}$, must contain factors of 
$b_{i, j}$ that, when viewed as graph edges between nodes $i$ and $j$,
connect all spin ``nodes'' in ${\cal C}$ fully.  In other words, each
non-constant term will involve all spins in ${\cal C}$ and may {\it not}
be factored into parts that involve disjoint, non-empty sets of spins (Fig.~\ref{connectedCluster}).
\end{lemma}

\begin{theorem}
If $L_{\cal S}$ is factorable under disconnected
interactions,
  $\tilde{L}_{\cal C} = \mbox{const.} + O(b_{i,
  j}^{k-1} t^{k-1})$ where $k = \|{\cal C}\|$ (the size of ${\cal C}$).
\end{theorem}
The above Theorem follows directly from the Lemma considering the
simple fact in graph theory that a set of $k$ nodes cannot
form a connected graph with less than $k-1$ edges; the Lemma is proven
in Appendix~\ref{ClusterFactorization}.
The lowest non-constant order will contain an additional factor of
$b_{i, j}$ under certain circumstances, so that $\tilde{L}_{\cal C} = \mbox{const.} + O(b_{i,
  j}^{k} t^k)$, but this is not an
important consideration for this discussion.
%in the case of the secular
%Hamiltonian with a bath density matrix composed of
%non-interacting product states (i.e., a bath without entanglement).
%This comes from the fact that, in each term of $L_{\cal S}$, each spin must
%be flipped an even number of times in order to return to its original
%state (e.g., $\langle \uparrow \vert \hat{S}_- \vert \uparrow
%\rangle = 0$, $\langle \uparrow \vert \hat{S}_- \hat{S}_+ \hat{S}_-
%\vert \uparrow \rangle = 0$, etc.).  In this case, Lemma~1 in 
%Appendix \ref{ClusterFactorization} will lead to the conclusion that 
%$\tilde{L}_{\cal C} = 1 + O(b_{i,j}^k)$, but this extra order of
%$b_{i, j}$ is not really important for this discussion.

The important point is that larger clusters will introduce higher order
corrections with respect to $b_{i, j}$.  This is the essence of the
CCE~\cite{Yang_CCE_PRB08} and
previous~\cite{Witzel_PRB05,Witzel_PRB06} cluster expansions.  
 However,
 $b_{i, j} t \ll 1$ is not a strict requirement for convergence; 
a localization effect in disordered systems has been demonstrated to 
extend convergence into the $b_{i, j} t > 1$ regime~\cite{Yang_CCE_PRB08}. 
This in fact happens in our canonical problem, albeit a few modifications of the CCE (explained in detail below) are necessary to capture this effect. In any case, the best practice is to test convergence by computing an extra order
in the expansion, providing an estimate of the error.

The CCE works extremely well out into the tail of the spin echo decay
in those central spin decoherence problems, such as nuclear-induced spectral diffusion,~\cite{Witzel_PRB05,Witzel_PRB06} in which the central spin is strongly coupled to many bath spins relative to the coupling strength among bath spins.  In this case, the decay timescale is typically short compared with interaction strength among the bath spins (i.e., the bath is slowly evolving).
Our canonical problem pushes us into the challenging regime of 
this perturbation  because the central 
spin has interactions to the bath spins that are comparable to the strength of the 
interactions among the bath spins.  This presents challenges to the cluster method, but we shall demonstrate effective techniques to address these challenges.

\subsection{Treatment of ``External'' Spins}

\begin{figure}
\includegraphics[width=3.5in]{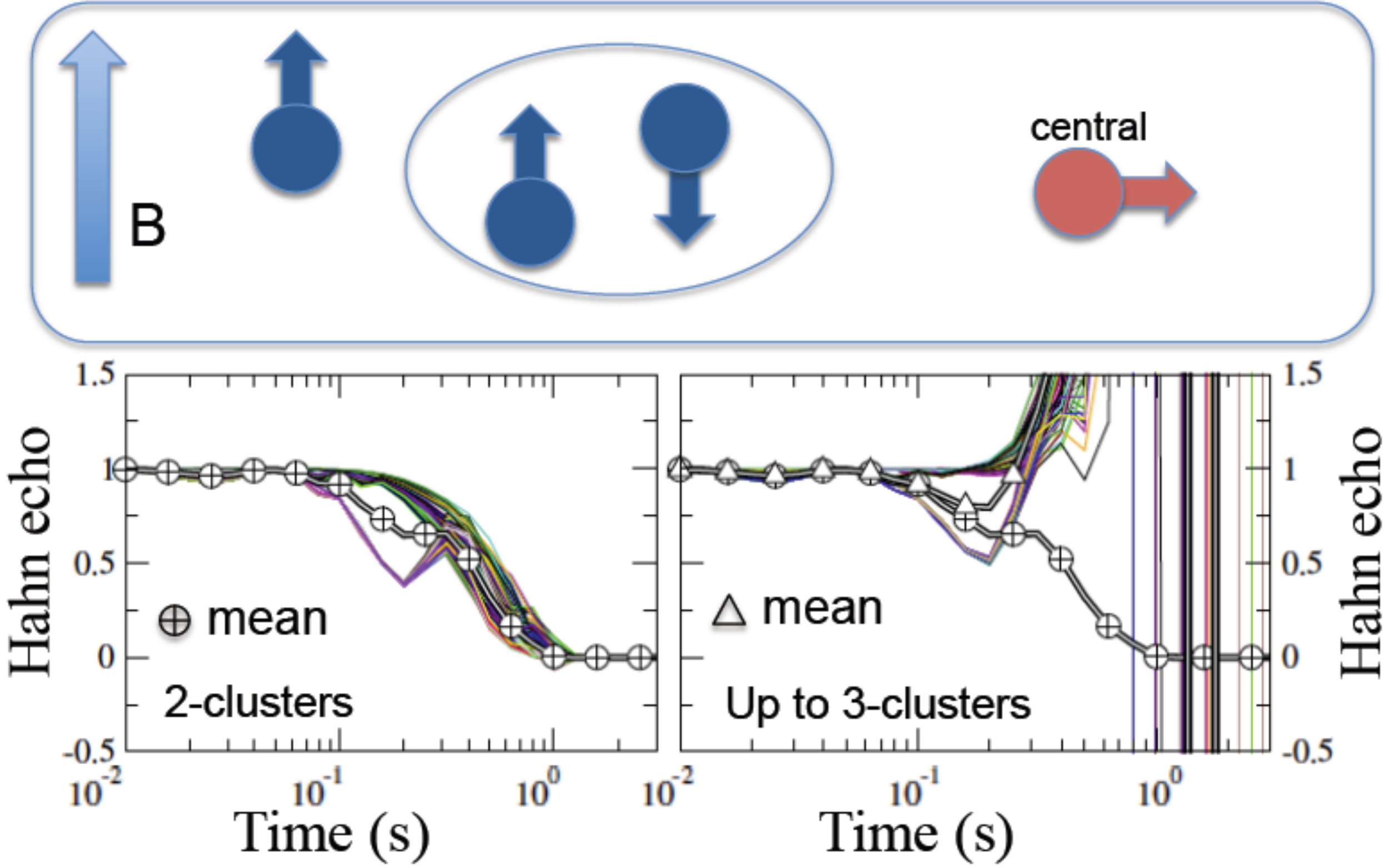}
\caption{
\label{Fig:NoExternalAwareness}
(Top) Depiction of a 2-cluster whose flip-flops are suppressed by the
magnetic field gradient generated by a third spin.  (Bottom) Spin echo
results from the CCE applied to spatial configuration instance {\bf A}
(see Fig.~\ref{Fig:SpatialConfigs}) when we don't define a cluster contributions to be ``externally aware.''  Each ``spaghetti'' strand is the result for a different random initial spin state (a product state of up or down for each spin) with the mean of $L_{\mbox{\scriptsize{CCE}}}^{(2)}$ [$L_{\mbox{\scriptsize{CCE}}}^{(3)}$] as encircled +'s [triangles].
The 3-cluster contributions must compensate for the lack of ``external awareness'' which leads to numerical instability at later times.
}
\end{figure}

The CCE that we described in Sec.~\ref{SecCCE} has some flexibility.
We have stated that $L_{\cal S}$ is to be defined in a manner that
involves external spins trivially.  The most simple way to define
$L_{\cal S}$ would be to ignore all the external spins entirely.  That
is, $L_{\cal S}$ is the result of $\rho_q^{+-}(t) / \rho_q^{+-}(0)$
when only including those bath spins into the problem that are
contained in set ${\cal S}$.  We show the CCE results from applying
this definition for the spatial configuration instance {\bf A}
(Fig.~\ref{Fig:SpatialConfigs}) in Fig.~\ref{Fig:NoExternalAwareness}.
At short times, the $L_{\mbox{\scriptsize{CCE}}}^{(3)}$ (up to
3-clusters) exhibits a small correction to
$L_{\mbox{\scriptsize{CCE}}}^{(2)}$ (2-clusters).  The CCE, in this
form, does not provide a robust solution to our canonical problem
except to study the initial part of the decay.
 Before significant decay occurs, $L_{\mbox{\scriptsize{CCE}}}^{(3)}$ blows up and becomes numerically unstable.  
We note that the primary contribution from 3-clusters near the onset of this numerical instability comes from the suppression of flip-flopping dynamics due to the 
magnetic field gradients generated by external spins as depicted in
the upper cartoon of Fig.~\ref{Fig:NoExternalAwareness}.  That is, the
2-cluster contributions are overestimated when we completely ignore
all of the external interactions, and this must be compensated at the
3-cluster level.  When we ignore all other bath spins, any pair of
spins is completely resonant and can freely flip-flop.  Considering the
presence of these other bath spins, they are generally off-resonant to
some degree.  That is, the Ising-like part of
the long-range dipolar interactions plays an important role regarding
whether or not a given pair of spins, for example, are resonant with each other
for flip-flopping.
Capturing this effect for small clusters is crucial for obtaining our computationally feasible and convergent theory of spectral diffusion in the canonical problem.

Ideally, one would want to define things in such a way that Lemma~1
would be applicable with respect to
just the flip-flop interactions and that the $\hat{S}^z_{i} \hat{S}^z_{j}$
interactions would come for free.  We have not found such an efficient
solution that achieves this,
 but a step in the right direction is to include these Ising-like 
interactions with spins outside of ${\cal S}$ for a given $L_{\cal S}$.
That is $L_{\cal S}$ excludes only the flip-flop interactions involving
 spins external to ${\cal S}$.
For each bath state $\lvert J \rangle$, let
 $L^{J} = \langle J^+(t) \vert J^-(t) \rangle$ where $J^{\pm}(t) =
\hat{U}(t) \lvert J, \pm \rangle$, using $\pm$ for the up/down central
spin states.  Now define $L_{\cal S}^{J} = \langle J_{\cal S}^+(t)
\vert J_{\cal  S}^-(t) \rangle$ where $\lvert J_{\cal  S}^{\pm}(t) \rangle
= \hat{U}_{\cal S}(t) \lvert J, \pm \rangle$;
here, $\hat{U}_{\cal S}$ gives the evolution which disregards all 
{\it except} the Ising-like $\hat{S}^z_{i} \hat{S}^z_{j}$ 
interactions with bath spins external to ${\cal S}$.
We thus have the CCE defined for each $J$ state
independently.  For a given density matrix $\rho = \sum_J p_J \lvert J
\rangle \langle J \rvert$, 
$L = \sum_J p_J L^{J} \equiv \langle L_{\cal S}^J \rangle_J$.  

\begin{figure}
\includegraphics[width=3.5in]{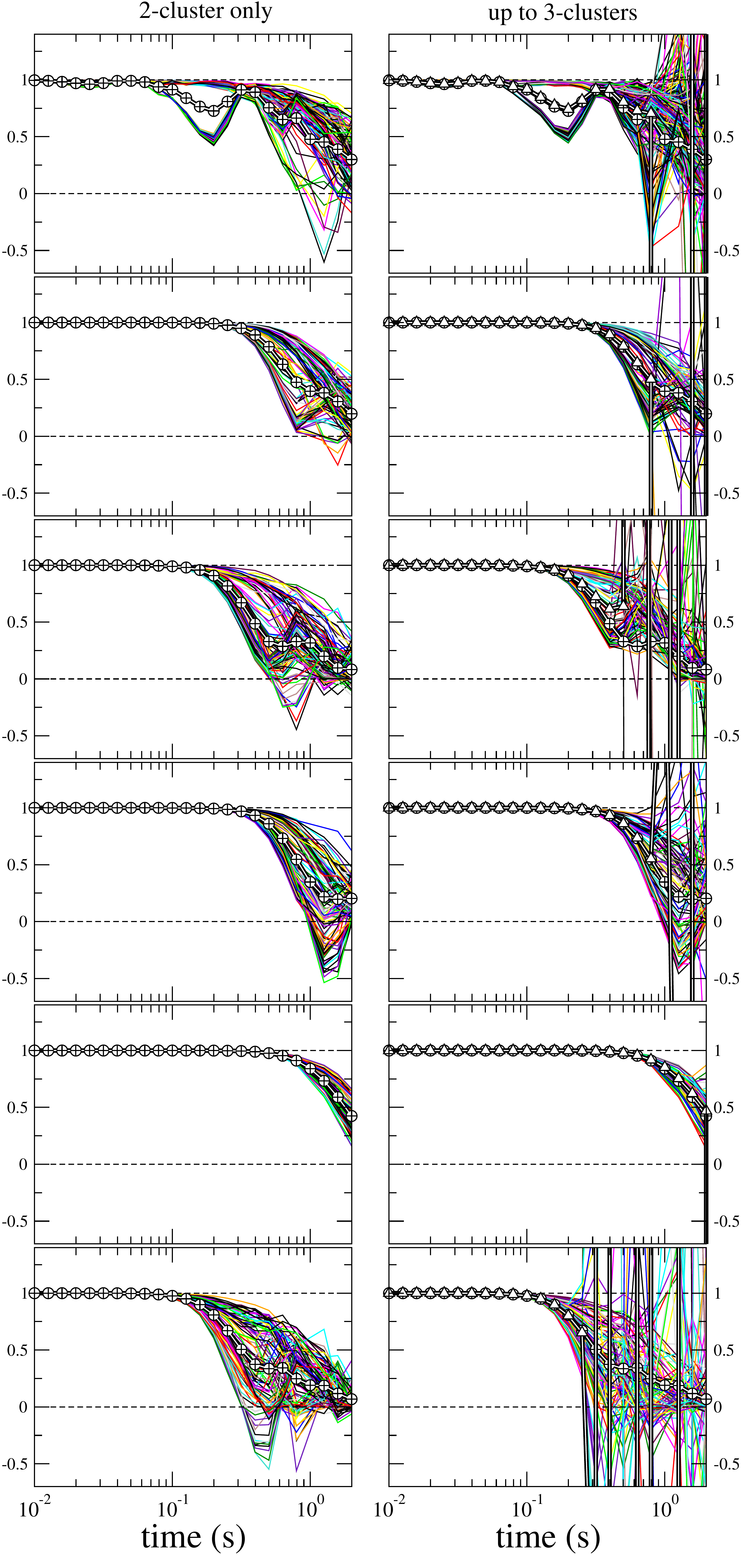}
\caption{
\label{Fig:NoIntAvgSpaghetti}
$L_{\mbox{\scriptsize{CCE}}}^{(2)}$ (left) and
$L_{\mbox{\scriptsize{CCE}}}^{(3)}$ spin echo results corresponding to
spatial configuration instances {\bf A}-{\bf F}
(Fig.~\ref{Fig:SpatialConfigs}) of our canonical problem using the ``externally aware'' treatment of clusters.  
Each ``spaghetti'' strand is the result for a different random initial spin state (a product state of up or down for each spin) with the mean of $L_{\mbox{\scriptsize{CCE}}}^{(2)}$ [$L_{\mbox{\scriptsize{CCE}}}^{(3)}$] as encircled +'s [triangles].
}
\end{figure}

We show the results of these CCE calculations, with ``external'' spin awareness
as described above, in Fig.~\ref{Fig:NoIntAvgSpaghetti} for each of the spatial configurations depicted in Fig.~\ref{Fig:SpatialConfigs}.  Comparing with Fig.~\ref{Fig:NoExternalAwareness}, considerable improvement is apparent.  However, these results, which still show strongly unstable behavior at longer times can be further improved.

\subsection{Interlaced spin state averaging}
\label{spinStateAvg}
As we have just discussed, disorder in a system may improve the CCE convergence due
to a localization.  It can also be problematic and create
numerical instability, however.  The problem arises when some
$L_{\cal C}$ attains, by chance of the disorder, a very small value for a time
that is short compared with the overall decoherence time.  In the
calculation of $\tilde{L}_{\cal S}$ for some supercluster 
${\cal S} \supset {\cal C}$, $L_{\cal C}$ may be a factor in its
denominator so that $\tilde{L}_{\cal S}$ attains a very large value,
disproportionate to its order, i.e., $\|{\cal S}\|$, in the CCE. 
We note this effect in the numerical instability observed in Fig.~\ref{Fig:NoIntAvgSpaghetti} for later spin echo times.

\begin{figure}
\includegraphics[width=3.5in]{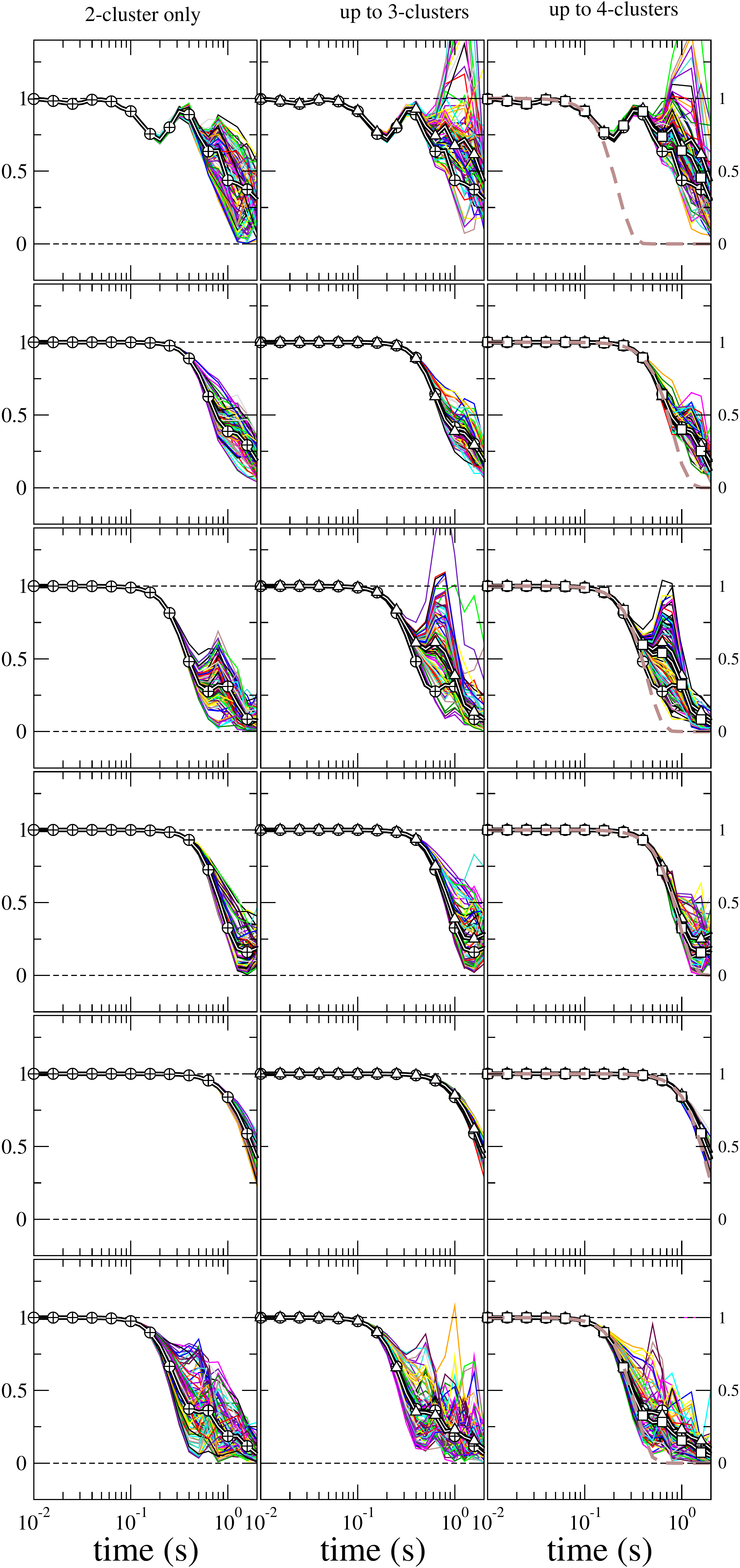}
\caption{
\label{Fig:WithIntAvgSpaghetti}
$L_{\mbox{\scriptsize{CCE}}}^{(2)}$ (left),
$L_{\mbox{\scriptsize{CCE}}}^{(3)}$ (middle), and
$L_{\mbox{\scriptsize{CCE}}}^{(4)}$ (right) spin echo results
corresponding to spatial configuration instances {\bf A}-{\bf F}
(see Fig.~\ref{Fig:SpatialConfigs}) of our canonical problem using ``external awareness'' {\it and} ``interlaced spin averaging'' 
in our implementation of the CCE as expressed in Eq.~(\ref{Eq:InterlacedCCE}).
Each ``spaghetti'' strand is the result for a different random spin state template $J$ [see Eq.~(\ref{Eq:InterlacedCCE})].  
The mean of $L_{\mbox{\scriptsize{CCE}}}^{(2)}$, $L_{\mbox{\scriptsize{CCE}}}^{(3)}$, and $L_{\mbox{\scriptsize{CCE}}}^{(4)}$ are represented as encircled +'s, triangles, and squares respectively.
The brown (color online) dashed curves of the form $\exp{(-t^3)}$ on the right panels are presented for comparison.
}
\end{figure}

We find that this ill effect from disorder may be mitigated very
effectively by defining the CCE to use ``interlaced'' spin state averaging.
 Rather than computing a separate $L$ for each
given bath state $J$, we can self-consistently average over the bath
states internal to the computation of $L_{\cal S}$.  
In terms of the definitions of
the previous sub-section, we may define 
$L_{\cal S} = \langle L_{\cal S}^J \rangle_J$, as an average over bath
states $J$, and use the standard CCE equations [see Eqs.~(\ref{CCEeqns}),
  (\ref{DefTildeL}), and (\ref{CCEtruncated})].  In the large cluster
limit, this will approach the correct solution.
Defined in this way, $L_{\cal S}$ is {\it not} factorable under
disconnected interactions (this is because $\langle \hat{S}^z \rangle^2 \neq
\langle (\hat{S}^z)^2 \rangle$), so the lemma and theorem of
Sec.~\ref{CCEconvergence} does not carry through.  However, we find the convergence behavior to be significantly improved (see Fig.~\ref{Fig:WithIntAvgSpaghetti}) by using this sort of strategy.

\begin{figure}
\includegraphics[width=3.5in]{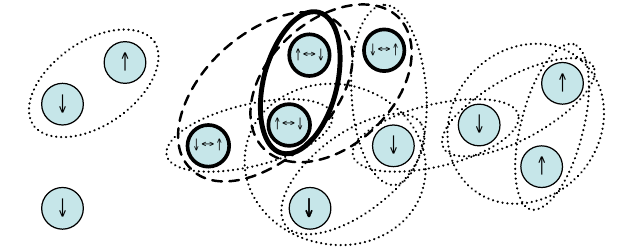}
\caption{
\label{Fig:SuperClusterSpinStates}
Illustration of ${\cal K}(J, {\cal C}, \Gamma)$
[see Eq.~(\ref{Eq:InterlacedSpinStates})].  Bath spins are denoted by the
circles; up/down arrows denote the spin state ``template'' $J$.
Clusters in $\Gamma$ (the included clusters)
are denoted by ovals that encompass multiple bath spins,
${\cal C}$ being the one with the solid, thick outline.  Superclusters
have dashed outlines and all other clusters have dotted outlines.  
${\cal K}(J, {\cal C}, \Gamma)$ are all spins states, including $J$,
that only differ 
from $J$ in spins within superclusters of ${\cal C}$, shown with the
thicker outline and $\uparrow \leftrightarrow \downarrow$ or $\downarrow
\leftrightarrow \uparrow$ inscriptions (the up/down arrow on the
left, say, denotes the state of the $J$ template, but all combinations
of these are included in ${\cal K}$).
}
\end{figure}

It is only feasible to approximate $L_{\cal S}$ with this spin state
averaged definition because explicit averaging over the state of all
external spins has exponential complexity.
In practice, we instead use the following formulation
which approximates interlaced spin averaging in a manner that corrects
itself as more clusters are included in the approximation.  
Let $\Gamma$ be a set of clusters (e.g., up to a certain size) that
we include to approximate the solution.  Let $J$ be some bath spin state, as a product state of up or down for each spin, that will serve as a template.
We now define 
\begin{equation}
\label{Eq:InterlacedCCE}
{\cal L}_{\Gamma}^J = \prod_{{\cal C} \in \Gamma}
\tilde{L}_{\cal C}^{{\cal K}(J, {\cal C}, \Gamma)},
\end{equation}
where ${\cal K}(J, {\cal C}, \Gamma)$ is the set of all spin
states that may differ from $J$ only for spins in
superclusters of ${\cal C}$ that are contained in $\Gamma$.
That is,
\begin{equation}
\label{Eq:InterlacedSpinStates}
{\cal K}(J, {\cal C}, \Gamma) = \{J'|~\exists {\cal C}' \in \Gamma,
{\cal C}' \supseteq {\cal C}, {\cal D}(\lvert J \rangle, \lvert J'
\rangle) \subseteq {\cal C}'\},
\end{equation}
where ${\cal D}(\lvert J \rangle, \lvert J' \rangle)$ is the set of
spins whose state differs between $\lvert J \rangle$ and $\lvert J' \rangle$:
\begin{equation}
{\cal D}\left(\bigotimes_n \lvert j_n \rangle, 
\bigotimes_n \lvert j_n' \rangle\right) = \{n|~
\lvert j_n \rangle \neq \lvert j_n' \rangle \}.
\end{equation}
An example for ${\cal K}(J, {\cal C}, \Gamma)$ is illustrated in Fig.~\ref{Fig:SuperClusterSpinStates}.
Then we define
\begin{equation}
\tilde{L}_{\cal C}^{\cal K} =
\langle L_{\cal C}^K \rangle_{K \in {\cal K}} / \prod_{{\cal C}'' \subset {\cal
    C}} \tilde{L}_{\cal C''}^{\cal K},
\end{equation}
where $L_{\cal C}^J$ solves the $L_{\cal C}$ problem for the given spin
state $J$.  Importantly, this yields the exact
spin state average solution for ${\cal L}_{\Gamma}^J$ in the
limit that $\Gamma$ includes all clusters ($J$ becomes
irrelevant).  Furthermore, it may be computed relatively efficiently if we are sufficiently selective in choosing the clusters of $\Gamma$ (see Appendix~\ref{clusterSampling}).
With proper bookkeeping, each Hamiltonian 
(for a given cluster and external spin state) 
need only be diagonalized once, and
each $L_{\cal C}^J$ need only be computed once and raised to the
proper power to be multiplied into the solution.

\begin{figure}
\includegraphics[width=3.5in]{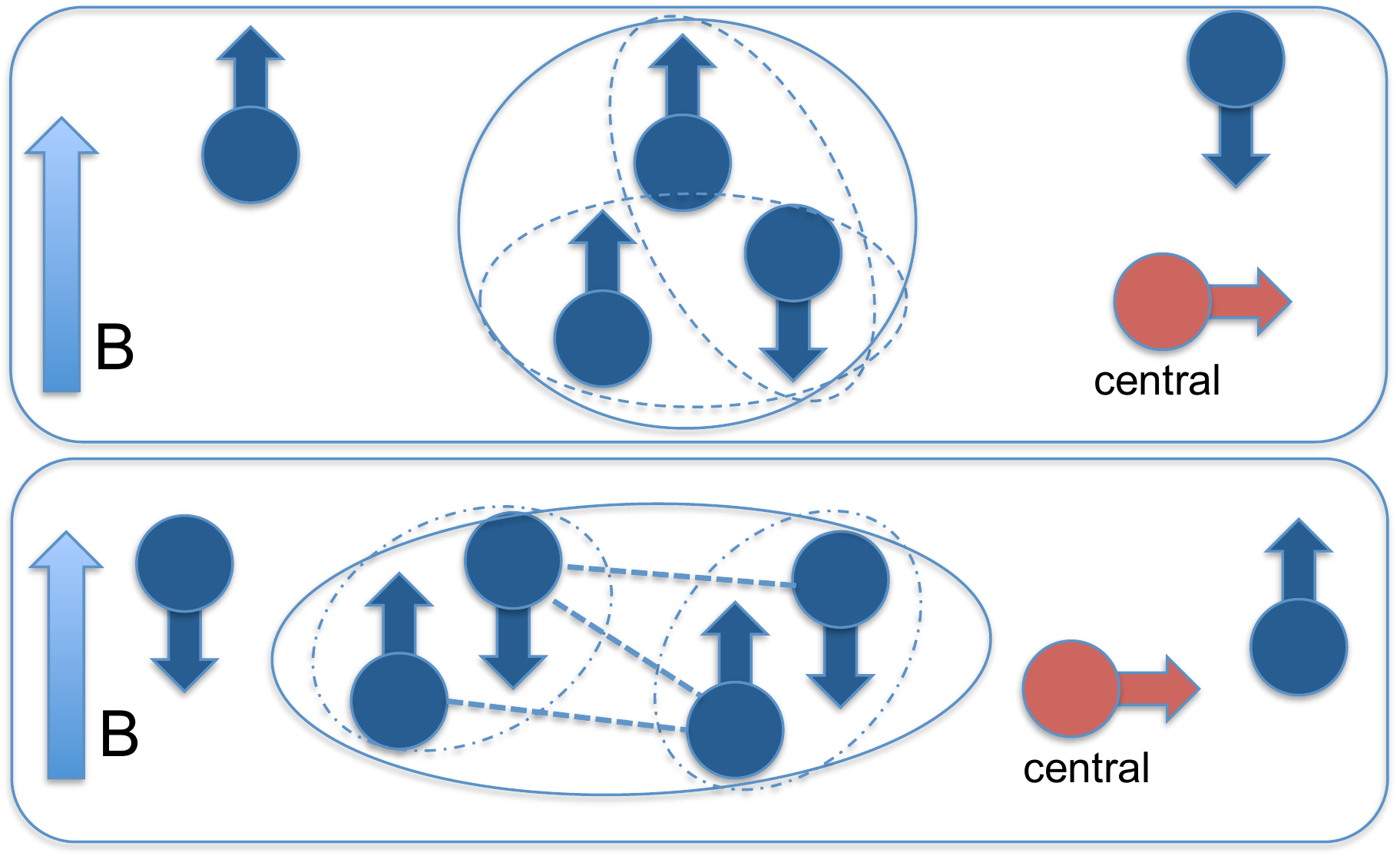}
\caption{
\label{Fig:Important3and4cluster}
Depiction of important 3-cluster (top) and 4-cluster (bottom) contributions in our adapted CCE with ``external awareness'' and ``interlaced spin averaging.''  At the 2-cluster level, all flip-flopping pairs are treated independently.  The 3-cluster level must compensate for overlapping flip-flopping pairs that cannot be independent because they share a bath spin.  The 4-cluster level must compensate for the fact that two separate flip-flopping pairs in close proximity influence each other magnetically (through the Ising-like $\hat{S}^z_{i} \hat{S}^z_{j}$ iteractions) as they flip-flop.  This can enhance spectral diffusion, for example, if the two pairs are off resonance independently, but on resonance (conserve energy) together as a simultaneous process.
}
\end{figure}

We present results using this revised CCE for our six respective spatial configurations (Fig.~\ref{Fig:SpatialConfigs}) in Fig.~\ref{Fig:WithIntAvgSpaghetti}.
The numerical instabilities we observe in
Fig.~\ref{Fig:NoIntAvgSpaghetti} have been removed, and the
convergence appears to be well behaved as we go from 
$L_{\mbox{\scriptsize{CCE}}}^{(2)}$ on up to $L_{\mbox{\scriptsize{CCE}}}^{(4)}$.
We still see some unphysical results (larger than unity) for some spin state templates, but not as widely ranging and erratic as before without interlaced spin averaging.  In principle, the CCE results should be exact when all clusters are included and any non-physical results would go away.  This trend toward the physically valid range is oberved in going from $L_{\mbox{\scriptsize{CCE}}}^{(3)}$ to $L_{\mbox{\scriptsize{CCE}}}^{(4)}$ ($L_{\mbox{\scriptsize{CCE}}}^{(2)}$ always gives results in the physical range).  
We also note, as an indication of convergence, that a split between
averaged 4-cluster and 3-cluster results in each case (particularly
visible in spatial configurations A, C, and F) occurs later in the
time parameter than the split between averaged 3-cluster and 2-cluster
results.  The convergence is not always great for this challenging problem, but a good fraction of the decay appears to be captured well.  As an aid to intuition about what is going on for the different cluster sizes, we depict important 3-cluster and 4-cluster contributions in Fig.~\ref{Fig:Important3and4cluster}.

We include decay curves of the form $\exp{(-t^3)}$ on the right panels
of Fig.~\ref{Fig:WithIntAvgSpaghetti} for comparison with the
calculations.  Such behavior is expected in the initial decay for
Ornstein-Uhlenbeck noise.  With exception to spatial configuration
{\bf A}, the $\exp{(-t^3)}$ form fits the calculated results very well
for the first $25\%$ to $50\%$ of the decay, confirming the results
from Refs.~\onlinecite{Hanson_Science08},
\onlinecite{Dobrovitski_PRL09}, and \onlinecite{deLange_Science10} using a calculation starting from a microscopic model of the dipolarly coupled bath. 

\subsection{Ensemble of spatial configurations}

\begin{figure}
\includegraphics[width=3.5in]{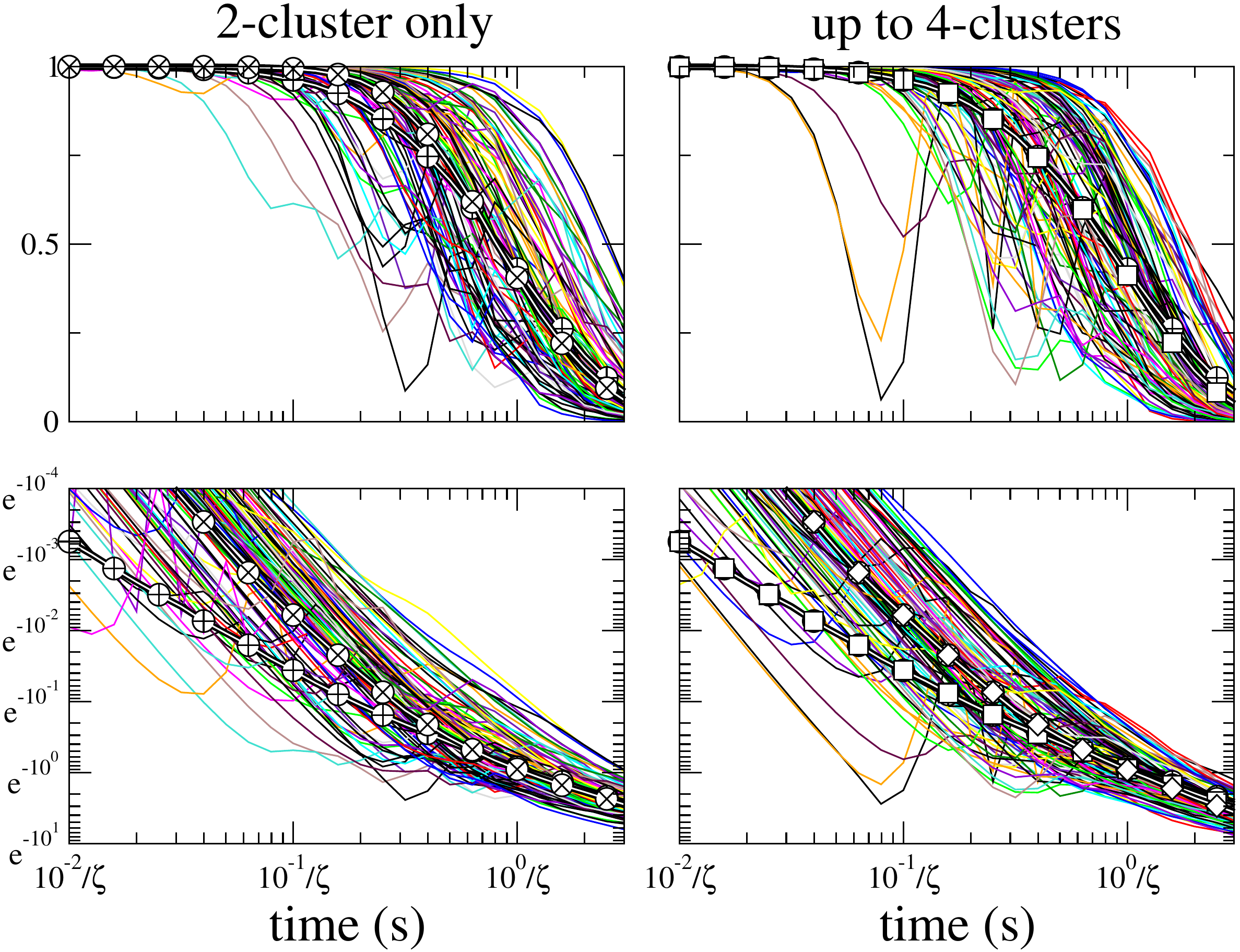}
\caption{
\label{Fig:SpatialEnsemble}
$L_{\mbox{\scriptsize{CCE}}}^{(2)}$ (left) and
$L_{\mbox{\scriptsize{CCE}}}^{(4)}$ (right)
 spin echo results of our canonical problem 
using ``external awareness'' {\it and} ``interlaced spin averaging'' 
in our implementation of the CCE as expressed in
Eq.~(\ref{Eq:InterlacedCCE}).
We use the scaling parameter $\zeta$:
time is scaled inversely with $\zeta$ and
$g^2 C_E = \zeta \times 4 \times 10^{13}/\mbox{cm}^{3}$
(i.e., $\zeta = 1$ is the result for $g=2$ and $C_E = 10^{13}/\mbox{cm}^{3}$).
Each ``spaghetti'' strand is the result for a different random spatial configuration averaged over a large number of spin state templates.  
The mean of $L_{\mbox{\scriptsize{CCE}}}^{(2)}$ and
$L_{\mbox{\scriptsize{CCE}}}^{(4)}$ are represented
as encircled +'s and squares respectively.
The median of $L_{\mbox{\scriptsize{CCE}}}^{(2)}$ and
$L_{\mbox{\scriptsize{CCE}}}^{(4)}$ are represented
as encircled x's, and diamonds respectively.
The bottom figures show the same spin echo results on a logarithmic
scale for the decay.
}
\end{figure}

We show results for a large ensemble of different spatial configurations in Fig.~\ref{Fig:SpatialEnsemble}.  We indicate the median and mean results for the 
ensemble (well converged to approximate the infinite ensemble).  Both
the median and mean are computed separately for the statistical
results at each point in time; that is, the median values at different
time points do not necessarily come from the same spatial
configuration result.  The median and mean differ drastically at short
times because the mean is dominated by rare cases where the spin echo
dips down early (for example, when a pair of bath spins happen to be
particularly close to the central spin).  Our CCE expansion, with
``external awareness'' and ``interlaced spin averaging'' exhibits very
good convergence as demonstrated most clearly in Fig.~\ref{Fig:CanonicalConvergence}; as
we increase the cluster size in even numbers (two, four, and six), we get slight corrections that push down the tails of the spin echo decay curves.  Let us stress again that the fact that only relatively small clusters of flip-flopping spins are enough to describe the SE decay is due to our use of a modified version of CCE.  Averaging of these cluster contributions over the states of external spins, while keeping the diagonal dipolar interactions between the spins from the cluster and from its outside, allows us to capture the effect of localization of spin dynamics over the duration of appreciable coherence.

\subsection{Convergence}

We demonstrate convergence of the ensemble average spin echo for our
canonical problem in Fig.~\ref{Fig:CanonicalConvergence}.  To be
confident in our results, we must ensure sufficient cluster sampling
(see Appendix~\ref{clusterSampling}) {\it and} the inclusion of
sufficiently large clusters.  The cluster sampling described in
Appendix~\ref{clusterSampling} relies upon radial cutoffs and cluster
counting cutoffs.  Our confidence in our results increases as we
increase either of the cutoffs.  We employ convenient methods that
automatically increase the cutoffs as needed by comparing results from
differing cutoff values; for example, if there is a significant
difference in results that include different numbers of clusters, we
increase the cluster count cutoff.

It is best to evaluate and analyze the effects of larger clusters in
terms of their correction to solutions that exclude them:
$L_{\mbox{\scriptsize{CCE}}}^{(k)} -
L_{\mbox{\scriptsize{CCE}}}^{(k-1)}$.  
For the most rapid convergence,
this correction (difference) should be averaged over the
ensemble of spatial realizations of the bath rather than the individual
$L_{\mbox{\scriptsize{CCE}}}^{(k)}$ results.
Such corrections, for successive cluster sizes, are presented in the
lower panel of Fig.~\ref{Fig:CanonicalConvergence}.
The
corrections are only significant toward the tail of the decay with an
onset time that increases with increasing cluster size.
At later times, larger cluster corrections will often become
numerically unstable, but the ealier parts of the decay are most
important for quantum information considerations.

We note that odd cluster sizes tend to increase the spin
echo decay time (enhanced coherence) and even cluster sizes tend to
decrease the spin echo decay time (furthering decoherence). Furthermore,
4-cluster corrections are the same order of magnitude
as 3-cluster corrections and 6-cluster corrections
are the same order of magnitude as 5-cluster corrections.
The reason for this even/odd trend relates the spin up/down
symmetry when regarding non-polarized baths and was
noted in Ref.~\onlinecite{Witzel_PRB06} for the case of
nuclear-induced spectral diffusion. A similar argument applies here. 
When averaging over a bath described by density operator proportional to unity (as it is the case at high temperatures and low magnetic fields), the operation of reversing the sign of the Hamiltonian leaves the expression for spin echo signal unchanged [see Eqs.~(60) and (61) in Ref.~\onlinecite{Witzel_PRB06}]. 
Therefore, in the non-polarized average,
only even perturbation order terms survive. Also
note that contributing processes must perform a full cycle
in state space, looping back to some initial state; that
is because the bath states are traced out in obtaining the reduced
density matrix of $L$. Thus, 3-clusters are fourth
order in such a perturbation expansion and 5-clusters are sixth
order.  For this reason, also, 3-clusters are not dominated
by the process of three flip-flops since this is canceled by the
symmetry of the unpolarized bath [this was also noticed in Ref.~\onlinecite{Saikin_PRB07}, see the discussion there above Eq.~(12)].  Instead they are dominated by
double flip-flops of two pairs sharing one spin in common as in the
top panel of Fig.~\ref{Fig:Important3and4cluster}.  This provides a
correction to the 2-cluster level which overcounts these as two
independent flip-flopping pairs; at the 3-cluster level, we see that
the shared spin cannot simultaneously flip-flop with two partners. 
Presumably, a similar situation occurs at the 5-cluster level, though
the picture is substantially more complicated.  As a side note, a nice aspect of the
CCE over the linked cluster expansion\cite{Saikin_PRB07} in practice
is that we do not have to decompose this complicated picture to obtain results.

\begin{figure}
\includegraphics[width=3.5in]{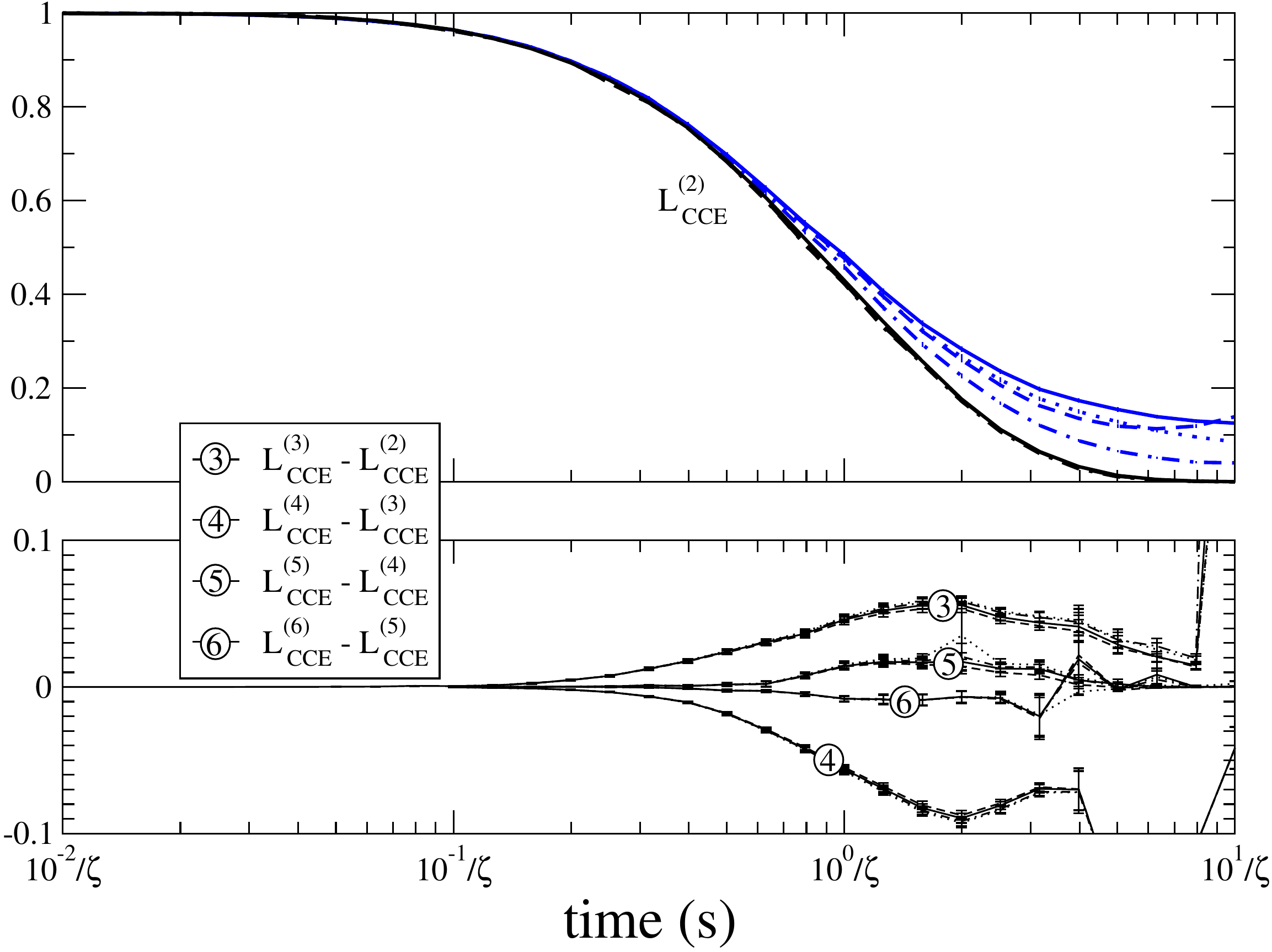}
\caption{
\label{Fig:CanonicalConvergence}
Ensemble-averaged spin echo results and corrections for our canonical problem within
various approximations using our adaptation of the CCE.  The upper
panel gives $L_{\mbox{\scriptsize{CCE}}}^{(2)}$ for various cutoffs
in our cluster sampling heuristics (Appendix~\ref{clusterSampling}).
The $\zeta$ scaling parameter is used as in Fig.~\ref{Fig:SpatialEnsemble}.
Relative to corresponding solid line curves, 
dotted lines include twice as many clusters ($N_k$),
dashed lines increase the radial cutoff ($R_C$) by $30\%$, and dot-dashed
lines do both.  The blue curves (color online) in the top panel
demonstrate cutoffs that are insufficient at later time scales ($N_2
= 200$ and $R_C = 600~\mbox{nm}$ for the solid blue line).  The black
curves use good cutoff values ($N_k \sim 30,000$ and $R_C \sim
1000~\mbox{nm}$); for this reason, it is difficult to distinguish
corresponding curves of differing line patterns.
The lower panel shows successive corrections as we increase the
maximum cluster size.
}
\end{figure}

\section{Variations of the Problem}
\label{Sec:Variants}

Our canonical problem was chosen for its simplicity as a problem to study in depth without complicating details to serve as distractions.  In the study of real-world problems, many variations of our canonical problem arise.  In this section, we consider some natural variations of this problem and discuss resulting trends in generality.  The canonical problem will provide a convenient reference point as we analyze these trends.  We will reference each of these variations in Sec.~\ref{Sec:Applications} where we discuss specific problems of interest in the realm of (solid state spin) quantum information.

\subsection{Coupling strengths, polarization, and resonances}
\label{Sec:VariedCouplingAndTemp}

In this section, we consider variations of the canonical problem with regard to coupling strengths, resonances among pairs of spins, and polarization of the bath spins.  

\begin{figure}
\includegraphics[width=3.5in]{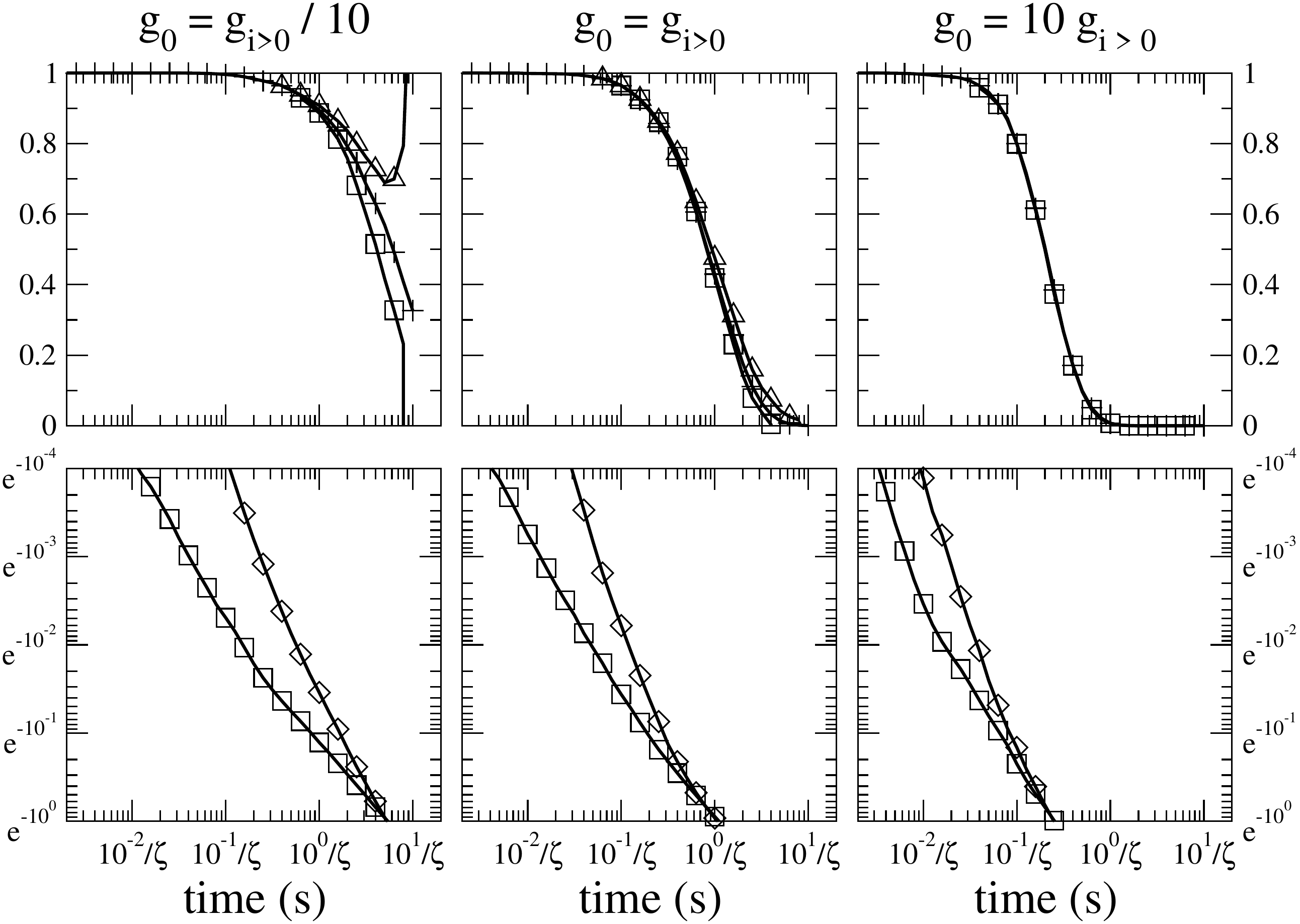}
\caption{
\label{Fig:CompareG0}
Comparison of Hahn echo results for different $g_0$ values (the
central spin g-factor) in the large magnetic field limit and off-resonant central spin treatment (e.g., the canonical problem otherwise).  
The $\zeta$ scaling parameter is used as in
Fig.~\ref{Fig:SpatialEnsemble} with $g_{i>0}^2 C_E = \zeta \times 4 \times 10^{13}/\mbox{cm}^{3}$.
The upper panels display $L_{\mbox{\scriptsize{CCE}}}^{(2)}$ (+'s),
$L_{\mbox{\scriptsize{CCE}}}^{(3)}$ (triangles) and
$L_{\mbox{\scriptsize{CCE}}}^{(4)}$ (squares) mean value results.  The
lower panels display $L_{\mbox{\scriptsize{CCE}}}^{(4)}$ 
median (diamonds) as well as mean (squares) value results on a
logarithmic scale.
}
\end{figure}

We first consider the effects of adjusting the coupling strengths by simply adjusting the g-factor of the central spin.   Only the relative strength of intra-bath interactions versus central spin-bath spin interactions are of qualitative importance (the absolute strengths only affect the scale of the time parameter).
By adjusting $g_0$ and holding $g_{i>0}$ constant,
we explore the transition from a central spin that is weakly coupled to the bath 
($g_0$ small) to one that is strongly coupled to the bath ($g_0$ large).  In relative terms, this is also an exploration of the transition from strong ($g_{i>0}$ large) to weak ($g_{i>0}$ small) coupling within the bath respectively, although this requires a translation of our timescales to reflect holding $g_0$ constant instead of holding $g_{i>0}$ constant.
We again treat the case in the large magnetic field limit where the central spin is off-resonant with the bath spins (either by a local magnetic field offset or the g-factor difference), as characterized by Eq.~(\ref{Eq:HeffOffRes}).  We compare three difference cases in Fig.~\ref{Fig:CompareG0}: $g_0 = g_{i>0} / 10$ (weak coupling to the bath), $g_0 = g_{i>0}$ (comparable interactions), and $g_0 = 10 g_{i>0}$ (strong coupling to the bath).  The time scaling factor is dependent upon the value of $g_{i>0}$.
Increasing $g_0$ increases the coupling of the central spin to the bath with causes a decrease in decoherence time.
It also makes sense that larger clusters become more important in the regime in which the coupling within the bath is relatively strong.  In our $g_0 = g_{i>0} / 10$ case, the cluster expansion is only well-behaved during the initial part of the decay.
Furthermore, the statistical spread of the results (exemplified by the discrepancy between the median and the mean) is reduced as we approach the weakly-coupled bath regime.

\begin{figure}
\includegraphics[width=3.5in]{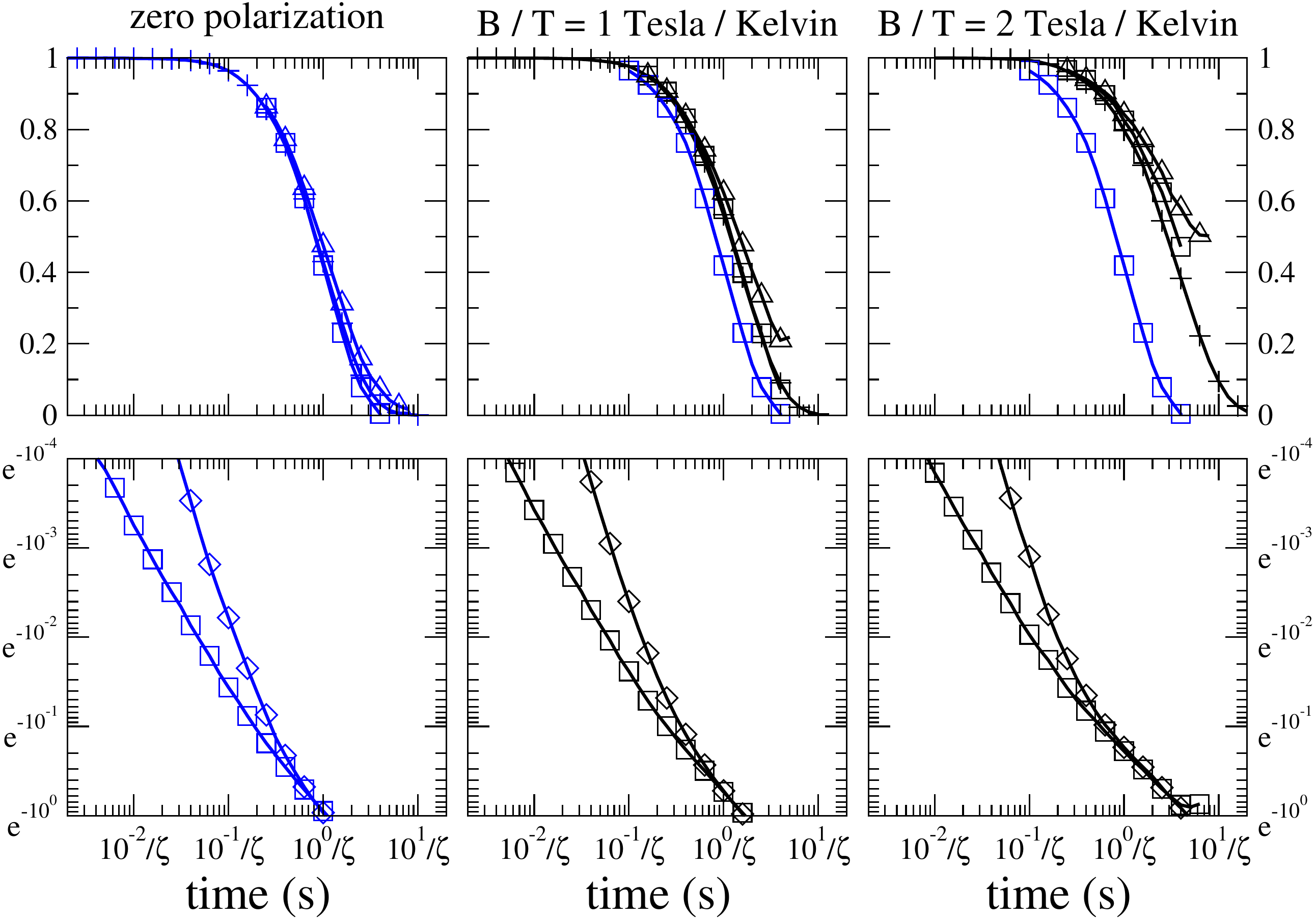}
\caption{
\label{Fig:Polarization}
Comparison of Hahn echo results for different degrees of thermal
polarization: zero polarization (left), $B / T = 1~\mbox{Tesla /
  Kelvin}$ (center), and $B / T = 2~\mbox{Tesla /
  Kelvin}$ (right).  
The zero polarization, infinite temperature results are
shown in blue (color online), provided for comparization along the
polarization columns. The upper panels display $L_{\mbox{\scriptsize{CCE}}}^{(2)}$ (+'s),
$L_{\mbox{\scriptsize{CCE}}}^{(3)}$ (triangles) and
$L_{\mbox{\scriptsize{CCE}}}^{(4)}$ (squares) mean value results.  The
lower panels display $L_{\mbox{\scriptsize{CCE}}}^{(4)}$ 
median (diamonds) as well as mean (squares) value results on a
logarithmic scale.  The $\zeta$ scaling parameter is used as in
Fig.~\ref{Fig:SpatialEnsemble}: $g^2 C_E = \zeta \times 4 \times 10^{13}/\mbox{cm}^{3}$.
}
\end{figure}

Thermal polarization of an electron spin bath is often feasible with
the temperatures and magnetic fields typically proposed in solid state spin 
quantum information processing.  At the standard $g=2$ value, the
electron Zeeman splitting corresponds to $1.3$ K per Tesla.  In
Fig.~\ref{Fig:Polarization}, we compare results for various polarized
versions of our canonical problem.  At the 2-cluster level, the
effects of polarization are fairly straightforward.  Since we are
taking the limit of large applied magnetic field and using the secular
approximation [see Eq.~(\ref{Eq:HeffOffRes})], the number of
pairs that may contribute at the 2-cluster level is proportional to
the probability of each spin being up times the probability of each
spin being down: $N_{\mbox{\scriptsize pairs}} \propto p_{\uparrow}
p_{\downarrow}$ with $p_{\uparrow / \downarrow} = \exp{(\pm E_z / 2
  k_B T)} / 2 \cosh{(E_z / 2 k_B T)}$ according to Boltzmann statistics where $E_z$ is the electron Zeeman energy splitting.
Since $L_{\mbox{\scriptsize{CCE}}}^{(2)}$ is a product over
contributing pairs, $\ln{(L_{\mbox{\scriptsize{CCE}}}^{(2)})} \propto
p_{\uparrow} p_{\downarrow}$.  Note, however, that contributions from
larger clusters play a more significant role earlier in the decay as we increase the polarization.

\begin{figure}
\includegraphics[width=3.5in]{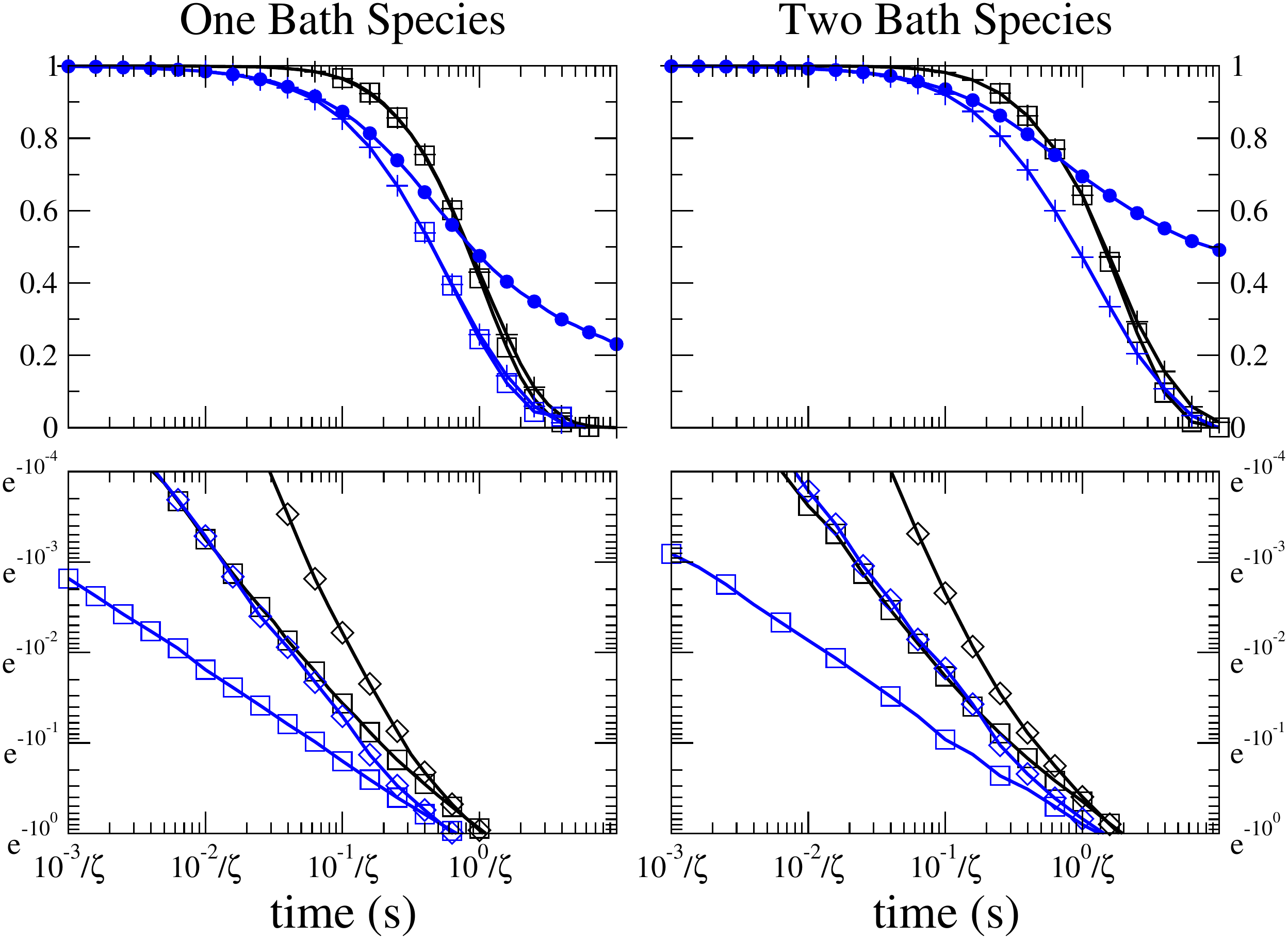}
\includegraphics[width=1.5in]{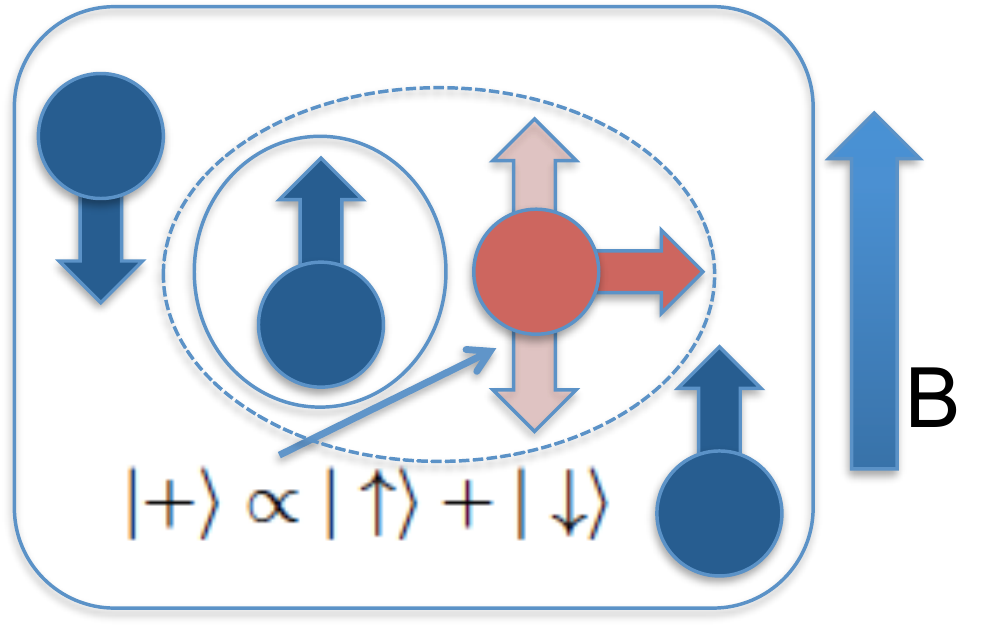}
\includegraphics[width=1.5in]{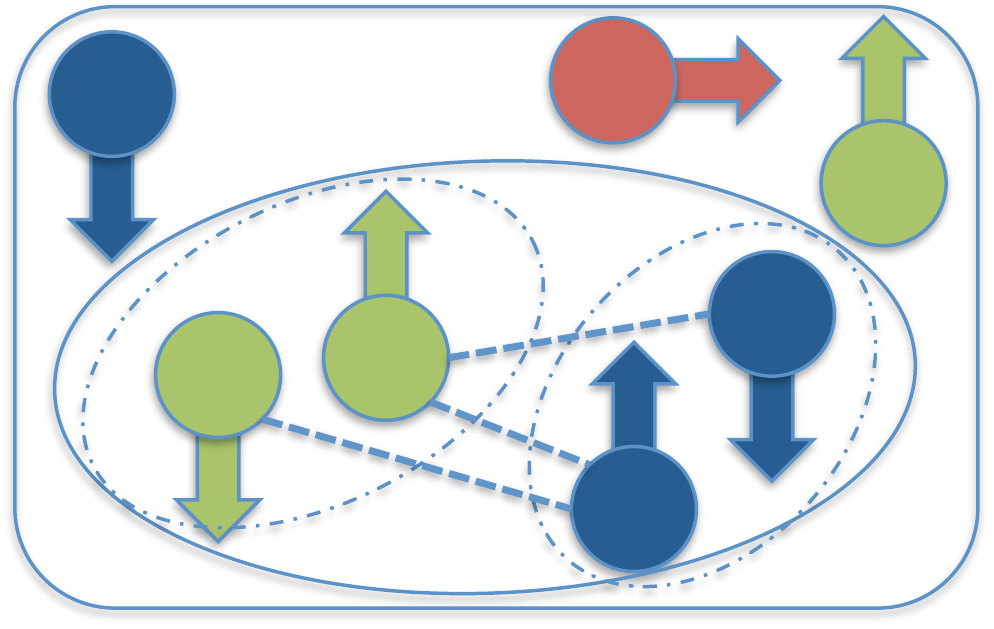}
\caption{
\label{Fig:Resonances}
(Top) Comparison of Hahn echo results for different resonance scenarios.  We
compare one bath species (left) and two bath species (right) with a
50/50 random mixture [see Eq.~(\ref{HeffTwoSpecies})].  Furthermore,
results when the central spin is resonant with bath spins
[see Eq.~(\ref{HeffOnRes})] are shown in blue.
The upper panels display $L_{\mbox{\scriptsize{CCE}}}^{(1)}$ (filled circles)
$L_{\mbox{\scriptsize{CCE}}}^{(2)}$ (+'s), and
$L_{\mbox{\scriptsize{CCE}}}^{(4)}$ (squares) mean value results.  
Below those, the panels display $L_{\mbox{\scriptsize{CCE}}}^{(4)}$ 
median (diamonds) as well as mean (squares) value results on a
logarithmic scale.  The $\zeta$ scaling parameter is used as in
Fig.~\ref{Fig:SpatialEnsemble}: $g^2 C_E = \zeta \times 4 \times 10^{13}/\mbox{cm}^{3}$.
(Bottom) Depiction of a 1-cluster process when the bath spins are resonant with the central spin (left) and of an important type of 4-cluster process when there are two bath species (right).
}
\end{figure}

Now we consider variations pertaining to resonances.  In our canonical problem, all spins except for the central spin are taken to be resonant with each other.  What happens when the central spin is resonant with the bath spins?  In this case, we replace Eq.~(\ref{Eq:HeffOffRes}) with
\begin{equation}
\label{HeffOnRes}
\hat{\cal H}_{\mbox{\scriptsize eff}} = \sum_{i, j} b_{i, j} \hat{S}_i^+ \hat{S}_j^-
- 2 \sum_{i, j} b_{i, j} \hat{S}_i^z \hat{S}_j^z,
\end{equation}
which allows flip-flops with the central spin.  
This induces both depolarization and dephasing together (i.e., where $T_2$ may be $T_1$ limited).
These direct flip-flops produce non-trivial 1-cluster contributions in
the CCE (depicted on the lower left of Fig.~\ref{Fig:Resonances}).  At short times,
this dominates the spin echo decay.  The spin echo that results in
this scenario actually exhibit oscillations at the Zeeman precessional
frequency (i.e., induced by the externally applied $B$ field).  These oscillations are extremely fast, about $30$~GHz  at $g=2$ and $B=1~$Tesla.  For numerical stability, we computed the CCE in this case by averaging over a few of these oscillation about each evaluated spin echo time; we define $L_{\cal S}$ in this manner.

Relevant to many of our applications, another variant is to have multiple bath species that are only resonant within respective species.  This applies to a bath of electrons bound to donor nuclei that have non-zero spin.  For example, phosphorus nuclei have a $1/2$ spin magnitude; this yields two bath species because up and down nuclei generate opposite hyperfine shifts for the electron spins.  
For this case, in the large applied magnetic field limit, we have
\begin{eqnarray}
\nonumber
\hat{\cal H}_{\mbox{\scriptsize eff}} &=& \sum_{i, j \in {\cal A}} b_{i, j} \hat{S}_i^+ \hat{S}_j^- + \sum_{i, j \in {\cal B}} b_{i, j} \hat{S}_i^+ \hat{S}_j^-
\\
\label{HeffTwoSpecies}
&& - 2 \sum_{i, j} b_{i, j} \hat{S}_i^z \hat{S}_j^z.
\end{eqnarray}
Note that this is not quite the same as the combined independent
effects of two baths at half concentration because there are
inter-species Ising-like interactions.  We compare various resonance
scenarios (one versus two species and the central spin being resonant
versus off-resonant with bath spins) in Fig.~\ref{Fig:Resonances}
along with schematic depictions of important process.

\subsection{Bath Geometry}
\label{Sec:BathGeom}

\begin{figure}
\includegraphics[width=3.5in]{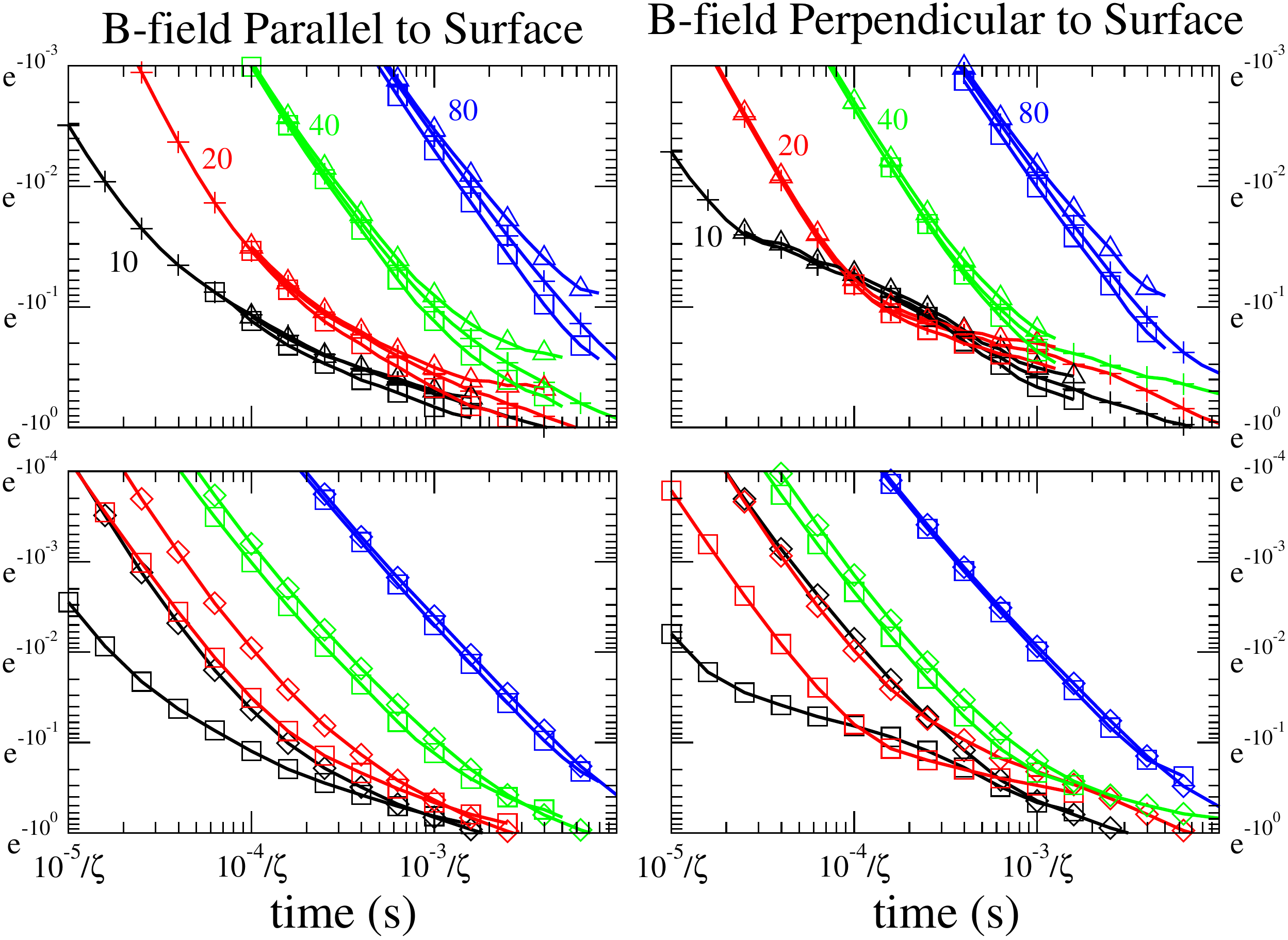}
\includegraphics[width=1.5in]{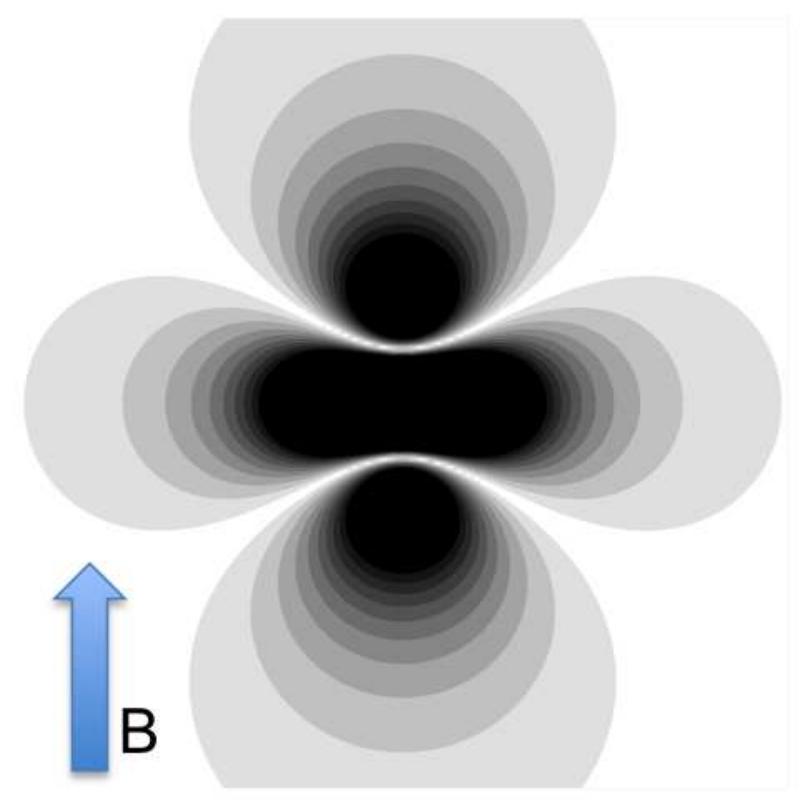}
\includegraphics[width=1.5in]{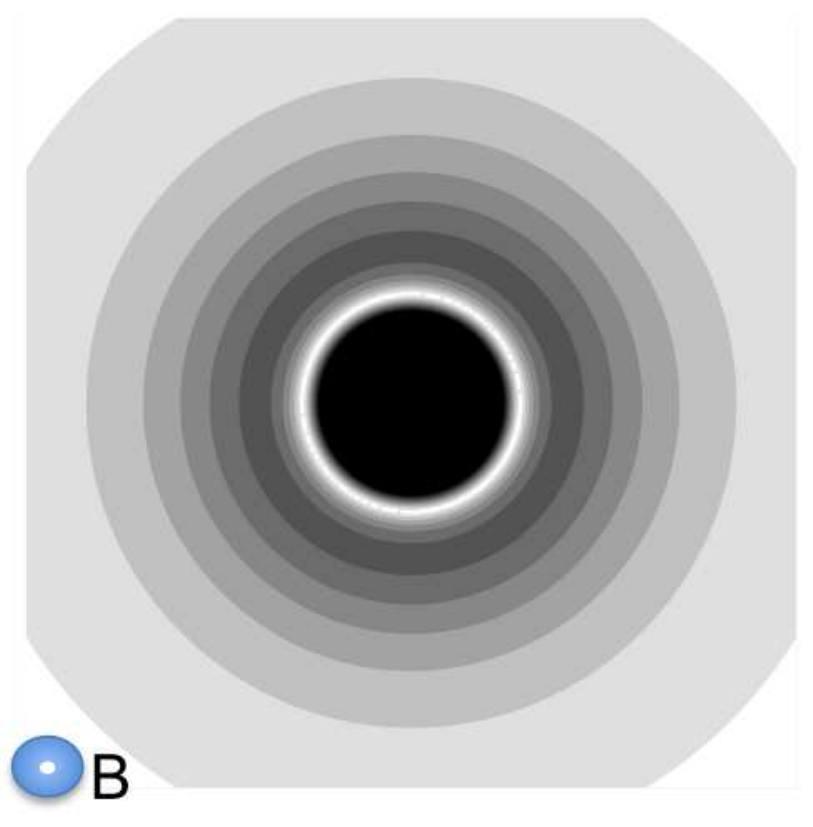}
\caption{
\label{Fig:BathGeom}
(Top) Hahn echo results on logarithmic scales for a central electron spin at various distances from a sheet of
random electron spins at a density of $\zeta^{2/3} \times 10^{11} / \mbox{cm}^2$ with a
magnetic field parallel (left) and perpendicular (right) to the
sheet. The distances are $\zeta^{-1/3} \times 10~\mbox{nm}$ (black), $\zeta^{-1/3} \times 20~\mbox{nm}$ (red), $\zeta^{-1/3} \times 40~\mbox{nm}$ (green), and $\zeta^{-1/3} \times 80~\mbox{nm}$ (blue), as labeled and generally from left
to right (central spins more distant from the surface have longer coherence times).  
The upper panels display $L_{\mbox{\scriptsize{CCE}}}^{(2)}$ (+'s), 
$L_{\mbox{\scriptsize{CCE}}}^{(3)}$ (triangles), and
$L_{\mbox{\scriptsize{CCE}}}^{(4)}$ (squares) mean value results.  
Below those, panels display $L_{\mbox{\scriptsize{CCE}}}^{(4)}$ 
median (diamonds) as well as mean (squares) value results.
Bottom: Contour plots show relative strengths of dipolar couplings to the central
spin from points on the sheet of bath spins (darker regions have
larger relative coupling strengths).
}
\end{figure}

High concentrations of impurity spins may occur at the interface
between materials.  For example, dangling chemical bonds may host
unpaired electrons.  As a variation of our canonical problem with this
in mind, we consider two-dimensional geometries of bath spins.  Our
central spin may be at various distances (depths) from this sheet of
bath spins.  We consider the limit of a large applied magnetic field,
but the results will depend upon the angle of this applied field
relative to the sheet of bath spins.  We show Hahn echo results in
Fig.~\ref{Fig:BathGeom} comparing various central spin depths for a
magnetic field that is parallel or perpendicular to a sheet of bath
spins.  The spin echo curves in the parallel case are fairly smooth compared
with the perpendicular case; this is due to differences in
the spatial dependence of the dipolar coupling to the central spin for
the different magnetic field angles (see the bottom of Fig.~\ref{Fig:BathGeom}).

Since our effective
Hamiltonian [Eq.~(\ref{Eq:HeffOffRes})] scales inversely with distance
cubed (dipolar interactions), we use a scale factor $\zeta$ for
rescaling time, concentration, and depth together appropriately.  $\zeta = 1$ is for
a bath concentration of $10^{11} / \mbox{cm}^2$.  These results are
therefore applicable to various bath concentrations at the appropriately
rescaled central spin depths.

\subsection{Finite spatial extent of the central spin electron wave function}
\label{Sec:CentralGeom}
Our canonical problem idealizes the central spin and bath spins as
being localized with zero extent (points) for the purposes of computing the dipolar interactions
[see Eq.~(\ref{Eq:Dip})].
This is a reasonable approximation for a sparse system of donor-bound
electrons such as Si:P.  However, electrostatically defined quantum dots
may have considerable lateral extent (e.g., roughly $50~\mbox{nm}$ in
Ref.~\onlinecite{Petta_Science05}).  To explore the impact of this
finite extent of the central electron's wave function, we use a simple Gaussian-shaped wave-function model in which
the relative probability of electron occupation is given by
\begin{equation}
\label{Eq:dotProbDensity}
P({\bf x}) \propto \exp{\left(-\frac{x_1^2 + x_2^2}{r_0^2}\right)}
\cos^2{\left(\pi \frac{x_3}{\delta} \right)},
\end{equation}
for $|x_3| < \delta/2$ and $P({\bf x}) = 0$ otherwise.
We use $r_0 = 50~\mbox{nm}$ as a defining lateral radius and $\delta =
5~\mbox{nm}$ as a defining thickness.  
The coordinates are labeled
with $1$, $2$, and $3$ so they will not be confused with the
coordinate system of Eqs.~(\ref{Eq:Ham}) and (\ref{Eq:HeffOffRes})
which define $z$ to point in the direction of the applied magnetic
field.  The $x_3$ direction is normal to an imagined surface forming a 
two-dimensional electron gas out of which the quantum dot is isolated.

\begin{figure}
\includegraphics[width=3.5in]{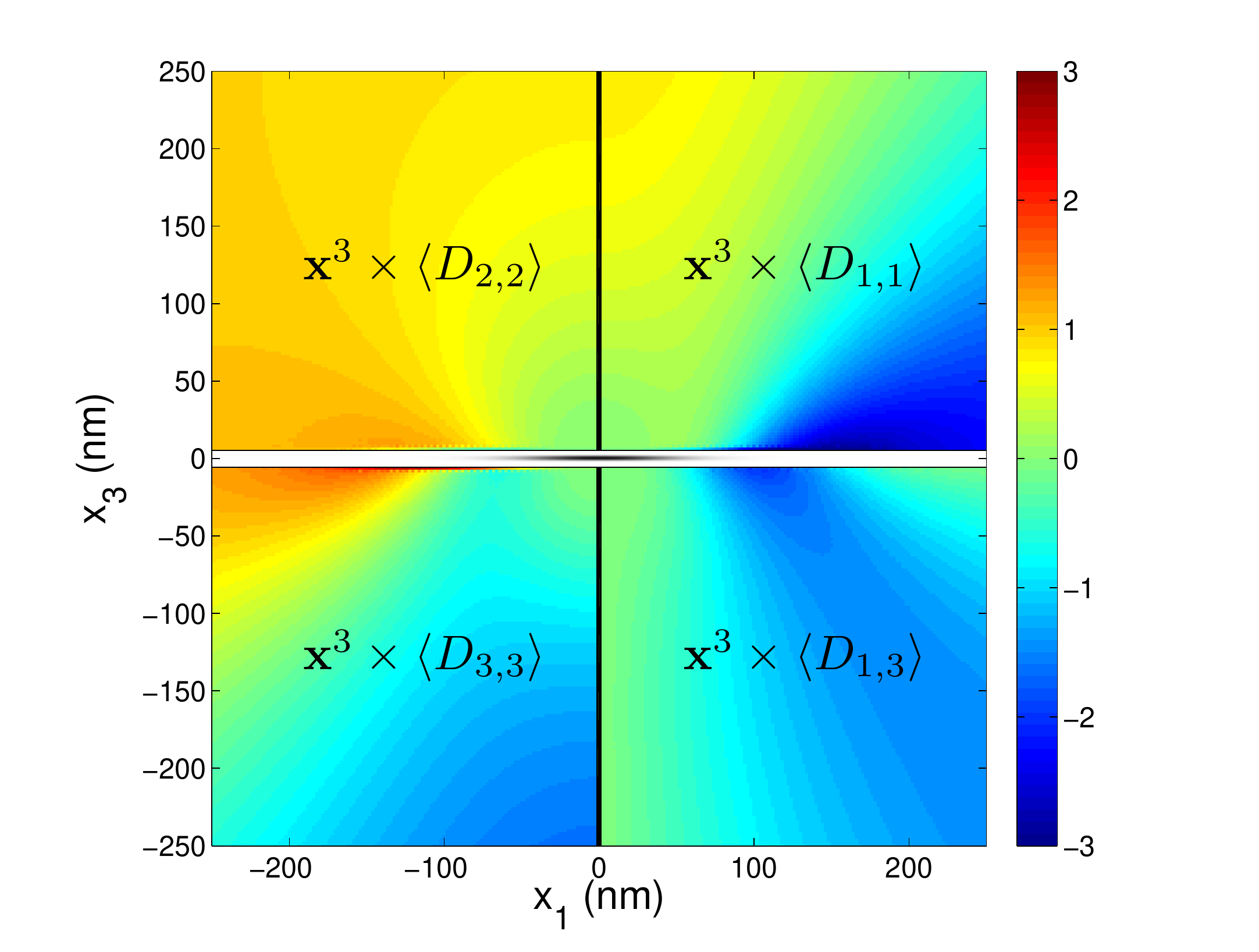}
\caption{
\label{Fig:PancakeDipData}
(color online) Color map representation of the non-zero dipolar tensor elements of Eq.~(\ref{Eq:QdotDipTensor}) in the $x_2 = 0$ plane
for interactions with a quantum dot electron geometrically defined by
Eq.~(\ref{Eq:dotProbDensity}).  By the $x_1$ versus $x_2$ symmetry,
the $x_2 = 0$ plane is sufficient to express the dipolar information.
Furthermore, due to the 
%reflexive 
reflection symmetry about the $x_1$ and
$x_3$ axes, one quadrant is sufficient for each of the non-zero tensor
elements.  We multiply the tensorial values by ${\bf x}^3 = (x_1^2 +
x_3^2)^{3/2}$ for convenience.  The probability density of the
electron in the $x_2 = 0$ plane is represented in a grayscale image
(black for high probability ranging to white for low probability) between the upper and lower quadrants.
}
\end{figure}

The finite extent of the electron's wave function impacts the dipolar interactions
between the central spin and any bath spin.  
We may still use the
effective Hamiltonian of Eq.~(\ref{Eq:HeffOffRes}), but must redefine
the $b_{0, i} = b_{i, 0}$ such that
\begin{equation}
b_{0, i} = \frac{-1}{4} \left \langle D_{z,z} \right \rangle({\bf r}_i),
\end{equation}
where ${\bf r}_i$ is the position of the $n$th bath spin relative to
the center of the quantum dot and we define
\begin{equation}
\label{Eq:QdotDipTensor}
\left \langle D_{\alpha,\beta} \right \rangle({\bf r}) = \int d {\bf
  x} P({\bf x}) D_{\alpha, \beta}({\bf r} - {\bf x}).
\end{equation}
Here, $D_{z, z}$ is an element of the dipolar tensor of Eq.~(\ref{Eq:Dip}) where $z$ refers to the direction of the applied magnetic field.
For full generality, we compute the $\left \langle D_{\alpha, \beta} \right \rangle$ tensor in the quantum dot coordinate system and then rotate as appropriate to obtain $\left \langle D_{z,z} \right \rangle$.  
Due to the symmetry between $x_1$ and $x_2$ in our round quantum dot
defined by Eq.~(\ref{Eq:dotProbDensity}), we may express all of the
tensorial information in the $x_2 = 0$ plane.  In this plane, it is
clear from Eq.~(\ref{Eq:Dip}) that $\left \langle D_{1, 2} \right
\rangle = \left \langle D_{2, 1} \right \rangle = \left \langle D_{2,
  3} \right \rangle = \left \langle D_{3, 2} \right \rangle = 0$.  The
remaining non-zero tensor elements, $\left \langle D_{1, 1} \right
\rangle$, $\left \langle D_{2, 2} \right \rangle$, $\left \langle
D_{3, 3} \right \rangle$, and $\left \langle D_{1, 3} \right \rangle =
\left \langle D_{1, 3} \right \rangle$ are displayed in color map form
in Fig.~\ref{Fig:PancakeDipData}.  Each tensor element is
represented in a different quadrant of the $x_2 = 0$ plane; due to
%reflexive 
reflection symmetry about the $x_1$ axis and the $x_3$ axis, one
quadrant each is sufficient to convey the information for all quadrants.

\begin{figure}
\includegraphics[width=3.5in]{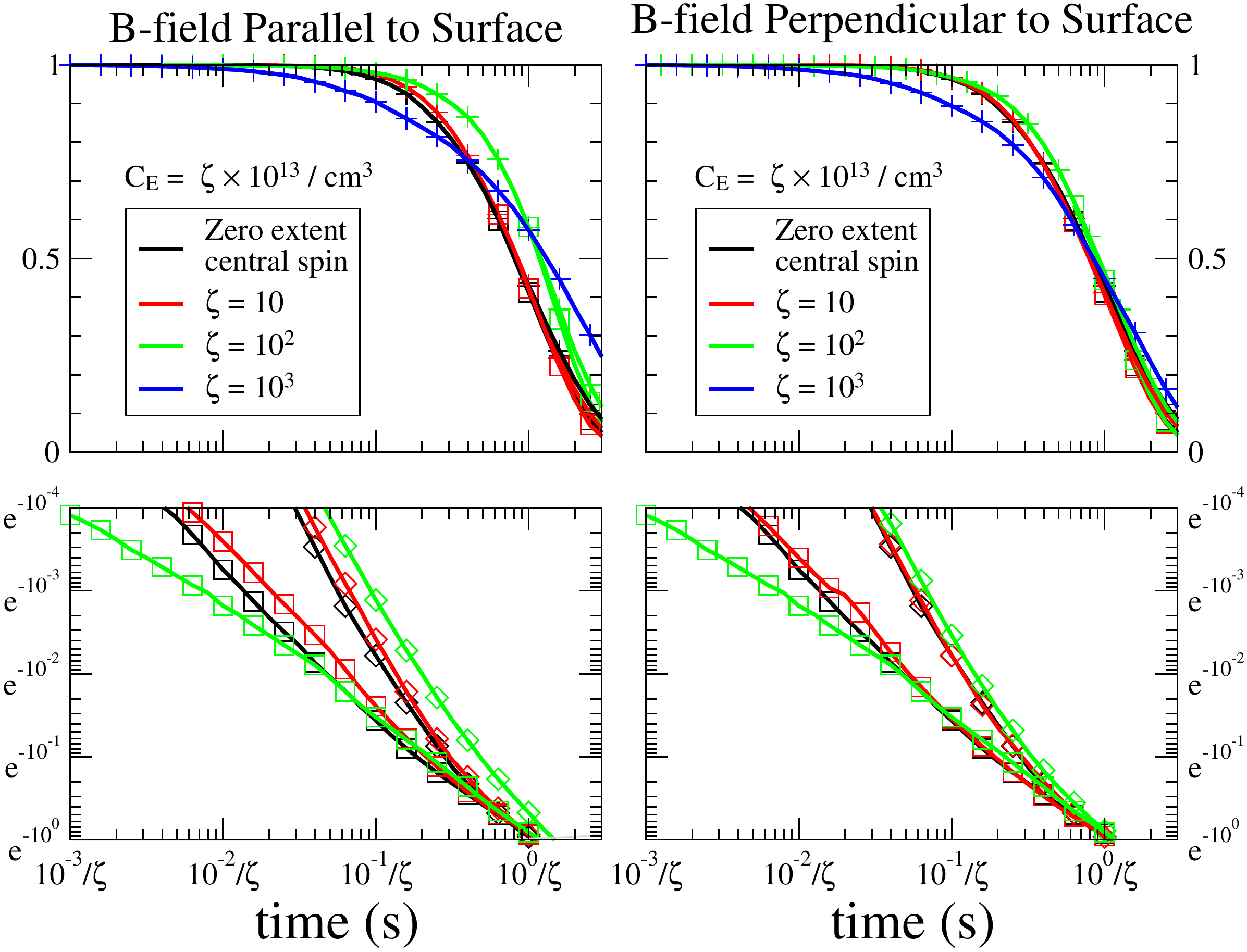}
\includegraphics[width=3.5in]{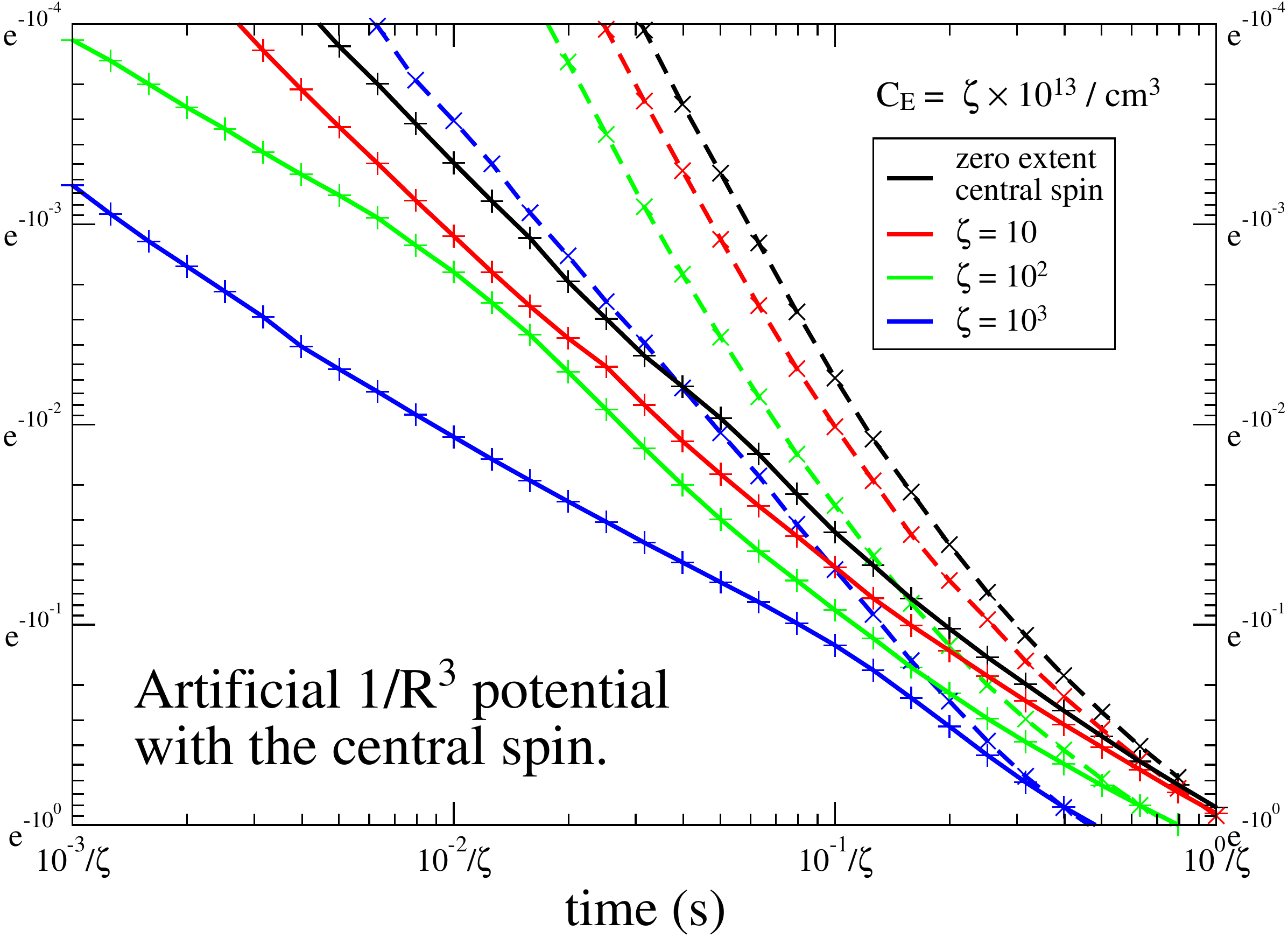}
\caption{
\label{Fig:CompareQubitGeom}
(Top) Comparison of Hahn echo results affected by the spatial extent of the 
central spin electron to various degrees.
Given a central spin ``quantum dot'' geometry defined by Eq.~(\ref{Eq:dotProbDensity}), 
we show results for various electron spin bath concentrations, $C_E$, and compare with the canonical results in which the central spin has zero extent.  
The applied magnetic field is parallel (left) or perpendicular (right) to the surface whose normal is the $x_3$ direction of Eq.~(\ref{Eq:dotProbDensity}).
We use $g=2$ ($g^2 C_E = \zeta \times 4 \times 10^{13}/\mbox{cm}^{3}$).
Using the scaling parameter $\zeta$, time is rescaled inversely with the bath concentration so that the ``zero extent'' curves are universal for all concentrations.
The upper panels display $L_{\mbox{\scriptsize{CCE}}}^{(2)}$ (+'s), 
and $L_{\mbox{\scriptsize{CCE}}}^{(4)}$ (squares) mean value results.  
Below those, panels display $L_{\mbox{\scriptsize{CCE}}}^{(4)}$ 
median (dashed line with diamonds) as well as mean (solid line with squares) value results.
(Bottom) Corresponding results when we use an artificial isotropic $1/R^3$ interaction with the central spin.  $L_{\mbox{\scriptsize{CCE}}}^{(2)}$ mean value (solid +'s) and median value (dashed x's) results are displayed.  Showing only 2-cluster results is sufficient to make our point: without the anisotropy of the interactions there is a consistent trend, monotonic in $\zeta$.
}
\end{figure}

We generated the information that is represented in
Fig.~\ref{Fig:PancakeDipData} from Eq.~(\ref{Eq:QdotDipTensor}) using
integration by Monte Carlo sampling at each point, ${\bf r}$, on a
two-dimensional grid of the $x_2 = 0$ plane.  Using data generated in
this fashion, and linear interpolation between grid points, 
we compute the Hahn spin echo of the quantum dot amongst
various concentrations of a bath of point dipole electron spins
in a 3-D bath (despite using a quantum dot typically defined from 2DEGs).
Results are shown in the upper panels of
Fig.~\ref{Fig:CompareQubitGeom} for two different magnetic field directions: parallel
to the surface (e.g., along $x_1$ or $x_2$), and perpendicular to the
surface (i.e., along $x_3$).
To ensure
adequate grid spacing and precision for our dipolar tensor data, we
compared our Hahn echo results with results using independent dipolar tensor 
data generated for larger grid spacing by a factor of two and found
there to be no significant difference.

In our canonical problem, we found that a rescaling of the
concentration of bath spins is equivalent to an inverse rescaling of
time.  In Fig.~\ref{Fig:CompareQubitGeom}, we show the deviation from
this behavior for our quantum dot qubit by plotting Hahn echo data
versus a time scale that adjusts inversely with the concentration of
bath spins.  This deviation is not terribly large up to
$C_E = 10^{15}/\mbox{cm}^3$.  At $C_E = 10^{16}/\mbox{cm}^3$, the spatial
extent of the central spin causes decoherence that has a significantly
faster initial decay than a point dipole central spin.  This can be
understood by considering the enhancement of dipolar interaction
strengths for bath spins that are near some part of 
the laterally extended quantum dot region but not particularly close 
to the center of this region.  However, in some regimes, such as at
later times, the lateral extent actually causes enhanced coherence.
This counterintuitive enhancement is an effect of the
anisotropy of the dipolar interactions.  It is erratic and differs for
the two cases of the differing $B$-field directions.  The lower panel of
Fig.~\ref{Fig:CompareQubitGeom} shows results using an artificial
$1/R^3$ potential for the interactions with the central spin, removing
the anisotropy, and we find a consistent decrease in the relative
coherence, quantum dot central spin versus point dipole central spin, as
the bath concentration is increased.  Thus, the intuitive
understanding is only thwarted, in some regimes and only slightly, by the anisotropy of the dipolar interactions. 

\subsection{Bath geometry and the wave function of the central spin}
\label{Sec:BathAndCentralGeom}

\begin{figure}
\includegraphics[width=3.5in]{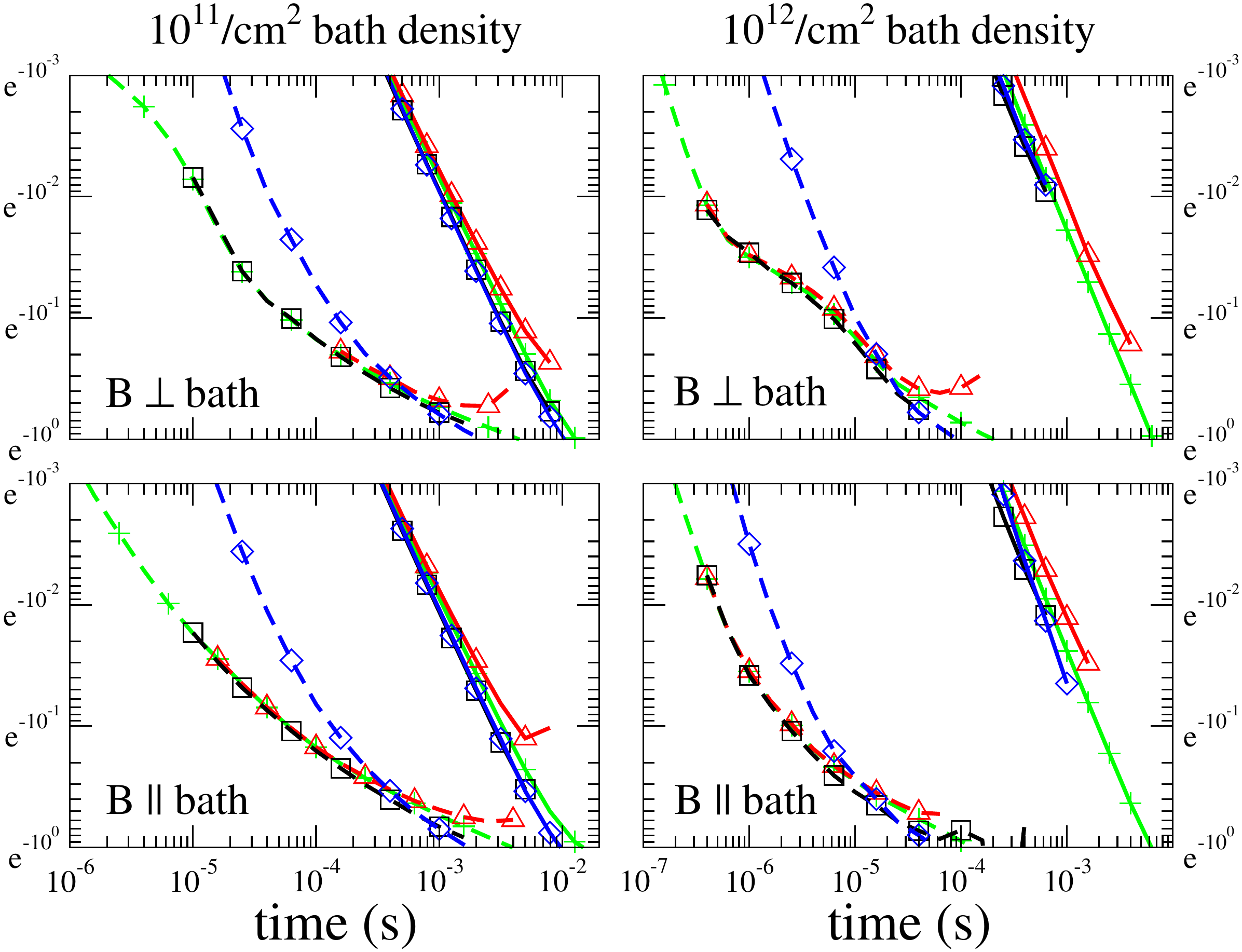}
\includegraphics[width=3.5in]{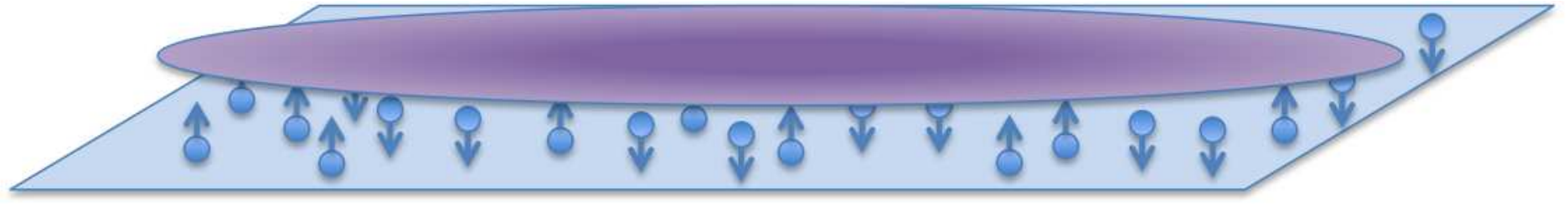}
\caption{
\label{Fig:3nm}
(Top) Hahn echo result of a Gaussian-shaped quantum dot defined by Eq.~(\ref{Eq:dotProbDensity}) with decoherence induced from a two-dimensional bath, in the $x_1, x_2$ plane, $5~\AA$ from its edge ($x_3 = 3~\mbox{nm}$). 
This was chosen as somewhat of a limiting case in terms of the closeness of the bath to the central spin quantum dot.  Dashed lines show results for a point-like central spin for comparison.  We show mean value $L_{\mbox{\scriptsize{CCE}}}^{(2)}$ (+'s), $L_{\mbox{\scriptsize{CCE}}}^{(3)}$ (triangles), and $L_{\mbox{\scriptsize{CCE}}}^{(4)}$ (squares) results, as well as median value $L_{\mbox{\scriptsize{CCE}}}^{(4)}$ (diamonds) results.  
Different colors (color online) are used to help make the curves more easily distinguishable.
Bath densities are $10^{11}/\mbox{cm}^2$ (left) and
$10^{12}/\mbox{cm}^2$ (right), and the applied magnetic field is
perpendicular to the bath and dot (upper) or parallel to the bath and
dot (lower).  (Bottom) Depiction of a laterally extended quantum dot in
the presence of sheet of bath spins.
}
\end{figure}

Combining bath geometry considerations of Sec.~\ref{Sec:BathGeom} with
finite extent of the central electron's wave function discussed  in Sec.~\ref{Sec:CentralGeom}, we
see a variety of different trends.   In Fig.~\ref{Fig:3nm}, we compare
the decoherence effects from a two-dimensional sheet of bath spins at
random positions for
a point-like central spin and a Gaussian-shaped quantum dot central
spin with the bath sheet at a distance of $3~\mbox{nm}$ from the
center of the central spin.  For the quantum dot central spin case,
this corresponds to $5~$\AA above the edge of 
the wave function, $x_3 = \delta/2 + 5~$\AA [Eq.~(\ref{Eq:dotProbDensity})]; we are considering this roughly as a limiting case for a sheet of
electron spins that is very close to the quantum dot but spatially independent.
Unlike the trend observed when changing the qubit's wave function
in the canonical three-dimensional sparse bath
(Fig.~\ref{Fig:CompareQubitGeom}), the coherence time for the quantum dot
central spin is actually increased relative to the point-like central
spin.  In the former case, more decoherence resulted from enhanced
dipolar interactions near the laterally extended quantum dot region.
Here, where we consider a 2-D bath at densities of 
$10^{11}/\mbox{cm}^2$ or $10^{12}/\mbox{cm}^2$, the extent of the quantum dot actually
reduced decoherence because it tends to reduce its sensitivity to
flip-flopping pairs of nearby bath spins.  To put this another way, the
important factor is the {\it difference} in the effective magnetic
field experienced by the central spin as bath spins flip-flop, not the
absolute magnitude of their interactions.  Thus, while the extent of
the quantum dot wave function tends to increase the strength of dipolar
interactions to the bath spins, in certain geometries, this difference
in interactions amongst nearby bath spins may be reduced.  The depiction at the bottom
of Fig.~\ref{Fig:3nm} helps to illustrate this effect.  It should
also be noted that, in going from the point-like limit to the
extended quantum dot wave function, we enter a regime in which the mean
and median of the Hahn echo decay are nearly identical; indeed, for a sufficiently dense bath,
we expect the decay to result from a large number of small cluster contributions to
govern the decoherence and, as an effect of the central limit theorem, most random instances of
the bath should and do yield similar results.

\begin{figure}
\includegraphics[width=3.5in]{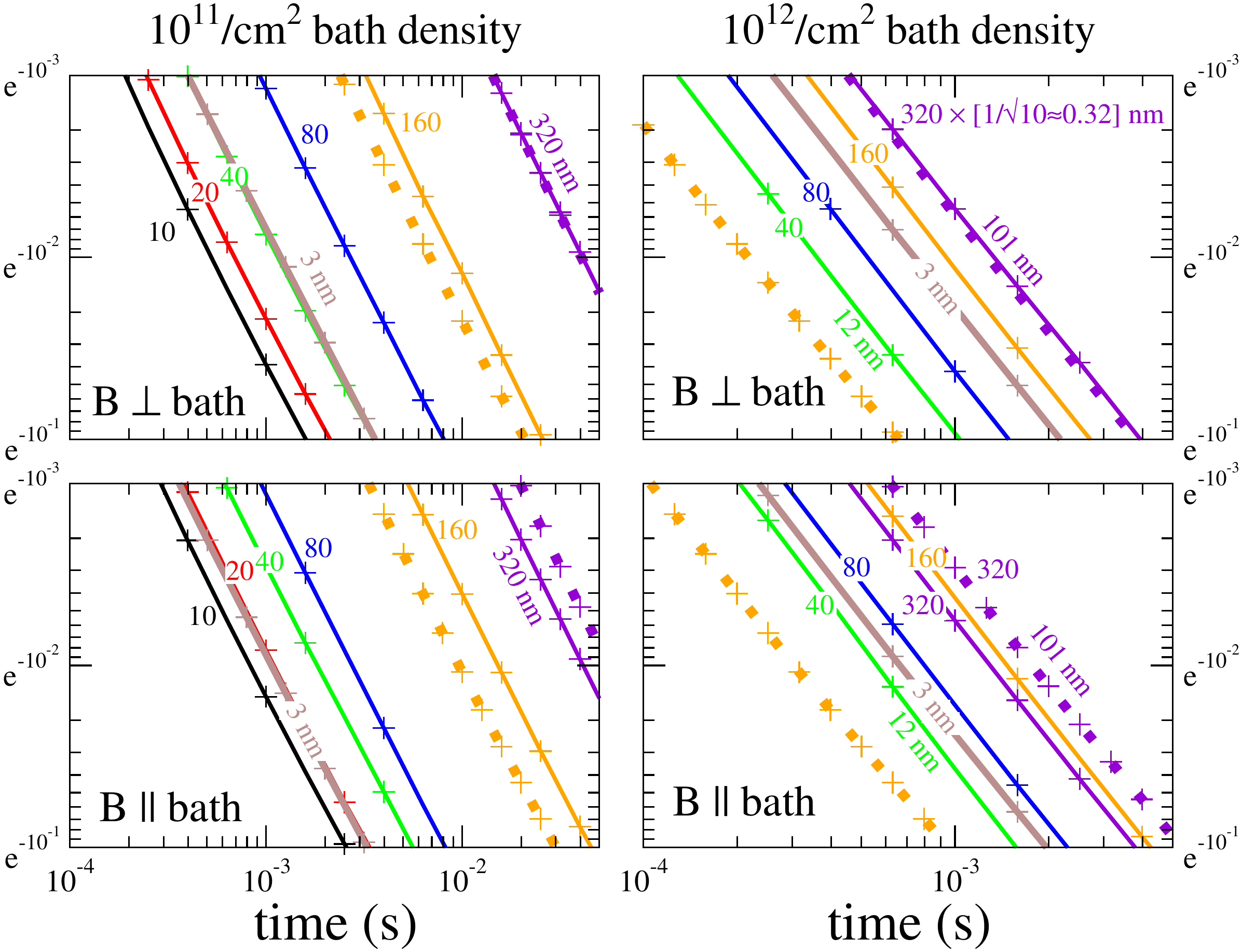}
\caption{
\label{Fig:DotVsPointManyDepths}
Hahn echo result of a Gaussian-shaped quantum dot defined by
Eq.~(\ref{Eq:dotProbDensity}) with decoherence induced from a
two-dimensional bath, in the $x_1, x_2$ plane, at various depth distances along $x_3$.
Dotted lines show results for a zero extent (point-like) central spin for comparison.  
%We show mean value $L_{\mbox{\scriptsize{CCE}}}^{(2)}$ (+'s), $L_{\mbox{\scriptsize{CCE}}}^{(3)}$ (triangles), and $L_{\mbox{\scriptsize{CCE}}}^{(4)}$ (squares) results, as well as median value $L_{\mbox{\scriptsize{CCE}}}^{(4)}$ (diamonds) results.  
%Different colors are used to help make the curves more easily distinguishable.
Bath densities are $10^{11}/\mbox{cm}^2$ (left) and $10^{12}/\mbox{cm}^2$ (right), and the applied magnetic field is perpendicular to the bath and dot (top) or parallel to the bath and dot (bottom).
As labeled, the depths (color online) are defined in correspondence to Fig.~\ref{Fig:BathGeom}: $\zeta^{-1/3} \times 10~\mbox{nm}$ (black), $\zeta^{-1/3} \times 20~\mbox{nm}$ (red), $\zeta^{-1/3} \times 40~\mbox{nm}$ (green), $\zeta^{-1/3} \times 80~\mbox{nm}$ (blue), $\zeta^{-1/3} \times 160~\mbox{nm}$ (orange), and , $\zeta^{-1/3} \times 320~\mbox{nm}$ (purple)
where $\zeta = 1$ for $10^{11}/\mbox{cm}^2$  and $\zeta = 10^{3/2}
\approx 32$ for $10^{12}/\mbox{cm}^2$ ($\zeta^{-1/3} = 10^{-1/2}
\approx 0.32$).  Additionally, the $3~\mbox{nm}$ depth results from
Fig.~\ref{Fig:3nm} are displayed in thick brown.
These are all results of $L_{\mbox{\scriptsize{CCE}}}^{(2)}$ mean values (+'s) as a reasonable short time approximation to understand trends.
}
\end{figure}

In Fig.~\ref{Fig:DotVsPointManyDepths}, we show decoherence induced by
2-D sheets of randomly located bath spins at various depths from a
quantum dot central spin.  We show only the 2-cluster results as a rough
approximation to 
understand general trends as we approach the large distance limit in which the
zero-extent approximation of the central spin is valid.  Some
peculiarities emerge.  First, in all of the cases we consider
[$10^{11}/\mbox{cm}^2$ (left) and $10^{12}/\mbox{cm}^2$ and different
magnetic field orientations], there is actually an initial decrease in
coherence time as we move beyond a $3~\mbox{nm}$ depth from the bath.
This is somewhat counterintuitive, but the effects of the change in
dipolar interactions with changing depth is not so straightforward due to the angular dependence of the dipolar interactions.
Beyond about $10~\mbox{nm}$, however, it follows the more intuitive
trend that increasing the depth causes coherence times to increase.
There is an exception for the case with a $10^{12}/\mbox{cm}^2$ bath density and parallel B
field in going from $\zeta^{-1/3} \times
160~\mbox{nm}$ to $\zeta^{-1/3} \times 320~\mbox{nm}$ depth
($51~\mbox{nm}$ to $101~\mbox{nm}$);
this is a relatively minor effect that is apparently related to
the anisotropy of dipolar interactions.  Generally we find that we
approach the large distance limit at a depth of about $\zeta^{-1/3}
\times 320~\mbox{nm}$ for the two densities we study.
In this limit, we can neglect the spatial extent of the central spin
wave function and use the results 
from Sec.~\ref{Sec:BathGeom}.

\subsection{Pulse Sequences}

\begin{figure}
\includegraphics[width=3.5in]{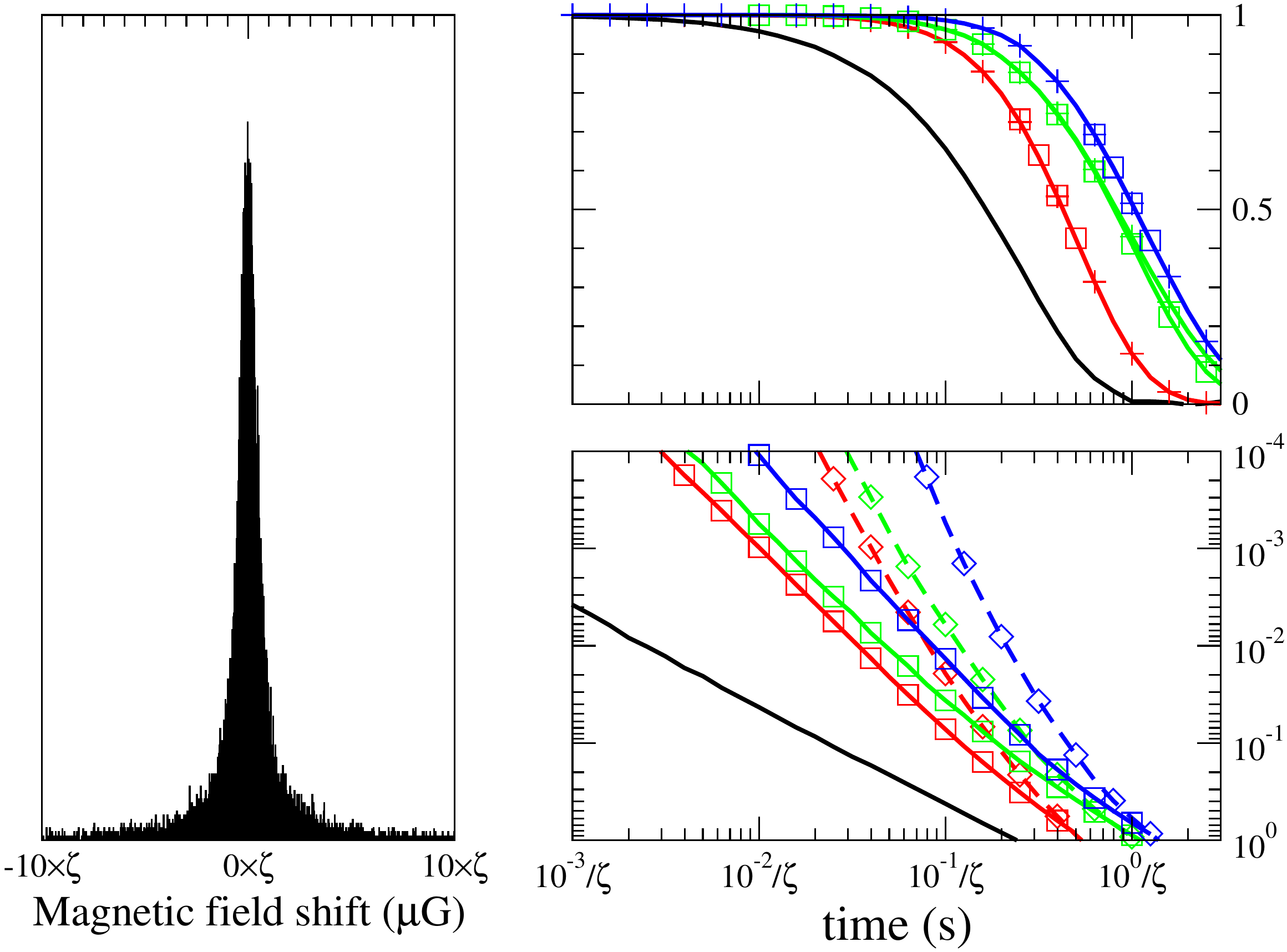}
\includegraphics[width=3.5in]{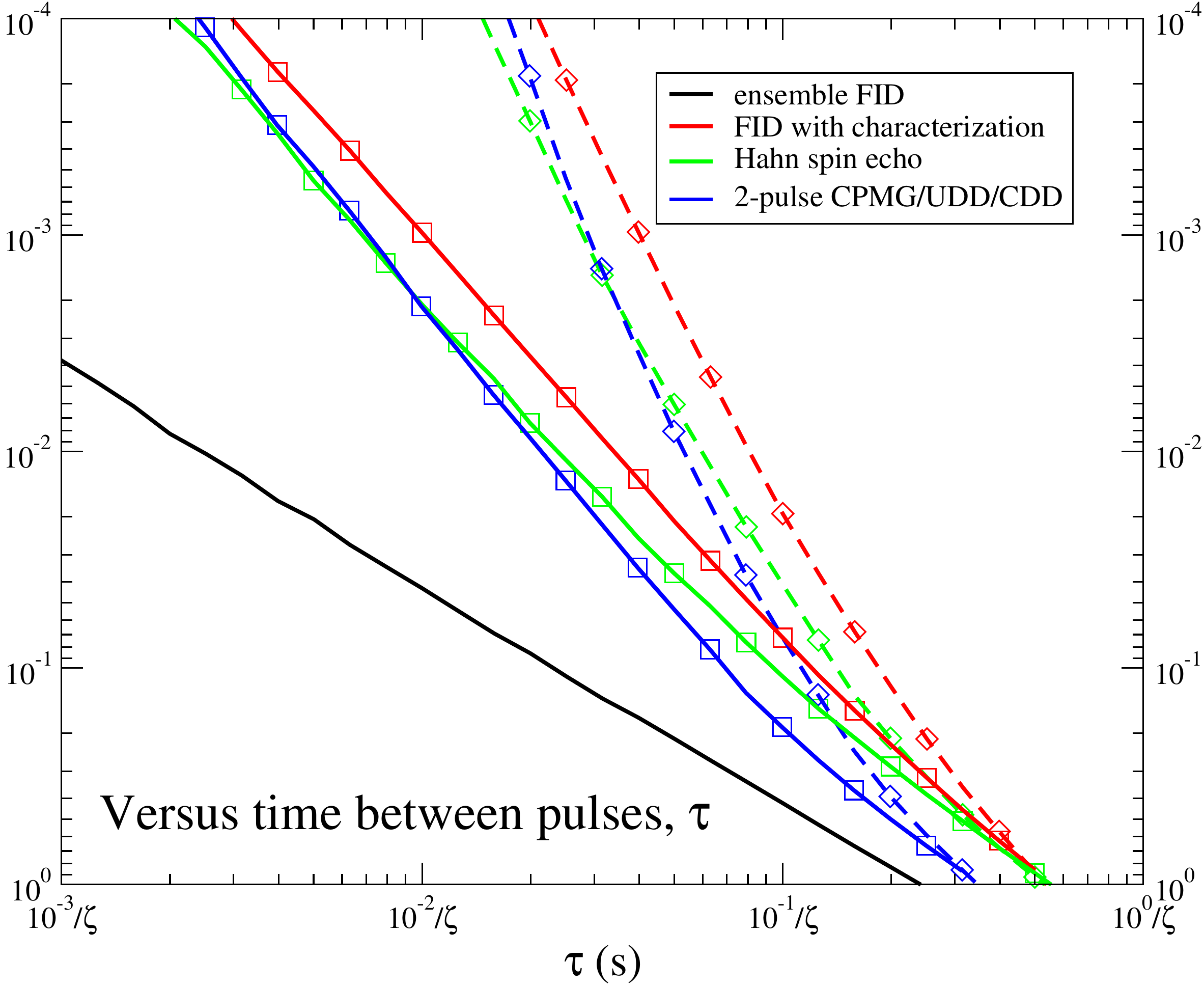}
\caption{
\label{Fig:PulseSeq}
Top left: probability distribution of magnetic field shifts due to background electron spins in the infinite temperature, point dipole limit.  The distribution is over different random
spatial realizations and spin states of the background spins.
Top right: coherence versus time for ensemble FID (black), FID with characterization (red), Hahn spin echo (green), and CPMG (blue), with a left to right trend, respectively.  We show mean values of $L_{\mbox{\scriptsize{CCE}}}^{(2)}$ (solid +'s) and $L_{\mbox{\scriptsize{CCE}}}^{(4)}$ (solid squares).  The bottom of the right panels displays results on a logarithmic scale and includes median values of 
$L_{\mbox{\scriptsize{CCE}}}^{(4)}$ (dashed diamonds).
Bottom: coherence versus $\tau$, the time between $\pi$ pulse for the same pulse sequences.  $\tau = t / 2$ for the Hahn spin echo and $\tau = t/4$ for the two-pulse CPMG.
The $\zeta$ scaling parameter is used as in
Fig.~\ref{Fig:SpatialEnsemble}: $g^2 C_E = \zeta \times 4 \times 10^{13}/\mbox{cm}^{3}$.
}
\end{figure}

As we discussed in Sec.~\ref{Sec:SpinEcho}, we chose an ideal Hahn
spin echo as the context of our canonical problem as a standard
approach of eliminating the effects of inhomogeneous broadening.
In this section, we consider other pulse sequence scenarios.  We first
examine the effects of inhomogeneous broadening itself which can
incur errors when refocusing pulses are not used.  We display the
probability distribution of magnetic field shifts of an ensemble, responsible for
inhomogeneous broadening, due to background electron spins in the infinite temperature, point dipole limit on the upper left side of Fig.~\ref{Fig:PulseSeq}.
On the upper right side of Fig.~\ref{Fig:PulseSeq}, we compare coherence 
versus time calculations in four different contexts: ensemble free
induction decay (FID), FID with characterization, Hahn
spin echo, and a two-pulse Carr-Purcell-Meiboom-Gill
(CPMG)~\cite{carr_effects_1954, meiboom_modified_1958} sequence.  

The ensemble FID corresponds to the previously described case of ensemble averaging of the signal which has no dependence upon bath dynamics whatsoever.
FID with characterization, or narrowed state
FID (NFID), takes the inhomogeneity out of the problem by accounting for a static
offset in the magnetic field characterized individually for each qubit; coherence of the central spin is lost,
however, due to dynamics of the bath.\cite{Yao_PRB06,Liu_NJP07} This corresponds to an experiment in which the central spin splitting is pre-measured, or prepared with a well fixed value.
The Hahn spin echo is simply
the case of the canonical problem.  In the context of ideal applied
pulses in a dephasing-only limit, the CPMG is equivalent to the Carr-Purcell sequence and is
represented as $(\tau \rightarrow \pi \rightarrow \tau)^{n}$ with $t
= 2 n \tau$.  The two-pulse CPMG sequence, $\tau \rightarrow \pi
\rightarrow 2 \tau \rightarrow \pi \rightarrow \tau$, is equivalent to the
two-pulse Uhrig dynamical decoupling~\cite{Uhrig_PRL07, Yang_PRL08} (UDD) sequence as well as
a first-level concatenation of the spin echo sequence~\cite{Khodjasteh_PRL05,Yao_PRL07}  Our CCE methods, as discussed in Sec.~\ref{Sec:Methods} work well for FID with characterization and for the two-pulse CPMG down to below $50\%$ of the decay.  For these cases, we do not show the tail of the decay for 
$L_{\mbox{\scriptsize{CCE}}}^{(4)}$ 
mean results because we fail to obtain convergent results.  However, the initial part of the decay is often most relevant for quantum computation in any case.

The coherence time improves successively in the FID-NFID-SE-CPMG$_{2}$ sequence of experiments.  In Refs.~\onlinecite{Witzel_CDD_PRB07} and
\onlinecite{Lee_PRL08}, CDD sequences were shown to work well for spectral diffusion in cases where the intra-bath coupling is weak compared with the qubit-bath coupling and UDD sequences work even better in cases where the relevant time scale is short compared with all of the couplings.  Neither of these perturbations are particularly relevant in the problem considered here.  If we were in the regime of weak intra-bath coupling, larger clusters (than pairs) might dominate the two-pulse CPMG results as observed in Refs.~\onlinecite{Witzel_CDD_PRB07} and \onlinecite{Lee_PRL08}.  This is not observed here.  If we were in the short time regime, we would see an $\exp{(-t^4)}$ decay for the Hahn spin echo and $\exp{(-t^6)}$ decay for the two-pulse CPMG.  We do not; in the logarithmic-scale plots of Fig.~\ref{Fig:PulseSeq}, we do not observe a significant change of slope between the Hahn spin echo decay and the two-pulse CPMG decay.  We do see some improvement with the two-pulse CPMG sequence in any case; although, as a function of the time between pulses $\tau$, we roughly break even relative to the Hahn spin echo (bottom of Fig.~\ref{Fig:PulseSeq}).  This is consistent with the calculations based on classical Ornstein-Uhlenbeck noise: the $1/\omega^{2}$ tail of the spectral density of this noise leads to $\exp(-t^{3})$ decay for any dynamical decoupling pulse sequence, and it leads to a sublinear $T_{2}\sim n^{2/3}$ scaling of the coherence time with the number of pulses. \cite{Cywinski_PRB08,deLange_Science10,medford_scaling_2012,Wang_PRB12}

\section{Applications}
\label{Sec:Applications}

Now that we have presented a detailed study of our canonical problem in Secs.~\ref{Sec:Canonical} and \ref{Sec:Methods} and looked at several variants of the problem in Sec.~\ref{Sec:Variants}, we now discuss specific applications.  We consider two crystalline material substrates, silicon and carbon (in diamond form), in which nuclear spins may be nearly eliminated through enrichment.  In both of these cases, however, impurity electron spins can become the predominant source of decoherence.  Such spin baths relate to our canonical problem and its variants.

\subsection{Donor in silicon}
\label{Sec:DonorApp}

Donors in silicon make a promising candidates to host quantum bits in
the form of electron spins.  In bulk, experiments indicate that
donor-bound electrons can maintain spin coherence on the timescale of a
second by using enriched silicon with few nuclear
spins,\cite{Tyryshkin_2011} and single-spin 
read-out\cite{morello_single-shot_2010} as well as coherent 
control\cite{pla-SiWorkshop, morello-QCPR, morello-APS-MM} has been
demonstrated.  In this section, we consider
common spin baths that may affect these qubits: $^{29}$Si, background
phosphorus donors, and electron spins at an interface.

Previous work\cite{Witzel_PRB05,Witzel_PRB06,Witzel_AHF_PRB07} has demonstrated remarkable
agreement of cluster expansion calculations for spin echo decay with
corresponding electron spin resonant measurements in natural
silicon~\cite{Tyryshkin_PRB03,Tyryshkin_JPC06} and accurately predicted the effects of
varying $^{29}$Si concentrations with experimental
measurements,\cite{Tyryshkin_PRB03,Tyryshkin_JPC06,Abe_PRB10}.
In Ref.~\onlinecite{Witzel_PRL10}, we examined the effects of
background phosphorus donors when the $^{29}$Si has been reduced to
very low concentrations through enrichment and demonstrated agreement
with experiments now published in Ref.~\onlinecite{Tyryshkin_2011}.

\begin{figure}
\includegraphics[width=3.5in]{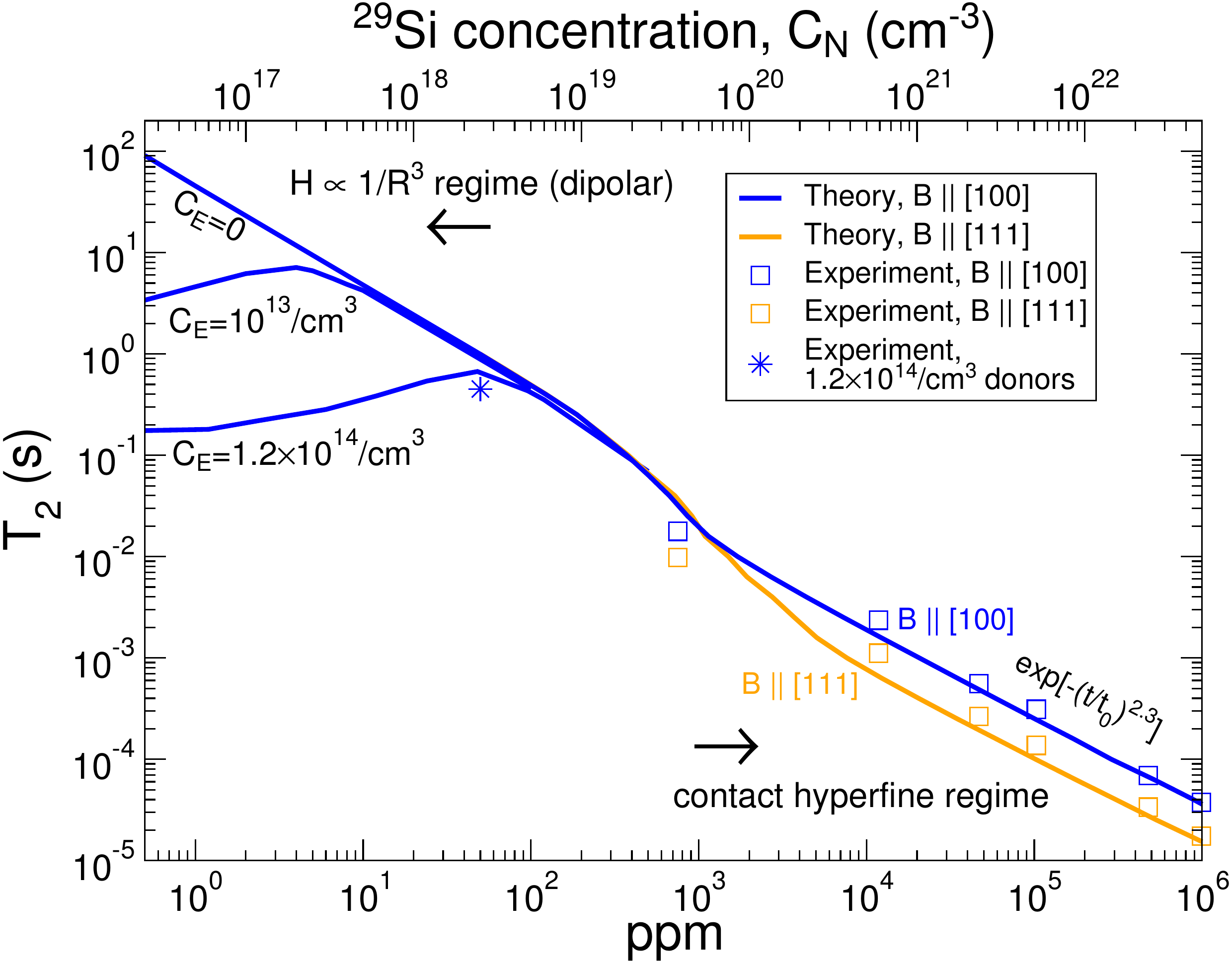}
\caption{
\label{Fig:T2vs29Si}
Hahn spin echo $T_2$ times, when the Hahn echo reaches a value of
$\exp{(-1)}$, for phosphorus donors as a function of the fraction of
$^{29}$Si donors, or corresponding concentration $C_N$, as well as the
concentration of background donors $C_E$.  At high $C_N$, contact hyperfine interactions
dominate and $T_2$ is dependent upon the magnetic field direction
relative to the lattice orientation.  At low $C_N$, $T_2$ is
dependent upon $C_E$, and eventually dominated only by
dipolar interactions (which includes 
dipolar-approximated electronuclear interactions).
Experimental results are
shown as square symbols, from Ref.~\onlinecite{Abe_PRB10}, and a star
symbol at 50~ppm $^{29}$Si, from Ref.~\onlinecite{Tyryshkin_2011}.  This figure is
slightly revised from that of Ref.~\onlinecite{Witzel_PRL10},
updating the $50~\mbox{ppm}$ $^{29}$Si experimental value to the published $T_2
= 450~\mbox{ms}$.}
\end{figure}

Ignoring the background phosphorus donors for a moment, we note that a
spin bath of $^{29}$Si in the low concentration limit, in which the spins act
as point dipoles, corresponds to the $g_{i>0} < g_0$ variant of
Fig.~\ref{Fig:CompareG0}.  In this case, $g_0 \approx 3000 \times
g_{i>0}$; the electron spin has a much larger magnetic moment than the
$^{29}$Si.  The background phosphorus donor spin bath
alone, in the low concentration limit, corresponds to the
two bath species variant of Fig.~\ref{Fig:Resonances} because of the
spin-$1/2$ phosphorus nucleus; to a good approximation, we can assume
that donors with differing nuclear polarizations do not flip-flop with
each other.  In Fig.~\ref{Fig:T2vs29Si}, we show Hahn spin echo $T_2$
times as a function of $^{29}$Si (nuclear spin) concentration, $C_N$,
in combination with various concentrations of phosphorus donors
(electron spins), $C_E$.  The $^{29}$Si bath induces decoherence through
its flip-flopping dynamics, but it also suppresses donor-induced
decoherence via Overhauser shifts that cause the donors to be
off-resonant with each other.  This effect is demonstrated in
Fig.~\ref{Fig:T2vs29Si} by the initial increase of $T_2$ as $^{29}$Si
is increased for the $C_E > 0$ cases.
However, this effect is not very prominent in the short time regime that is of most
interest for meeting quantum error correction thresholds.\cite{Witzel_PRL10}

\begin{figure}
\includegraphics[width=3.5in]{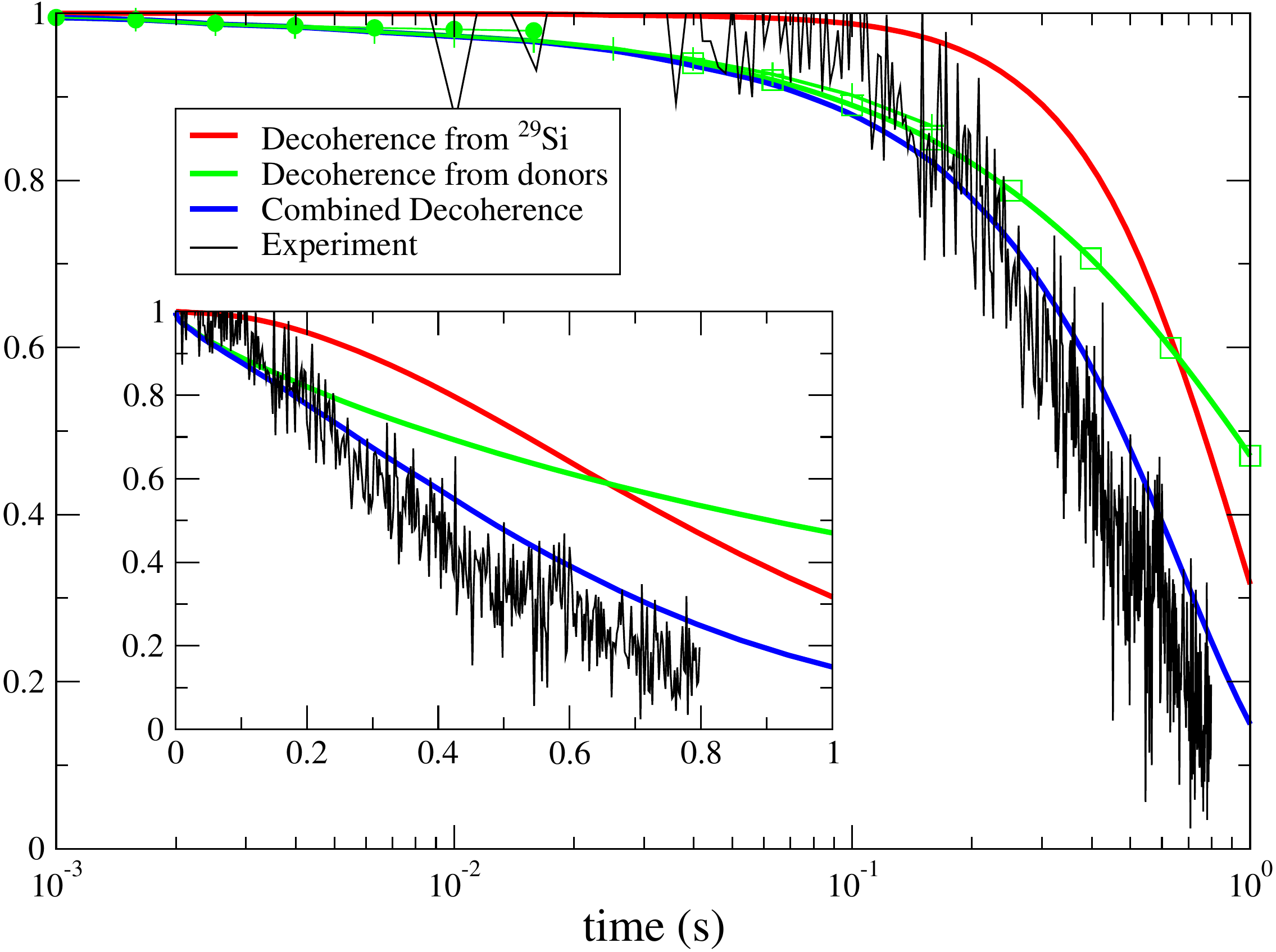}
\caption{
\label{Fig:comp450msExp}
Hahn spin echo versus time corresponding with the 50~ppm $^{29}$Si and
$C_E = 1.2 \times 10^{14} / \mbox{cm}^{3}$ experiment of
Ref.~\onlinecite{Tyryshkin_2011}.  Time is shown on a logarithmic
scale in the main plot and a linear scale in the inset.
The theory exhibits a $T_2 \approx
600~\mbox{ms}$, reasonably close to the $450~\mbox{ms}$ $T_2$ fit of
the experiment~\cite{Tyryshkin_2011}.
The short-time (first few percent
of the decay) behavior is dominated by 1-cluster contributions shown
with filled circle symbols; these are direct flip-flops between the
central spin and other donors.  Indirect flip-flop processes then dominate the
decay, 2-clusters (+'s) then 4-clusters (squares).  In the $T_2$
regime near $600~\mbox{ms}$, the decoherence is a combined effect from
$^{29}$Si flip-flops as well as donor flip-flops.
}
\end{figure}

In Fig.~\ref{Fig:comp450msExp}, we examine the Hahn spin echo
corresponding with the $T_2 \approx 450~\mbox{ms}$ experiment  from Ref.~\onlinecite{Tyryshkin_2011}.  This is a bulk experiment in which all the 
donors are subject to the spin echo refocusing pulses.  As
a result, the measured decoherence signal is dominated by
instantaneous diffusion, the inhomogeneous broadening effects of the
donors that are mutually flipped by the refocusing pulse.  In other
words, the spin echo is only able to cancel the effects of
inhomogeneous broadening from the parts of the bath that remain
unchanged by the refocusing pulse.  However, by performing a series of
measurements in which the angle of the refocusing pulse is varied
(i.e., smaller than $\pi$ pulses), they extrapolate the effective
single-spin decoherence.  Our calculations apply to this extrapolated
result.  In order to confirm that the dominant decoherence is due to 
flip-flopping donors, Ref.~\onlinecite{Tyryshkin_2011} presents measurements of donor decoherence in the presence of a significant magnetic field gradient, prolonging $T_2$ to an extrapolated value of about $10~\mbox{s}$.  Such a gradient suppresses donor flip-flopping by shifting them off resonance from each other.

Two types of decoherence induced by donor flip-flopping are distinguished in Ref.~\onlinecite{Tyryshkin_2011} as indirect flip-flops (flip-flops among bath spins) and direct flip-flops (flip-flops with the central spin).
In Ref.~\onlinecite{Witzel_PRL10}, we did not consider
the effects of direct flip-flops between the central spin and
background donor spins.  This effect, which we
examine in Fig.~\ref{Fig:Resonances} (where the central spin is
resonant with bath spins), results in 1-cluster contributions that
dominate in the short time regime.  We include both direct and indirect flip-flops in Fig.~\ref{Fig:comp450msExp} and note that the direct flip-flops have a negligible effect on the $T_2$ time but significant effect on the short time behavior.

\begin{figure}
\includegraphics[width=3.5in]{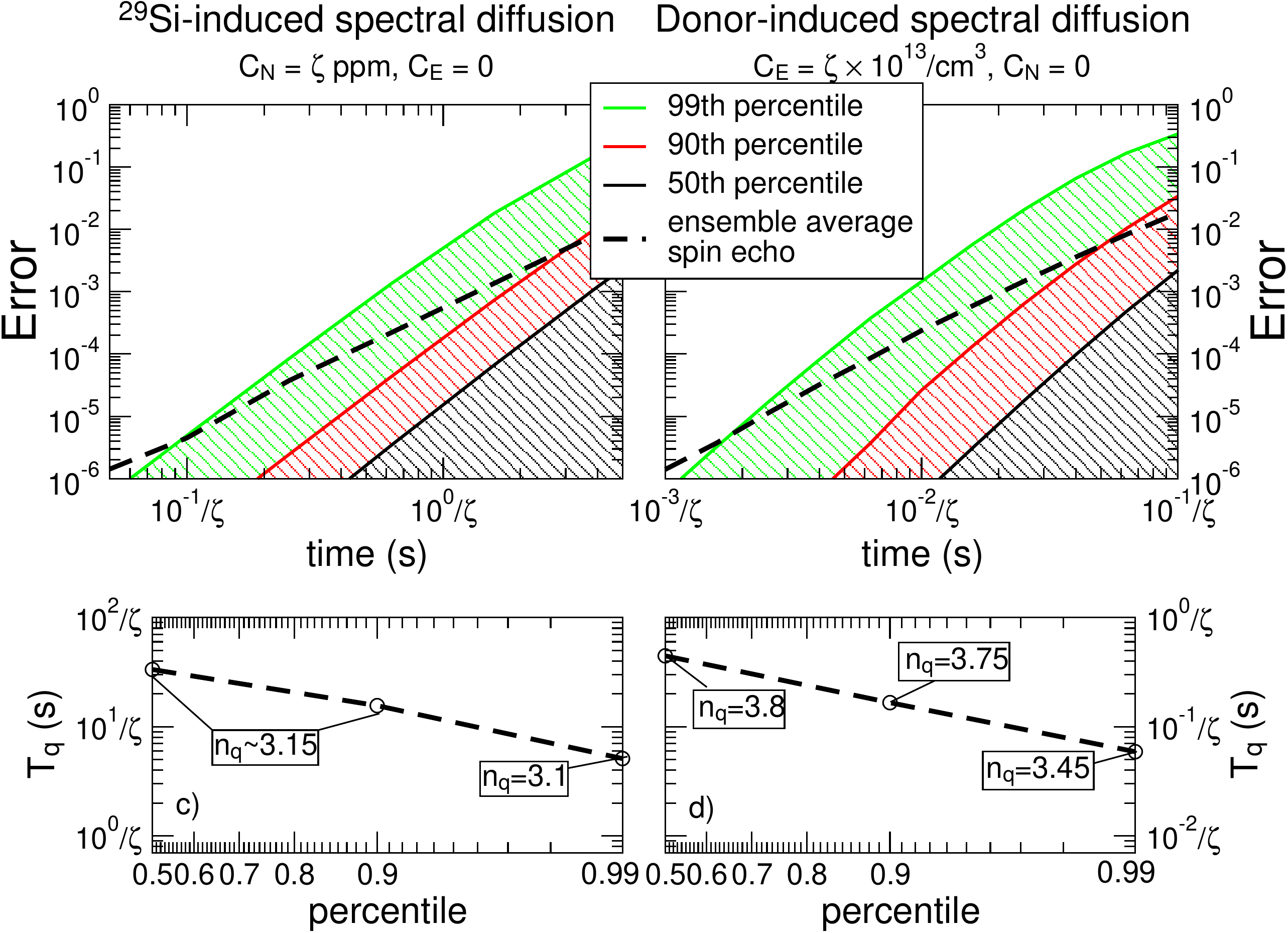}
\caption{
\label{DonorStatistics}
Statistical dependence of the Hahn spin echo upon various random
instances of spin baths, $^{29}$Si (left) or donors (right), in
the low concentration limit.  Top: Hahn echo error (one minus the
echo) for various percentiles, computed at each time point
independently, and the ensemble averaged spin echo.  Bottom: fits of
the error to $1 - \exp{\left[-(t / T_q)^{n_q} \right]} \approx (t /
T_q)^{n_q}$ in the $10^{-4}$ error regime (motivated by common
fault-tolerance thresholds) for various percentiles.  The $\zeta$
scaling parameter applies simultaneously to the time scale and concentration scale,
having an inverse relationship in the low concentration limit.
}
\end{figure}

It is important to keep in mind that there will be large statistical
variations for different spatial realizations of spin baths when we are in the
low concentration regime.\cite{Dobrovitski_PRB08}  This has been indicated by differences in
the median and mean values throughout this paper.  The statistics are
examined in greater detail in Fig.~\ref{DonorStatistics} for $^{29}$Si
spin baths and for background donor spin baths separately.

In addition to $^{29}$Si and background donors, interfaces that play
important roles in semiconductor technology may introduces additional
baths of spins that may induce decoherence.  Experiments~\cite{schenkel_electrical_2006} have
demonstrated that donors closest to an interface have shorter
coherence times which have thus far proven to be much shorter than the
coherence times observed in the bulk.  We refer to
Fig.~\ref{Fig:BathGeom} with spin echo calculations for decoherence
induced by a 2-D bath of electron spins.  This is applicable where we
can approximate spin interactions as point dipole interactions and
where bath spins are resonant with one another.  Dangling bond spins
at an interface may have $g$-factor variations that cause them to be
off-resonant with one another and suppress flip-flopping
noise.  The experiments of
Ref.~\onlinecite{schenkel_electrical_2006} exhibit faster spin echo decay than we
would expect from calculations along the lines of
Fig.~\ref{Fig:BathGeom}, even assuming no $g$-factor variation.  A
different theory of dangling bond spin -induced decoherence 
proposed in Ref.~\onlinecite{deSousa_PRB07}, involving phonons and
spin-orbit interactions, matches with Ref.~\onlinecite{schenkel_electrical_2006} but
it requires dangling bond concentrations as high as $10^{13} /
\mbox{cm}^2$, which might be unrealistic.  There is no consensus at this time as to the true cause
of the decoherence in Ref.~\onlinecite{schenkel_electrical_2006}. 
One suspicion is that the noise is induced by exchange-coupled parasitic dots at the interface such as observed in Ref.~\onlinecite{shankar_spin_2010}.

\subsection{Nuclear spin qubit}

\begin{figure}
\includegraphics[width=3.5in]{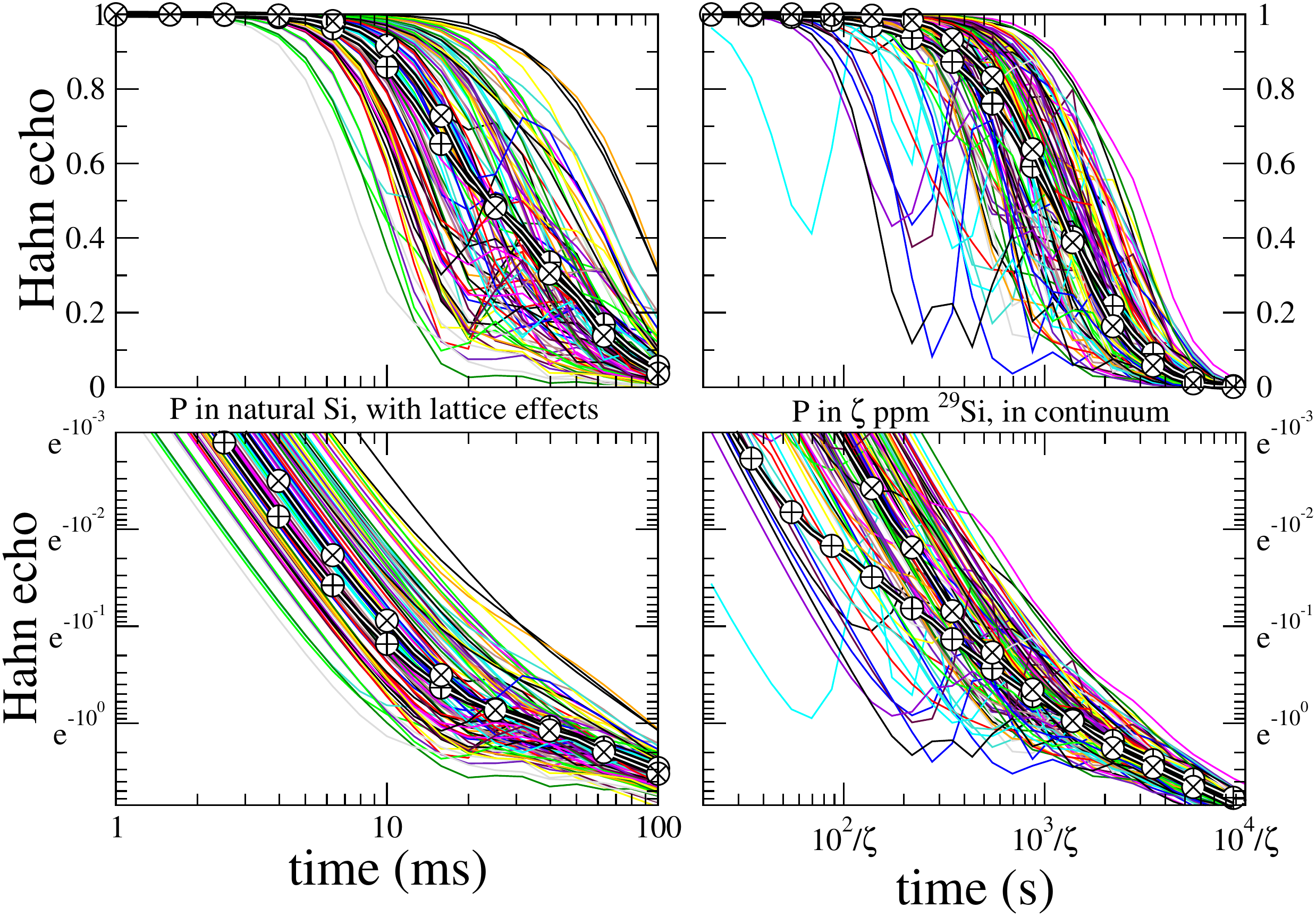}
\caption{
\label{Fig:PinSi}
$L_{\mbox{\scriptsize{CCE}}}^{(2)}$ Hahn spin echo results for a phosphorus
nuclear spin in a bath of $^{29}$Si.  Mean values
are encircled +'s and median values are encircled x's.  
The left panels are for natural Si with $4.67\%$ $^{29}$Si and includes lattice effects.
The right panels are for varied low concentrations of $^{29}$Si in the continuum limit.
The bottom panels present the same data as respective top panels but in a logarithmic scale.
}
\end{figure}

It has been proposed\cite{kane_architecture} to use donor nuclei for quantum memory storage in between quantum information processing.
Nuclear spins have much weaker magnetic moments than electrons and are
thus less susceptible to magnetic field noise, leading to  longer
coherence times.  It is interesting to note that the scenario of a
central nuclear spin in a nuclear spin bath, where the interactions
strengths with the central spin and among the bath are comparable,
is essentially the same as a central electron spin in an electron spin
bath.  Thus, our methods are applicable to nuclear spin decoherence that is
caused by other nuclei.
In Fig.~\ref{Fig:PinSi} we present calculated results for
the Hahn echo decay of a nuclear spin of a phosphorus donor in silicon
with $^{29}$Si as bath spins.  The left panels are for natural Si and 
include lattice effects while the right panels apply to varied $^{29}$Si concentrations in the continuum limit.
The range of $T_2$ values for different spatial configurations of the $^{29}$Si are 
consistent with recently reported single P nucleus $T_2$ measurement of roughly 
$60~\mbox{ms}$ (see Ref.~\onlinecite{morello-QCPR}) and $30~\mbox{ms}$
(see Refs.~\onlinecite{dreher-SiWorkshop} and \onlinecite{dreher-PinSi}).
The right panels are analogous to the 2-cluster part of
Fig.~\ref{Fig:SpatialEnsemble}; however, it differs from the canonical
problem in that the central spin gyromagnetic ratio is roughly a
factor of two larger than that of the bath spins.
The variant that is explored in Fig.~\ref{Fig:CompareG0} of
Sec.~\ref{Sec:VariedCouplingAndTemp} is applicable here.  For that variant, we found that
the cluster expansion has better convergence when the central spin has
a larger gyromagnetic ratio (e.g., g-factor) than the bath spins.  For
that reason, the decay is well approximated with 2-clusters.

\subsection{Quantum dot in silicon}
\label{Sec:DotApp}

Quantum dot electron spins are promising for qubit realizations.  In particular, using a singlet/triplet encoding for two electrons in double quantum dots, demonstrated in GaAs~\cite{Petta_Science05,Foletti_NP09,Bluhm_NP11}, fast single-qubit operations with electrical controls and readout\cite{Elzerman_Nature04,Barthel_PRL09} are possible.  In silicon, quantum dot spin qubits are starting to be realized,\cite{shaji_spin_2008,Lim_APL09,Nordberg_APL09,Lai_SR11,Simmons_PRL11,thalakulam_single-shot_2011,shi_tunable_2011,borselli_pauli_2011,maune_coherent_2012} and long coherence times are possible, particularly with isotopic enrichment of Si\cite{Wild_APL12} (and also Ge in Si/SiGe quantum dots\cite{Witzel_SiGe}).

Even with very high isotopic enrichment, as with the silicon donor
qubits discussed in Sec.~\ref{Sec:DonorApp}, coherence times will be
limited by a background of impurity phosphorus donors.  In
Sec.~\ref{Sec:CentralGeom}, we found that the effects of the spatial
extent of a quantum dot electron in such a spin bath are negligible up
to concentrations of $10^{14}/\mbox{cm}^3$.  So at very achievable
levels of silicon purity, the wave function extent is insigificant and the background phosphorus induced decoherence problem is essentially the same as that of Sec.~\ref{Sec:DonorApp} and the two-species variant (due to the two P donor nuclear polarizations) of the canonical problem as presented in Fig.~\ref{Fig:Resonances}.

In addition to background phosphorus donors, 2-D electron spin baths
may be present.  For example, Si/SiGe quantum dot structure may employ
modulation doping layers with some fraction of un-ionized donors.  For
these effects, we refer to Sec.~\ref{Sec:BathAndCentralGeom} where we
study electron spin decoherence for a quantum dot extended wave function with a 2-D electron spin bath at various distances.  This study has important implications for tolerated densities and distances of such a bath in order to achieve desired decoherence times.

\begin{figure}
\includegraphics[width=3.5in]{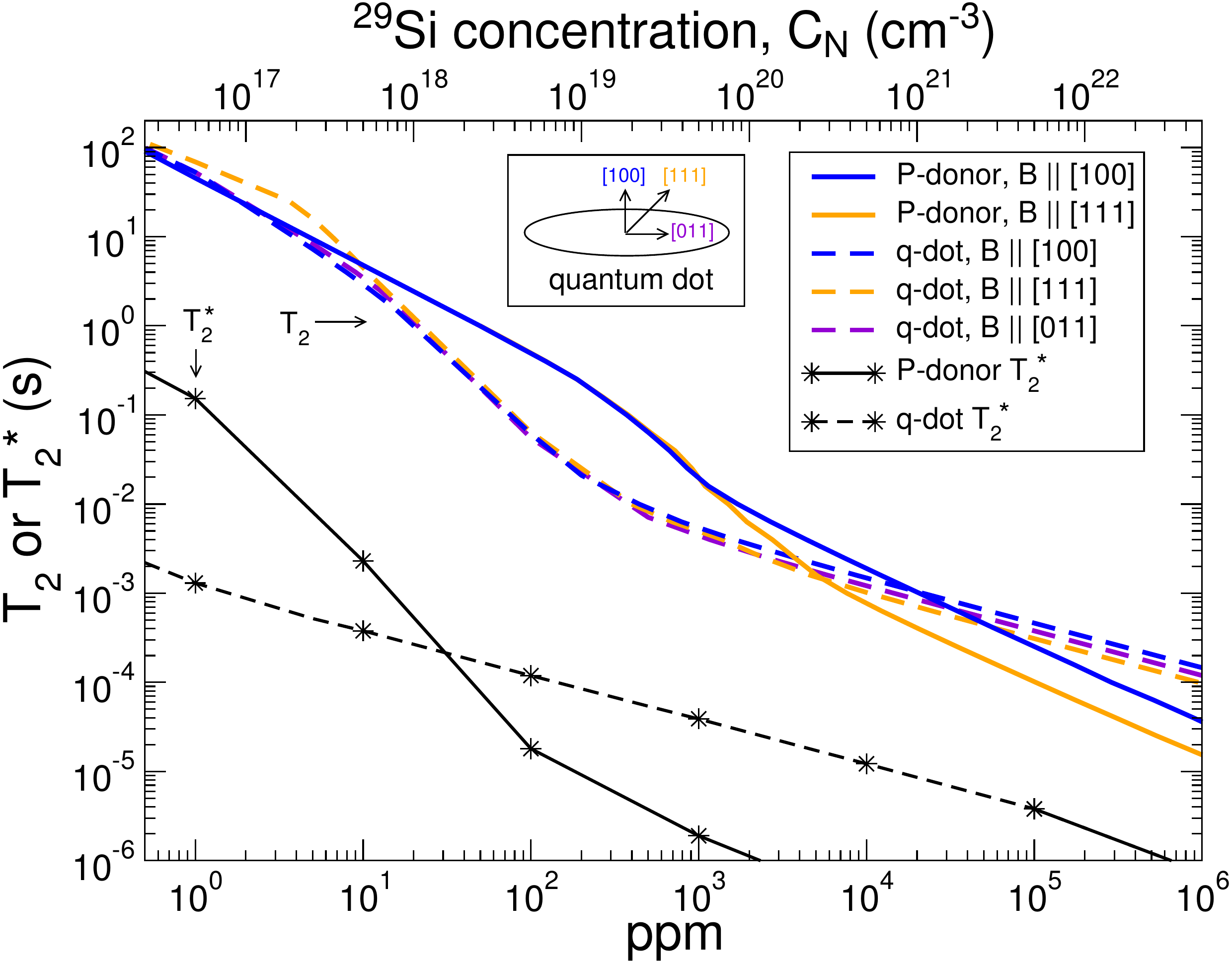}
\caption{
\label{Fig:qDotT2vs29Si}
Computed Hahn spin echo $T_2$ times, when the Hahn echo reaches a values of
$\exp{(-1)}$, as well as $T_2^*$ times, when the ensemble FID decay
reaches $\exp{(-1)}$, for P donors as well as quantum dots (q-dots)
defined with the probability density of Eq.~(\ref{Eq:dotProbDensity})
in Si.  We consider decoherence induced by various concentrations of
$^{29}$Si.  For our quantum dot model, we chose the lattice
orientation such that [100] corresponds with the $x_3$ direction
(normal to the confinement well) as depicted. 
}
\end{figure}

It is also important to understand how the effects of various $^{29}$Si concentrations may differ for a quantum dot compared with the donor qubit discussed in Sec.~\ref{Sec:DonorApp}.
In Fig.~\ref{Fig:qDotT2vs29Si}, we present decoherence time versus
$^{29}$Si concentration in contrast with results for the donor qubits
and observe the effects of the difference in the qubit wave-function
shape.  At very low concentrations, the wave-function shape is
irrelevant and a point-like model works for either type of qubit.  At
moderate concentrations, the laterally extended quantum dot qubit has
increased coupling to a number of bath spins so that decoherence times
are faster than those of the donor qubit.  This is the same effect
observed in Fig.~\ref{Fig:PancakeDipData} of
Sec.~\ref{Sec:CentralGeom} for an extended central electron spin in an
electron spin bath at $10^{15}/\mbox{cm}^3$.  At higher
concentrations, the situation is reversed.  The effect here is
analogous to what was observed in Fig.~\ref{Fig:3nm} of
Sec.~\ref{Sec:BathAndCentralGeom} for the effects of a laterally
extended central spin near a 2-D bath; the common cause is a reduced
sensitivity to flip-flopping of neighboring bath spins.

At higher concentration, the decoherence is affected by the magnetic
field angle relative to the lattice orientation, having the shortest
coherence times when the $B$ field is aligned with the nearest neighbor
direction.  For the quantum dot wave function, this effect is less
pronouned compared with P donors but is still appreciable.  
Around 2 to 4 ppm of $^{29}$Si, the quantum dot case with $B$ parallel to [111]
is affected by the anisotropy of the dipolar interactions in an
unusual way that causes $T_2$ to be long compared with other $B$-field
directions.  With just these few exceptions, however, $T_2$ is not
strongly dependent upon the $B$-field direction.

We also plot $T_2^*$ in Fig.~\ref{Fig:qDotT2vs29Si} for P donors and
quantum dots.  Above roughly $30$ ppm $^{29}$Si, the quantum dot exhibits longer $T_2^{*}$ times 
than the P donors as expected from the central limit theorem that
predicts longer $T_2^{*}$ for wave functions with greater spatial extent: $T_2^{*} \propto \sqrt{N}$.  The central limit theorem, however, does not apply well to the donor case, particularly at low densities.  In fact, the magnetic field shift probability distribution for point dipoles in the top left of Fig.~\ref{Fig:PulseSeq} fits a Lorenzian distribution much better than a Gaussian distribution.  In the low concentration regime, the roles are reversed; $T_2^{*}$ is shorter for the quantum dot 
that has a few strong $^{29}$Si contact hyperfine interactions than for the donor with negligible $^{29}$Si contact hyperfine interactions.
At low enough concentrations, the extent of the qubit
wave function should be negligible in either case and the $T_2^*$ times
should be the same, but that regime is well below the range we
present.  It is interesting to note that we do approach that regime for
$T_2$ but not $T_2^*$.  To understand this, consider that all bath
spin give direct contributions to $T_2^*$ while, in our large $B$-field
limit, bath spins contribute to $T_2$ only indirectly via near-resonant
flip-flopping pairs.

The effect of the magnetic field angle is negligible for the $T_2^{*}$ data 
we present.  This is expected where the isotropic contact hyperfine 
interaction dominates.  It is also expected in the low concentration limit 
where the extent of the wave function has a negligible upon $T_2^{*}$; in the presented range, this limit is approached for the donor but not the quantum dot.

\subsection{NV Center in Diamond}

Nitrogen vacancy (NV) defect centers in diamond form remarkable qubits
which may be coherently controlled at room
temperature.\cite{Hanson_PRB06, Wrachtrup_JPC06,Childress_Science06,Hanson_Science08,takahashi_quenching_2008,Balasubramanian_NM09,deLange_Science10,buckley_spin-light_2010,huang_observation_2011}
Electron spins of nitrogen atoms known as P1 centers in typical
diamond are a major source of decoherence.  At concentrations of below
200 ppm ($3.5\times 10^{19}/\mbox{cm}^{3}$), spin echo decoherence times of about
$T_2 \sim 3\mu$s have been observed at room temperature.\cite{deLange_Science10}
High-purity diamond with low concentrations of nitrogen decohere due
to $^{13}$C nuclear spins with spin echo times of $T_2 \sim
13~\mu$s.\cite{Childress_Science06}  In isotopically enriched high-purity
diamond with 0.3\% $^{13}$C and paramagnetic defects below
$10^{13}/\mbox{cm}^3$, spin echo coherence as long as 
$T_2 = 1.8~\mbox{ms}$ has been measured.\cite{Balasubramanian_NM09}

\begin{figure}
\includegraphics[width=3.5in]{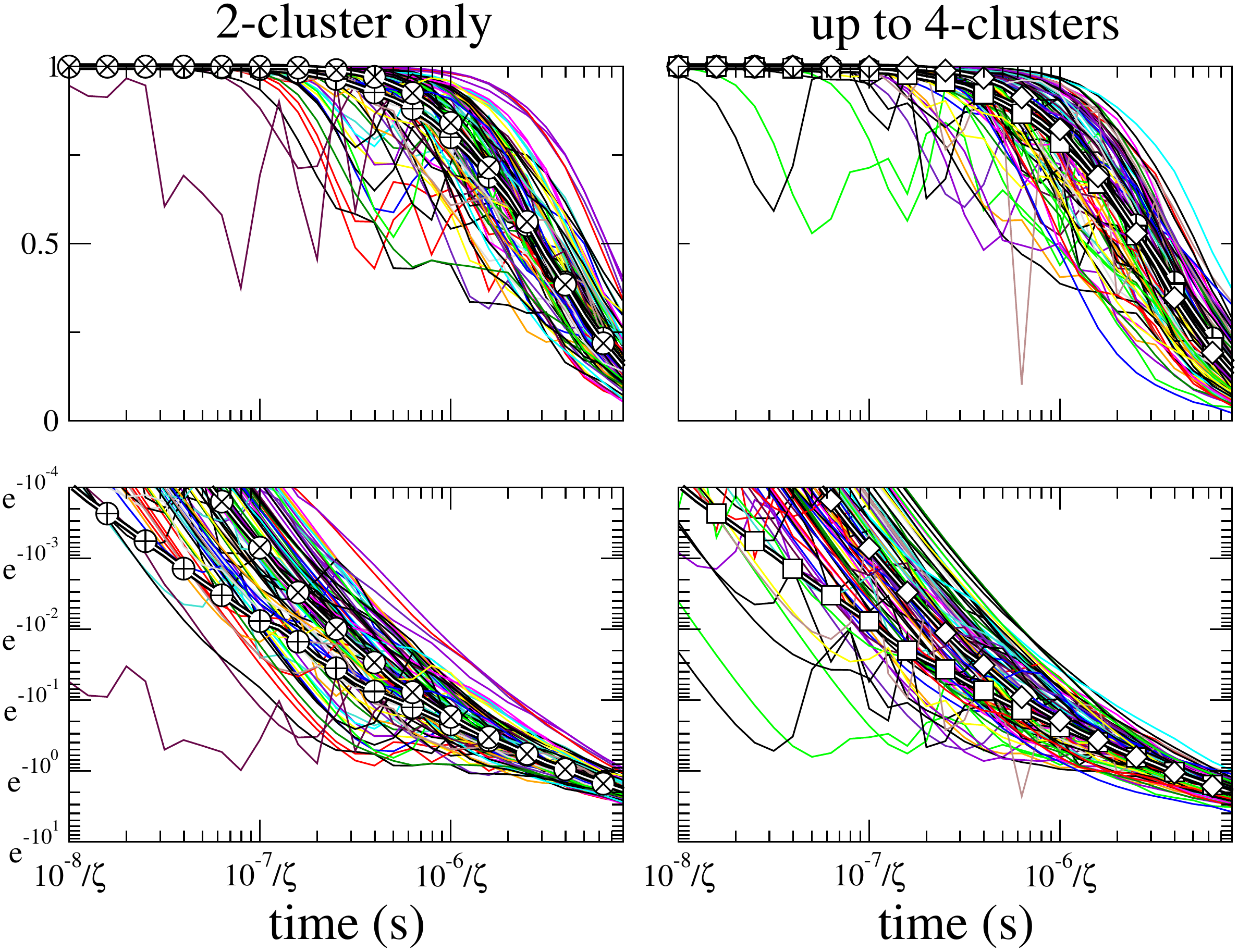}
\caption{
\label{Fig:NVinP1Bath}
Analogous to Fig.~\ref{Fig:SpatialEnsemble} but with the spin echo
results for an NV center in a bath of P1 centers at a concentration
of $C_E = \zeta \times 10^{19}/\mbox{cm}^{3}$.
Each ``spaghetti'' strand is the result for a different random spatial
configuration averaged over a large number of spin state templates.  
The mean of $L_{\mbox{\scriptsize{CCE}}}^{(2)}$ and
$L_{\mbox{\scriptsize{CCE}}}^{(4)}$ are represented
as encircled +'s and squares, respectively.
The median of $L_{\mbox{\scriptsize{CCE}}}^{(2)}$ and
$L_{\mbox{\scriptsize{CCE}}}^{(4)}$ are represented
as encircled x's, and diamonds, respectively.
The bottom figures show the same spin echo results on a logarithmic
scale for the decay.
}
\end{figure}

The NV decoherence problem in the presence of P1 centers in a large
magnetic field is similar to our canonical problem.  There are a few
important differences.  Details of the interactions of the NV and P1
center system appear in the supplemental online material of
Ref.~\onlinecite{Hanson_Science08}.  The important features for our
consideration are the following.  The NV center is treated as a
localized spin with $S_0 = 1$ and $g=2$ (the free electron g-factor).
The two-level qubit system in Ref.~\onlinecite{Hanson_Science08}, and
in our calculation, is the $m_s = \{0, -1\}$ subspace.
The P1 centers are spin 1/2 electrons (with $g=2$) bound to spin 1
nitrogen nuclear spins.  Each center has a delocalization axis
indicating the neighboring carbon atom is sharing the electron with
the nitrogen atom.  This delocalization axis changes over a time that
is much longer than the characteristic time of a single experimental
run in Ref.~\onlinecite{Hanson_Science08}.
In the large magnetic field regime, e.g.,
$B = 740~\mbox{G}$, as in Ref.~\onlinecite{Hanson_Science08}, the delocalization axis
simply determines a hyperfine energy coupling of the P1 centers: $A_1=114$~MHz
for the $[111]$ axis and $A_1 = 86$~MHz for $[\bar{1}11]$, $[1\bar{1}1]$, or
$[11\bar{1}]$ where $A_1 S_k^z I_k^z$ is the hyperfine energy shift
with $S_k$ as a P1 electron spin operator and $I_k$ as its nitrogen
spin operator.  We therefore have five different species in our spin
bath corresponding to five different hyperfine shifts with various
population percentages: $1/12$ fraction each for $A_1=114$~MHz and
$I_k^z = \pm 1$, $1/4$ fraction each for $A_1=86$~MHz and $I_k^z = \pm
1$, and $1/3$ fraction for $I_k^z = 0$.  For our high-field limit
calculation, we neglect flip-flopping between spins of different
species; these are off-resonant with each other.  
We can also safely neglect flip-flopping between the NV
center and any bath spin.  Our results in Fig.~\ref{Fig:NVinP1Bath}
show $T_2 \sim 3~\mu$s for a P1 center concentration of $10^{19} /
\mbox{cm}^3$, consistent with measurements reported in
Refs.~\onlinecite{Hanson_Science08} and \onlinecite{deLange_Science10} where $10^{19} - 10^{20} / \mbox{cm}^3$ concentrations are estimated.

\section{Conclusion}
\label{Sec:Conclusion}
We have presented a detailed account of a cluster-based theory of spin echo decoherence of a central spin (qubit) interacting dipolarly with a bath of spins of the same kind (i.e.~we focused on the case of the qubit-bath coupling being the same as the intrabath coupling). While the previously developed cluster theories were proven to be very successful in the case of the intrabath coupling being much weaker than the qubit-bath coupling, the all-dipolar problem with symmetric couplings requires the cluster-based approach to be nontrivially modified. We have shown that the decay of the spin echo signal can be calculated reliably by solving for evolution of finite groups (clusters) of bath spins coupled to the central spin, provided that the offsets of the splittings of these spins cause by dipolar interactions with the rest of the bath are properly taken into account. In a sparse bath the disorder in energy splittings of bath spins leads to localization of flip-flop dynamics, i.e.~at the timescale at which the qubit's coherence decays it is enough to consider the dynamics of still rather small clusters (up to 4-6 spins). This result does not follow from any kind of simple perturbative argument, and while it could have been suspected, the existence of such an effect had to be checked by careful numerical simulations. 

Our theory allows for quantitative evaluation of decoherence in an
all-dipolar system of spins.  
It presents a microscopic (i.e.~derived from the Hamiltonian of the system) solution to the original spectral diffusion problem, which has been approached by phenomenological or semi-phenomenological stochastic theories for more than fifty years. We have presented a broad selection of realistic applications of this theory, including calculations of spin echo decay for (1) electrons bound to phosphorus donors in isotopically purified silicon (reported previously in Ref.~\onlinecite{Witzel_PRL10} and recently confirmed experimentally in Ref.~\onlinecite{Tyryshkin_2011}); (2) nuclear spin qubits in silicon; (3) quantum dots in isotopically purified silicon; and (4) nitrogen-vacancy (NV) centers in diamond, in the case in which the nitrogen spins are the dominant source of decoherence. 
Although we have mainly emphasized Hahn
spin echo decoherence, our technique is very general and can be
applied to any quantum control context in principle.  
Together with previous works on cluster theories of decoherence due to dipolar interactions among the bath spins\cite{Witzel_PRB05,Witzel_PRB06,Yao_PRB06} (in which the qubit-bath hyperfine coupling was much stronger than the intra-bath interaction), the theory from this paper completes a body of work devoted to realistic calculations of decoherence in systems in which the dipolar interactions within the bath play a dominant role.

The temporal nature of these decoherence problems is an important
aspect for allowing cluster expansions to succeed.  The perturbative
arguments always have a factor of time accompanying interaction
energies.  As time increases, the effective perturbation parameter
increases, reducing the performance of the perturbative (cluster)
expansion.  At long times, we observe that the expansion fails
entirely.  We consider a cluster expansion successful, however, when
it is well-behaved and convergent on the timescale of the decay.
Furthermore, only the initial part of the decay is of interest for
typical quantum computing applications.  Although we do not
prove formal convergence in the cluster expansion or our selection
heuristics, we demonstrate good convergence in practice going out to
the 6-cluster order of the expansion provided that simulation times
are sufficiently short.

While a standard desktop computer is capable of producing many of the results that we present here in a reasonable amount of time (hours for a 2-cluster spin echo decay of a typical scenario with good accuracy), we made significant use of Sandia's high performance resources to acquire accurate results for the wide range of scenarios in our study as well as a lot of experimentation with our methods and heuristics.  Runtimes increase significantly with increasing cluster size.  The cluster selection heuristics we use are critical for making larger cluster calculations feasible.  Fortunately, the calculations are easy to parallelize with each processor treating a particular instance of a random bath instantiation.

Possible future directions will be to probe spectral densities of the
bath directly using cluster techniques and to answer questions
regarding conditions under which the bath may be treated independently
from qubit control (i.e., a classical bath).  Spectral density
descriptions are very convenient and are commonly probed in
experiment. Whether or not a bath acts classically has important
implications for quantum control.  The connection to
Ornstein-Uhlenbeck noise in our all-dipolar model that others have
noted and we confirm does imply that a classical noise model is applicable in
some manner (i.e., for a particular spatial configuration averaged
over spin states).

\section*{Acknowledgements}

We acknowledge Viatcheslav Dobrovitski, Andrea Morello, Alexei Tyryshkin, Steve Lyon,
Kevin Young, Rajib Rahman, Erik Nielsen, and Richard Muller 
for valuable discussions
and insights.  Sandia National Laboratories is a multi-
program laboratory operated by Sandia Corporation, a
wholly owned subsidiary of Lockheed Martin Corpora-
tion, for the U.S. Department of Energy's National Nu-
clear Security Administration under contract DE-AC04-
94AL85000. S.D.-S. and {\L}.C. acknowledge an LPS-NSA-CMTC grant; {\L}C acknowledges funding from the Homing Programme of
the Foundation for Polish Science supported by the EEA
Financial Mechanism.

\appendix

\section{Cluster Sampling Heuristics}
\label{clusterSampling}
The relative importance of clusters may be judged, heuristically, by
the following two factors: the strength of coupling between the central
and bath spins of the cluster, and the strength of coupling among the
bath spins of the cluster.  We address the first factor with a radial
cutoff.  Since interactions decrease with increasing distance, we'll
ignore clusters that are fully outside of some cutoff radius, $R_C$ relative
to the central spin.  
For the remaining clusters, those with at least
one bath spin within $R_C$ from the central spin, we
apply a heuristic to address the second consideration (the coupling
strength among spins).  Motivated by Lemma~1 of
Sec.~\ref{CCEconvergence}, which is not completely valid when using
the interlaced spin state averaging of Sec.~\ref{spinStateAvg} but
should still have some approximate validity, we want clusters that may
be fully connected with sufficiently strong interactions.  We will
therefore assign a heuristic strength of a cluster to be the smallest
interaction neccessary to complete the connectivity of the cluster (as
in Fig.~\ref{connectedCluster}).  Algorithmically, we may compute this by
successively picking off the strongest interaction among bath cluster
spins until the cluster has full connectivity from the picked
interactions; the last interaction picked (the smallest necessary
interaction) is the heuristic strength of the cluster.

In addition to $R_C$, we also employ a cutoff for the
maximum number, $N_k$, of clusters to be
considered for each cluster size, $k$.  
For each CCE computation (which is typically
averaged over bath spin locations and initial polarizations) and for
each cluster size $k$ (up to some maximum), we
select the $N_k$ clusters of the highest heuristic strength for
evaluation.
In the donor spectral diffusion application, we use one more cutoff in
which we only consider, for the sake of cluster selection,
intereractions among donors with comparable Overhauser shifts; this is
a resonance energy cutoff, $E_C$.

\begin{figure}
\includegraphics[width=2.5in]{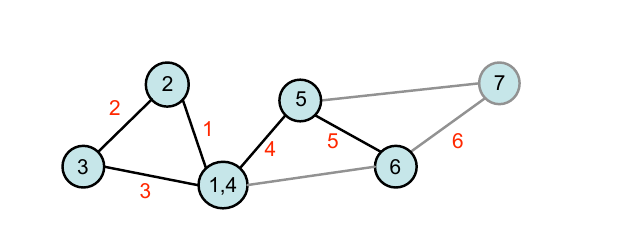}
\caption{
\label{lemma2Illustration}
Illustration of the constructive proof for Lemma 2, showing a path,
with numbered edges and vertices, that traverses the entire graph 
${\cal G}$.  A connected sub-graph ${\cal H}$ is formed from the edges and
vertices of the path with the exclusion of the final edge and vertex.
The edges and vertex that are excluded from ${\cal H}$ are in gray.
}
\end{figure}

The key to an efficient algorithm for finding each set of $N_k$ clusters,
one that avoids iterating over the potentially vast number of clusters
not be evaluated, is to recognize that a valid
cluster (containing a spin that is within $R_C$) of some heuristic
strength may be built by adding a single spin to a valid sub-cluster
of at least this same strength.  This is formalized in the following theorem:
\begin{theorem}
For any cluster, ${\cal S}$, with a particular heuristic strength $s$
and any spin in that cluster, $i$, 
there exists a
sub-cluster ${\cal C}$ containing $i$ 
of size $\|{\cal C}\| = \|{\cal S}\| - 1$ with
a heuristical strength $c \geq s$.
\end{theorem}
This theorem follows from the heuristic strength definition ensuring
that  ${\cal S}$ may be connected by
interactions of strength $s$ or greater and the following Lemma:
\begin{lemma}
For any connected undirected graph, ${\cal G}$, and contained vertex,
$v$,
there exists a connected
sub-graph, ${\cal H}$, containing $v$ but having one fewer vertex.
\end{lemma}
Lemma 2 has a simple constructive proof.  Starting from $v$, 
traverse ${\cal G}$ until all vertices are visited; it
doesn't matter if any vertices or edges are traversed multiple times
as long as we stop at the point at which all vertices have been visited.
Since ${\cal G}$ is connected, this must be possible. Because our
stopping point comes once all vertices have been visited, the last
vertex must have been visited only once.  A sub-graph, ${\cal H}$,
with the desired property from Lemma 2 is formed by the vertices and
edges of the traversed path excluding the final traversed edge and
final visited vertex.  This is illustrated in Fig.~\ref{lemma2Illustration}.

\begin{figure}
\includegraphics[width=3.5in]{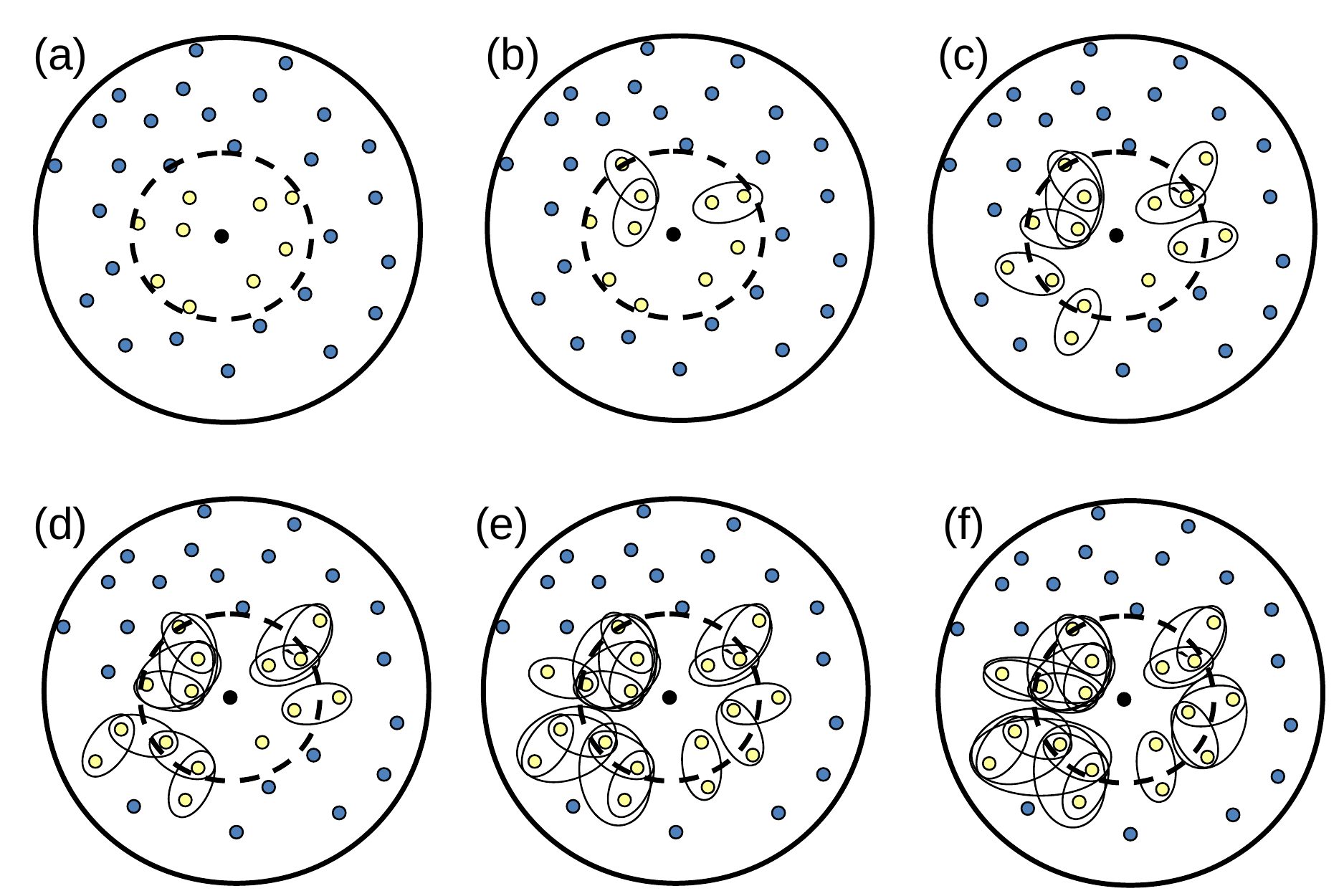}
\caption{
\label{Fig:GrowingClusters}
Schematic illustration of our algorithm to find the strongest
clusters, the set of desired clusters $D$,
with $N_k$ quotas (in this example, 13 2-clusters, 7 3-clusters, and 2
4-clusters).  The dashed circle denotes the $R_C$ cutoff.  The
black dot in the center represnent the central spin and other dots
represent bath spins.  (a) shows
the initial population of 1-clusters within $R_C$ that must seed all
selected clusters.  (a)-(f) shows a progression of cluster selections,
skipping a few selections at a time.  The clusters grow from
previously considered clusters (which may are may not be in $D$
depending on the $N_k$ quotas) based upon the interactions strengths
of the spin system being represented.
}
\end{figure}

Given Theorem 2, we may find the clusters of size $k$ with the highest
heuristic strength by constructively building from clusters
of size $k-1$ with the highest heuristic strength.
The algorithm works with three different lists: strongest clusters
$S$,
potential clusters $P$, and the desired clusters $D$.  
 The algorithm starts by adding into $S$ all clusters
of size one for each of the spins within the $R_C$ cutoff
and proceeds as follows:
\begin{enumerate}
\item Take the strongest cluster ${\cal C}$ off of list $P$
(which should be kept sorted) and add it to $S$.
\item If $D$ does not yet contain $N_k$ clusters of size $k = \|{\cal
  C}\|$, add ${\cal C}$ to this $D$ list as well.
\item Add into $P$ any new cluster ${\cal C}'$ (not already contained in $S$)
  that may be generated by extending ${\cal C}$ by one spin.  
\item Optionally, to minimize memory usage, remove all clusters that
  are too weak to be relevant (i.e., clusters that cannot compete or
  be built upon to compete for one of the $N_k$ spots of strongest
  $k$-clusters for any $k$).
\item While $\|D\| < \sum_k N_k$, repeat from step 1.
\end{enumerate}
At the end of this process, $D$ will contain, for each $k$, the $N_k$
strongest clusters of size $k$.  A schematic example of this process is shown in Fig.~\ref{Fig:GrowingClusters}.

\section{Proof of Lemma~1}
\label{ClusterFactorization}
We will prove Lemma~1 using induction and contradiction.  As the
base case of our induction, note that $L_{\emptyset}$ and
$\tilde{L}_{\emptyset}$ are constants, which follows from our prerequisite
that $L$ and $\tilde{L}$ are constant when all coupling constants are
taken to be zero.
For convenience, let us denote the bath coupling power series as $L_{\cal S} =
f_{\cal S}(\{b_{i \in {\cal S}, j \in {\cal S}}\}) = f_{\cal S}$ and 
 $\tilde{L}_{\cal S} = \tilde{f}_{\cal S}(\{b_{i \in {\cal S}, j \in {\cal
    S}}\}) = \tilde{f}_{\cal S}$.  The $b_{i, j}$ parameters of $f_{\cal S}$ and
$\tilde{f}_{\cal S}$ are occasionally dropped for convenience but are
still implied (they should be regarded as power series functions of
$b_{i, j}$).
By Eqs.~(\ref{CCEeqns}) and (\ref{DefTildeL}),
\begin{eqnarray}
\label{CCEinF}
f_{\cal S} &=&
\prod_{{\cal C} \subseteq {\cal S}} \tilde{f}_{\cal C}, \\
\label{DefTildeF}
\tilde{f}_{\cal S} &=& 
f_{\cal S} / \prod_{{\cal C} \subset {\cal S}} \tilde{f}_{\cal C}.
\end{eqnarray}

By induction, let us assume that $\tilde{f}_{\cal C}(\{b_{i \in {\cal
 C}, j \in {\cal C}}\})$ obeys the connected graph lemma for all
 ${\cal C}$ whose size is less than $k$.  
For the contradictory part of the proof, we
 assume that the lemma does not hold for some 
$\tilde{f}_{\cal S}(\{b_{i \in {\cal S}, j \in {\cal S}}\})$ with $\|{\cal
 S}\| = k$.  Thus, there exists in $\tilde{f}_{\cal S}(\{b_{i \in
   {\cal S}, j \in {\cal S}}\})$ some non-constant term, $g(\{b_{i \in {\cal S}, j \in
 {\cal S}}\})$, and non-empty disjoint sets ${\cal X}, {\cal Y}
 \subset {\cal S}$, such that $g(\{b_{i \in {\cal S}, j \in
 {\cal S}}\})$ does not depend on any $b_{i \in {\cal X}, j \in
 {\cal Y}}$.  Such a term is therefore unaffected if we impose that
 all $b_{i \in {\cal X}, j \in {\cal Y}}$ be zero.  By our assumed
 factorability property of $L_{\cal S} = f_{\cal S}(\{b_{i \in {\cal S}, j \in {\cal S}}\})$, then
\begin{equation}
\label{factoredF}
\left. f_{\cal S}(\{b_{i \in {\cal S}, j \in {\cal S}}\}) \right|_{b_{i \in {\cal X}, j \in
 {\cal Y}} = 0} =
f_{\cal X} f_{\cal Y}. 
\end{equation}
By inductive reasoning, for all ${\cal C} \subset {\cal S}$ such that
${\cal C} \cap {\cal X} \neq \emptyset$ and ${\cal C} \cap {\cal Y} \neq \emptyset$,
\begin{equation} 
\label{constTildeF}
\left. \tilde{f}_{\cal C}(\{b_{i \in {\cal C}, j \in {\cal C}}\}) \right|_{b_{i \in {\cal X}, j \in {\cal Y}} = 0}  = \mbox{const}
\end{equation}
since all non-constant terms would contain 
$b_{i \in {\cal X}, j \in {\cal Y}}$ factors that are taken to be zero.
From Eqs.~(\ref{DefTildeF}), (\ref{factoredF}), and (\ref{constTildeF}),
\begin{equation}
\left.\tilde{f}_{\cal S}(\{b_{i \in {\cal S}, j \in {\cal S}}\})\right|_{b_{i \in {\cal X}, j \in
 {\cal Y}} = 0} \propto 
\frac{f_{\cal X} f_{\cal Y}}
{\prod_{{\cal C}_1 \subseteq {\cal X}} 
\tilde{f}_{{\cal C}_1} \prod_{{\cal C}_2 \subseteq {\cal Y}}
 \tilde{f}_{{\cal C}_2}}.
\end{equation}
Applying Eq.~(\ref{CCEinF}) for ${\cal S} = {\cal X}$ and ${\cal S} =
{\cal Y}$, the numerator and denominator above will
cancel and we are left with a constant.
This is a contradiction since this function, by the
contradiction-proof assumption, should contain the
non-constant $g(\{b_{i \in {\cal S}, j \in {\cal S}}\})$ term that is
not affected by imposing that $b_{i \in {\cal X}, j \in {\cal Y}} = 0$.

%\bibliography{refs_quant_SD}
\bibliography{refs_quant,refs_Si_qd}
\end{document}